\documentclass[a4paper,11pt]{article}
\pdfoutput=1 % if your are submitting a pdflatex (i.e. if you have
\usepackage{jinstpub} % for details on the use of the package, please
\usepackage{subfigure}
\usepackage{amsmath}
\usepackage{gensymb}
\usepackage{multirow}

\title{The CALOCUBE project for a space based cosmic ray experiment: design, construction, and first performance of a high granularity calorimeter prototype.}

\author[a,b]{O.~Adriani,}
\author[c,d]{S.~Albergo,}   
\author[e,d]{L.~Auditore,}
\author[f,g]{A.~Basti,}
\author[a,b]{E.~Berti,}
\author[h,f]{G.~Bigongiari,}
\author[b]{L.~Bonechi,}
\author[a,b]{M.~Bongi,}
\author[i]{V.~Bonvicini,}
\author[b]{S.~Bottai,}
\author[h,f]{P.~Brogi,}
\author[c,d]{G.~Cappello,}
\author[j]{G.~Carotenuto,}
\author[k,b]{G.~Castellini,}
\author[l]{P.W.~Cattaneo,}
\author[h,f]{R.~Cecchi,}
\author[m,n]{C.~Checchia,}
\author[a,b,1]{R.~D'Alessandro, \note{Corresponding author.}}
\author[b]{S.~Detti,}
\author[o,p]{M.~Fasoli,}
\author[q,b]{N.~Finetti,}
\author[d]{A.~Italiano,}
\author[a,b]{P.~Lenzi,}
\author[d]{M.G.~Pellegriti,}
\author[h,f]{P.~Maestro,}
\author[b]{M.~Manetti,}
\author[h,f]{P.S.~Marrocchesi,}
\author[b]{N.~Mori,}
\author[f]{F.~Morsani,}
\author[b]{M.~Olmi,}
\author[f]{A.~Orsini,}
\author[i]{G.~Orzan,}
\author[a,b]{L.~Pacini,}
\author[b]{P.~Papini,}
\author[l]{A.~Rappoldi,}
\author[k,b]{S.~Ricciarini,}
\author[r]{ A.~Sciuto,}
\author[b]{P.~Spillantini,}
\author[b]{O.~Starodubtsev,}
\author[h,f]{L.~Stiaccini,}
\author[h,f]{F.~Stolzi,}
\author[h,f]{A.~Sulaj,}
\author[h,f]{J.E~Suh,}
\author[a,b]{A.~Tiberio,}
\author[c,d]{A.~Tricomi,}
\author[e,d]{A.~Trifir\`o,}
\author[e,d]{M.~Trimarchi,}
\author[b]{E.~Vannuccini,}
\author[o,p]{A.~Vedda,}
\author[i]{G.~Zampa,}
\author[i]{N.~Zampa}

\affiliation[a]{Department of Physics and Astronomy, University of Florence,  Via G. Sansone 1, I-50019 Sesto Fiorentino (Firenze), Italy}
\affiliation[b]{INFN Sezione di Firenze, Via B. Rossi 1, I-50019 Sesto Fiorentino (Firenze), Italy}
\affiliation[c]{Department of Physics and Astronomy, University of Catania, Via S. Sofia 64, I-95123  Catania, Italy}
\affiliation[d]{INFN Sezione di Catania, Via S. Sofia 64, I-95123   Catania, Italy}
\affiliation[e]{MIFT, University of Messina, Viale F. Stagno d'Alcontres 31, I-98166  Messina, Italy}
\affiliation[f]{INFN Sezione di Pisa, Largo Bruno Pontecorvo 3, I-56127 Pisa, Italy}
\affiliation[g]{Department of Physics, University of Pisa, Largo Bruno Pontecorvo 3, I-56127 Pisa, Italy}
\affiliation[h]{Department of Physical Sciences, Earth and Environment, University of Siena, I-53100 Siena, Italy}
\affiliation[i]{INFN Sezione di Trieste, Padriciano 99, I-34149 Trieste,  Italy}
\affiliation[j]{IPCB-CNR, P.le Enrico Fermi 1, Portici, I-80055 Napoli, Italy}
\affiliation[k]{IFAC-CNR,Via Madonna del Piano 10, I-50019 Sesto Fiorentino (Firenze), Italy}
\affiliation[l]{INFN Sezione di Pavia, Via Agostino Bassi 6, I-27100 Pavia,  Italy}
\affiliation[m]{Department of Physics and Astronomy, University of Padova, Via F. Marzolo 8, I-35131 Padova, Italy}
\affiliation[n]{INFN Sezione di Padova, Via Francesco Marzolo 8, I-35121  Padova, Italy}
\affiliation[o]{Department of Materials Science, University of Milano-Bicocca, via Cozzi 55, I-20125 Milan, Italy}
\affiliation[p]{INFN Sezione di Milano-Bicocca, Piazza della Scienza, 3, I-20126 Milano, Italy}
\affiliation[q]{Department of Physics and Chemistry, University of L'Aquila, Via Vetoio 40, I-67100 L'Aquila, Italy}
\affiliation[r]{IMM-CNR, Str. VIII Zona Industriale, I-95121 Catania, Italy}

\emailAdd{candi@fi.infn.it}

\abstract{Current research in High Energy Cosmic Ray Physics touches on fundamental questions regarding the origin of cosmic rays, their composition, the acceleration mechanisms, and their production.
Unambiguous measurements of the energy spectra and of the composition of cosmic rays at the "knee" region could provide some of the answers to the above questions. So far only ground based observations, which rely on sophisticated models describing high energy interactions in the earth's atmosphere, have been possible due to the extremely low particle rates at these energies. 

A calorimetry based space experiment that could provide not only flux measurements but also energy spectra and  particle identification, would certainly overcome some of the uncertainties of ground based experiments. Given the expected particle fluxes, a very large acceptance is needed to collect a sufficient quantity of data, in a time compatible with the duration of a space mission. This in turn, contrasts with the lightness and compactness requirements for space based experiments. 

We present a novel idea in calorimetry which addresses these issues whilst limiting the mass and volume of the detector. In this paper we report on a four year R\&D program where we investigated materials, coatings, photo-sensors, Front End electronics, and mechanical structures with the aim of designing a high performance, high granularity calorimeter with the largest possible acceptance. Details are given of the design choices, component characterisation, and of the construction of a sizeable prototype (Calocube) which has been used in various tests with particle beams.}


\begin{document}
\maketitle
\flushbottom

\section{Introduction and detector concept}
\label{sec:intro}
\begin{figure}[htbp]
  \centering
  \includegraphics[height=.7\textwidth]{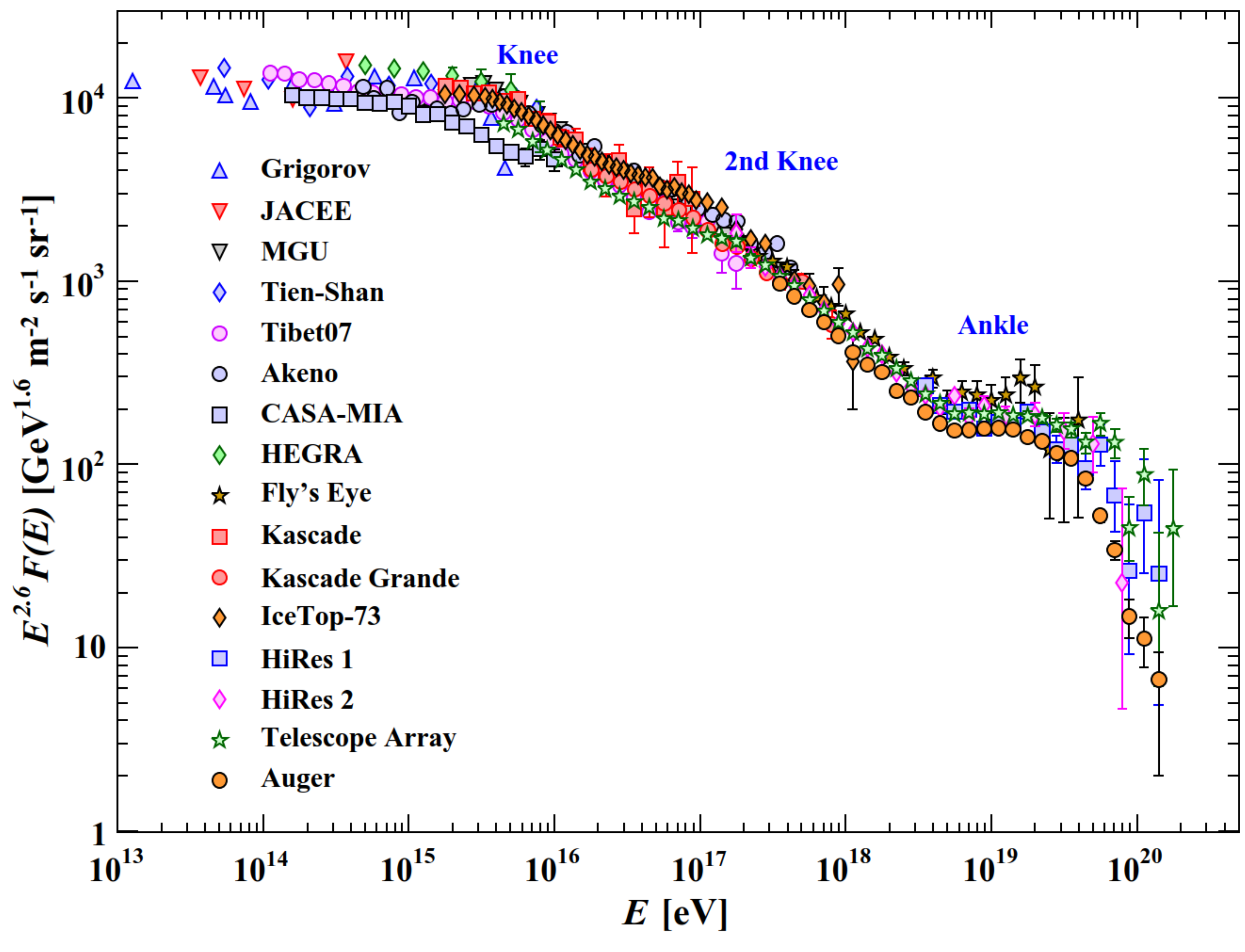}
%  \subfigure[\label{fig:knee02}]{\includegraphics[height=.4\textwidth]{Images/Cosmic_ray_spectra_02_new.pdf}}
  \caption{\label{fig:Knee} Cosmic ray energy spectrum. In evidence the slope changes (knee, ankle) and expected particle fluxes .}
\end{figure}

Protons and nuclei energy spectra from cosmic rays, show a "knee"
structure at around 1 PeV (see figure\ref{fig:Knee}, taken from~\cite{pdg2018}), which could signal a change of regime due to various possible causes, i.e. production mechanisms, an 
upper energy limit of galactic accelerators, an energy dependant cosmic ray composition, the presence of cutoffs in the transportation of cosmic rays. 
Spectral measurements in the knee region are derived from data collected by ground-based shower detectors 
\cite{Antoni:2005,Amenomori:2011,Bartoli:2014,Bartoli:2015,Auger:2011,Hires:2010,Auger:2013,Agasa:2005}
whose reliability would benefit greatly if direct, above the earth's atmosphere, measurements in the PeV region were to be made available.  As shown in figure~\ref{fig:Knee}, such a measurement faces enormous challenges due to the extremely low flux of expected particles at those energies (1 particle per m$^2$ per year). Thus an orbiting detector must provide not only a very good energy resolution for hadrons ($<$ 40\%) and a way to identify the cosmic ray atomic number, but must also have a large acceptance (few m$^2$sr). This last requirement is in direct antithesis to the constraints of space missions where volume and mass are at premium~\cite{Adriani:2013}. 
The design we present in this paper has a high granularity with a homogeneous segmentation both laterally and depthwise. This unique design allows to achieve an excellent energy resolution using shower shape reconstruction algorithms, an enhanced distinction between hadrons and electrons, and a large acceptance obtained by maximising the number of entrance windows to the detector thus keeping a reasonable mass and volume budget (1.6 tons, $<$1 m$^3$). In order to faithfully reconstruct the most energetic showers while maintaining sensitivity to Minimum Ionizing Particles (MIPs), a dynamic range on the order of 10$^7$ is required. This has been achieved with a dual photodiode readout and a novel Front End electronics design (see following sections).

From the start, we have proceeded with a calorimeter design optimised for space applications. Thus volume and mass constraints have always been considered of paramount importance (i.e. acceptance increases should rely only on novel approaches that do not necessarily involve a volume increase). 
The fundamental tenet of the design is a cubic detector in which 5 of the 6 sides are capable of particle detection and measurement in a totally symmetric and interchangeable way. Such a design effectively translates in a five fold acceptance increase while keeping the overall volume constant. \\
Mass constraints, on the other hand, translate in a maximum number of interaction lengths ($\lambda_{\rm{I}}$) which, depending on the material used, can be at most 2-3. A high granularity approach to the design, allows us to reconstruct a detailed image of each interacting particle shower profile. This not only helps us achieve a high discrimination between electrons and protons/nuclei, but also a relatively high energy resolution for hadrons using partial shower profile fits, given that for obvious limits, hadronic showers will not be fully contained.
\begin{figure}[htb]
\centering
\includegraphics[width=0.42\linewidth]{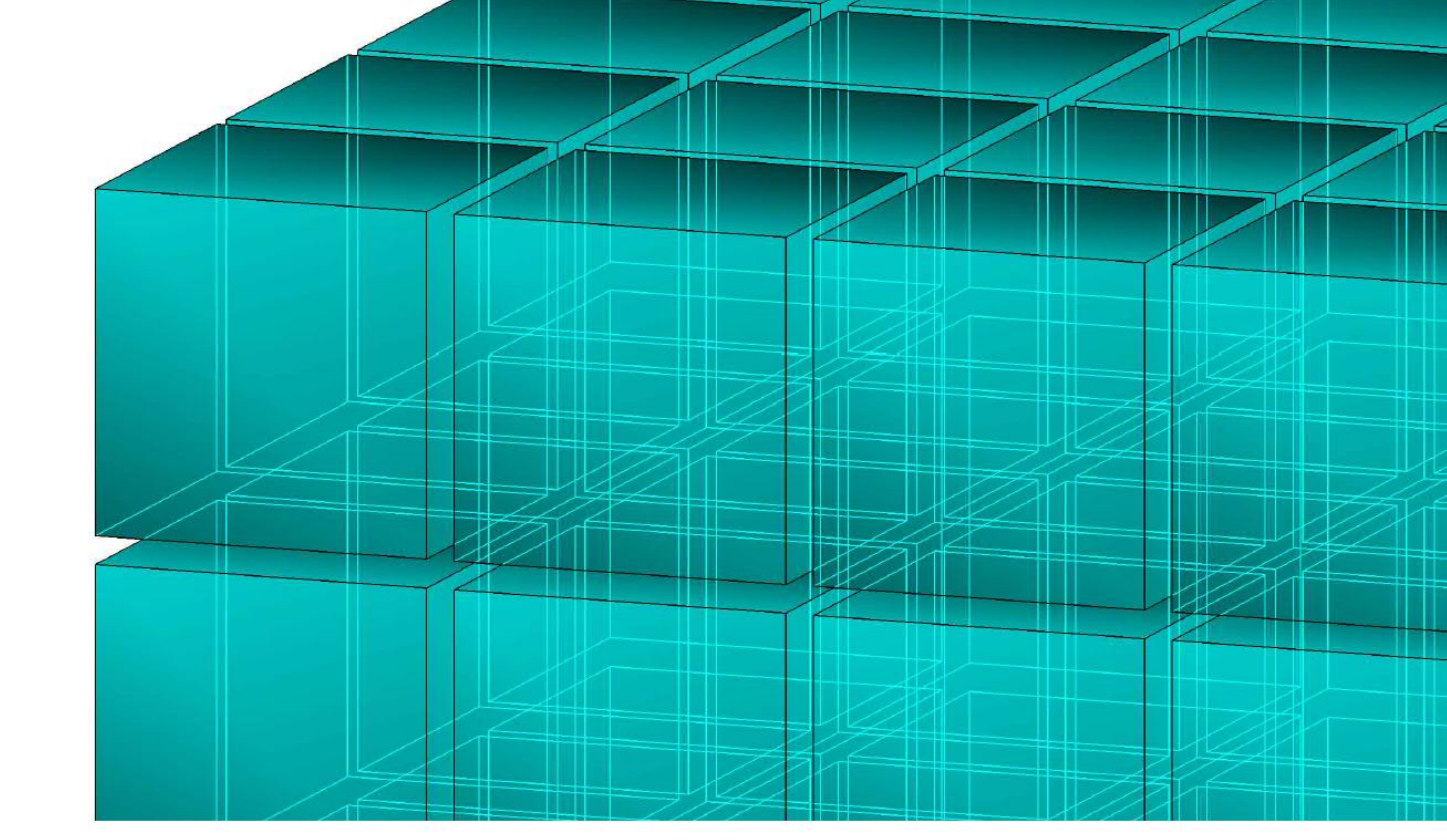}\ \ \ \ \ \ \ \ \ \ 
\includegraphics[width=0.37\linewidth]{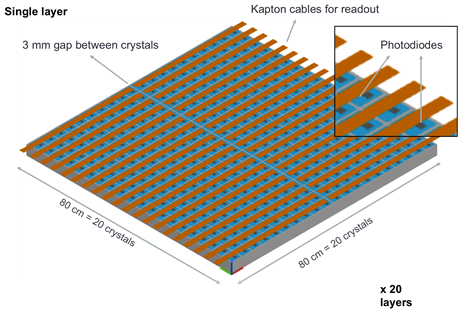}
\caption{Conceptual mesh design for the Calocube prototype (left). Drawing of a tray holding together a matrix of 20x20 crystals (right). Also depicted the Kapton PCBs bringing out the photodiodes signals.}
\label{fig:Calocube01}
\end{figure}

Conceptually the Calocube calorimeter consists of a mesh of cubic scintillator crystals (see figure \ref{fig:Calocube01}) 
arranged as a stack of trays similar to the one depicted in the figure  holding together a matrix of small crystals, each small cube having an edge of roughly one Moliere radius. The total number of radiation lengths (X$_0$) is of the order of 40 with a corresponding number of $\lambda_{\rm{I}}$  between 1 and 2 depending on the final design depth and scintillator material choice.
The gaps between each crystal are kept as constant as possible to avoid any non homogenous detector response. Not only are the crystals packed as closely as possible but also the trays. 

We have realised a large size prototype (using roughly 700 cubic crystals), with which we have taken data using  particle and ion beams at CERN. During its construction, various choices on detector material, mechanical structure, readout electronics, and diode sensors were investigated and validated  with the aim of a final design submission for a full scale detector to be flown on the chinese space station (HERD Mission~\cite{HERD:2019}).
Translating this design concept into a viable project that could be prototyped has not been an easy passage. To this end, our collaboration has investigated many solutions. We have performed detailed mechanical simulations on a number of design choices~\cite{calosimu}, we have characterised various scintillator materials, evaluated readout solutions for the scintillators, developed custom chips for the Front End electronics, and constructed a $\sim$700 crystal protoype that has been tested with particle beams.
The  following sections will detail these aspects of our R\&D with the results obtained, highlighting the ensuing design choices.

\section{Photodiode and scintillator choice}
\label{sec:photodiode}
\subsection{Introduction}

We based our scintillator choice mainly on the obtainable electromagnetic energy resolution, but also considering both the radiation and interaction lengths of the material. In fact, the need to provide even partial hadronic shower containment, while at the same time keeping mass, volume and costs at a reasonable scale, narrows down the available choices. 
To improve the energy resolution for hadronic showers, we also considered Cherenkov light detection in the same crystal, not unlike the DREAM \cite{DREAM} project at CERN. Unfortunately while we have presented results \cite{elena} \cite{CLASSIC} showing an improvement in the hadronic energy resolution, we have found that this becomes significant only in the case of full shower containment which is not our case. We will show some tests we performed in section \ref{sec:DualR}.

As described in the following sections, we performed several studies and tests on inorganic scintillators which we considered as plausible candidates for the Calocube calorimeter. Consequently we also tested the light collection efficiency for various  wrapping schemes both reflective and diffusive. 
We used commercial photodiodes (PDs) as photosensors (see sub-section below). Our single crystal tests relied on a high accuracy spectrometer by Amptek~\cite{Amptek:A250}, consisting of the charge-sensitive preamplifier A250 coupled to the digital pulse analyser PX5. For these tests, a $^{241}$Am source was used for both calibration and signal generation while some measurements were also performed with MIPS using atmospheric cosmic rays. In addition some further calibrations involved the use of LEDs at different wavelengths to illuminate directly the PDs. 

\subsection{Photodiodes}
\begin{figure}[htbp]
  \centering 
  \subfigure{\includegraphics[width=.32\textwidth]{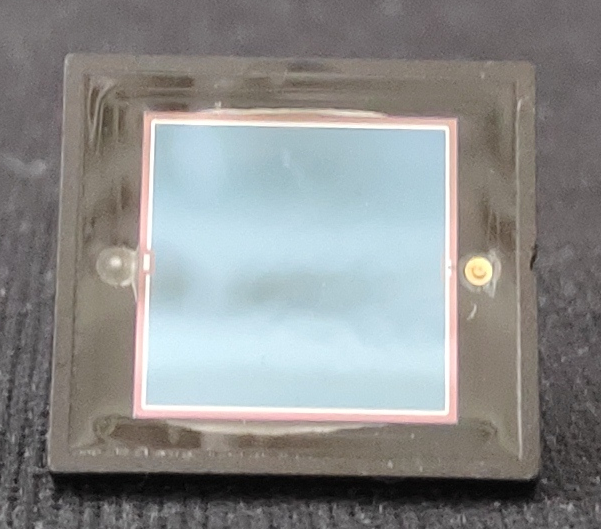}}~
  \subfigure{\includegraphics[width=.32\textwidth, trim=2 42 2 0,clip]{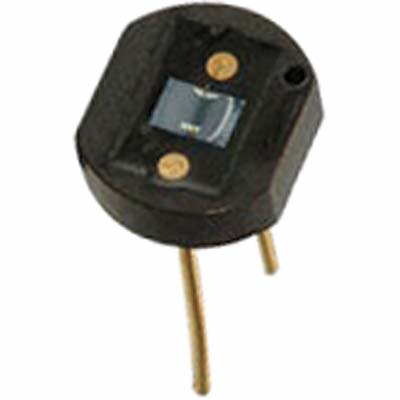}}
  
  %\hfill  
  \subfigure{\includegraphics[width=.32\textwidth]{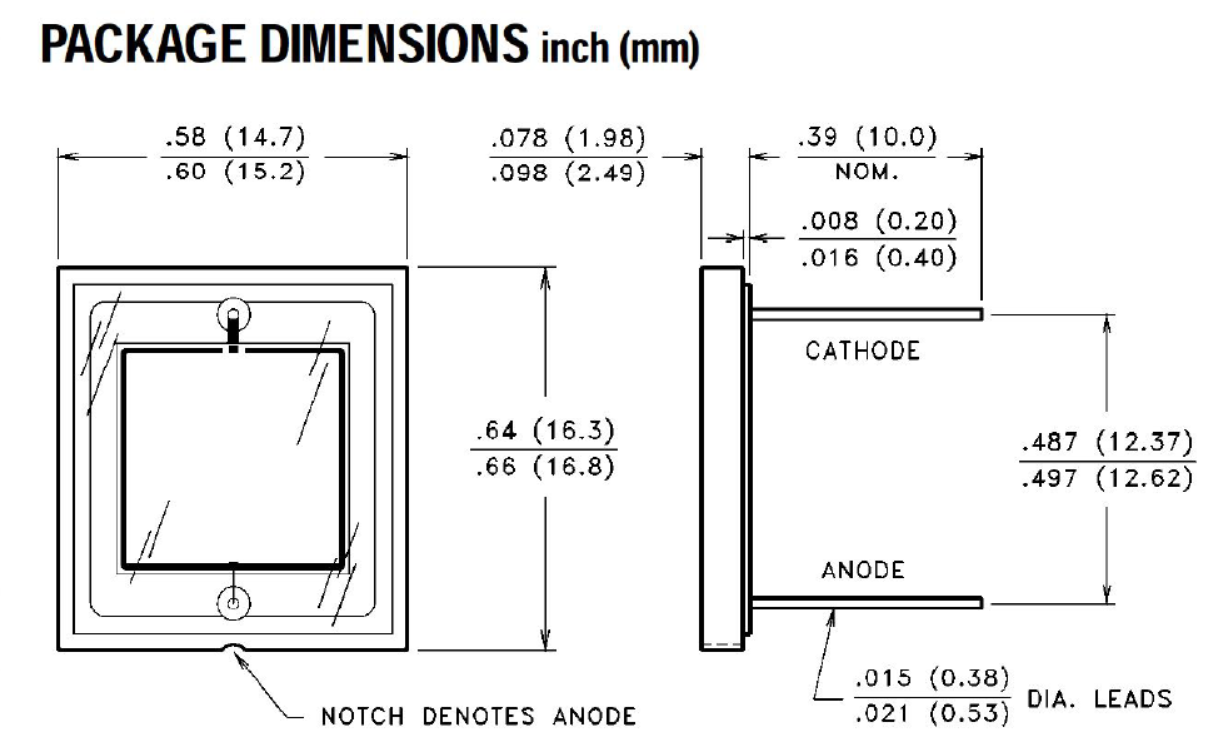}}~
  \subfigure{\includegraphics[width=.32\textwidth]{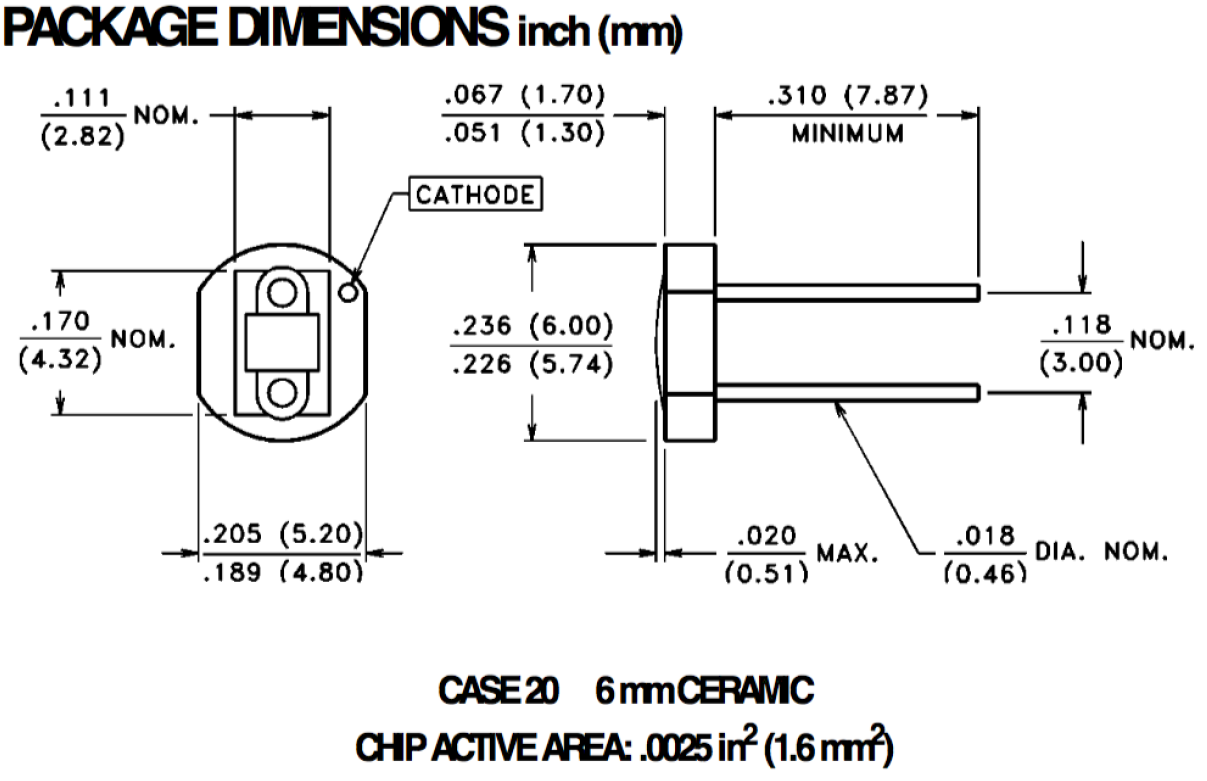}}
  \caption{\label{fig:photoPD}Excelitas VTH2090 and VTP9412 diodes with their dimensions.  }
\end{figure}
There are not that many options available when choosing a photon sensor to convert the scintillator light output to electrical signals. Considering that this calorimeter should be an instrument orbiting in space, we have then stringent overall mass constraints, power requirements and dissipation, and also require compactness of the ancillary calorimeter components. We thus opted for a silicon photodiode (PD) solution, which satisfies all the above criteria and provides a very stable photon to electron conversion efficiency versus temperature and voltage biasing variations.
Given the physics scope of the calorimeter, namely the need to correctly measure a MIP signal for calibration purposes (typically 10 MeV per crystal), while mantaining linearity even for energy deposits of up to 100 TeV in one crystal, we opted for a solution with two PDs each with different active area.
After an extensive market analysis, we chose the following Excelitas~\cite{EXCELITAS} PDs: for small signals (such as those from MIPs) a large area silicon PIN photodiode VTH2090 (figure \ref{fig:photoPD}), and a small area VTP9412  for large signals from showers. The latter is a standard PD with an area roughly a fifty times smaller that will not saturate the F.E. electronics even in the presence of the most energetic deposits.
\begin{figure}[htbp]
  \centering
  \subfigure[\label{fig:VTH}]{\includegraphics[width=.4\textwidth]{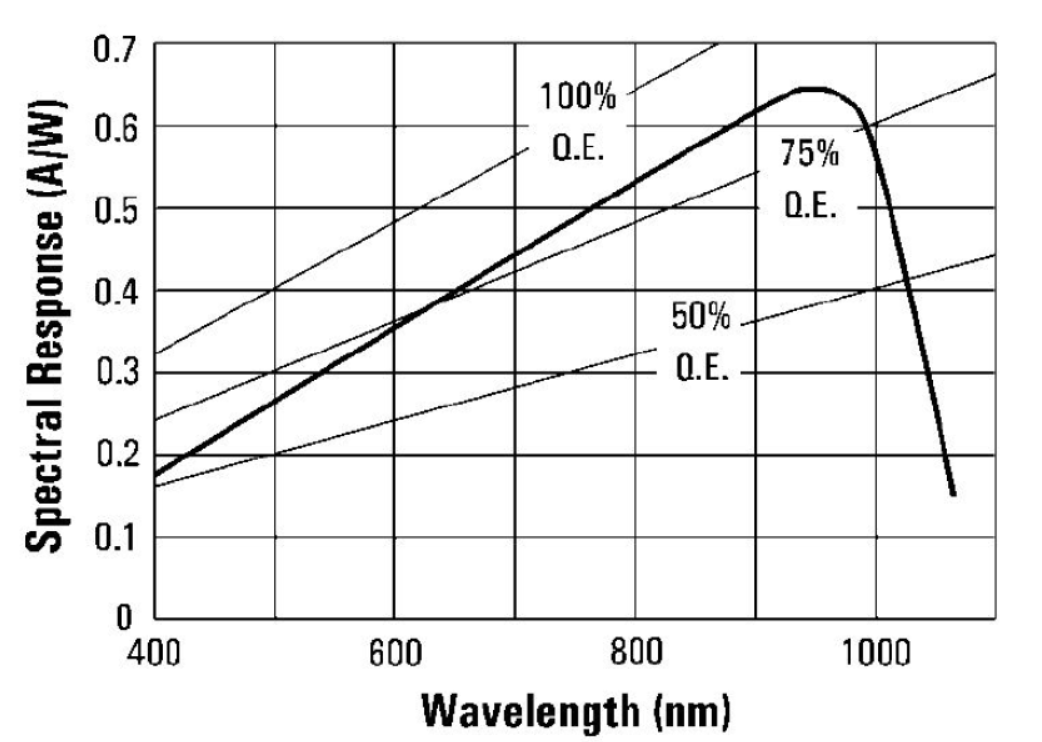}}
  \qquad
  \subfigure[\label{fig:VTP}]{\includegraphics[width=.4\textwidth]{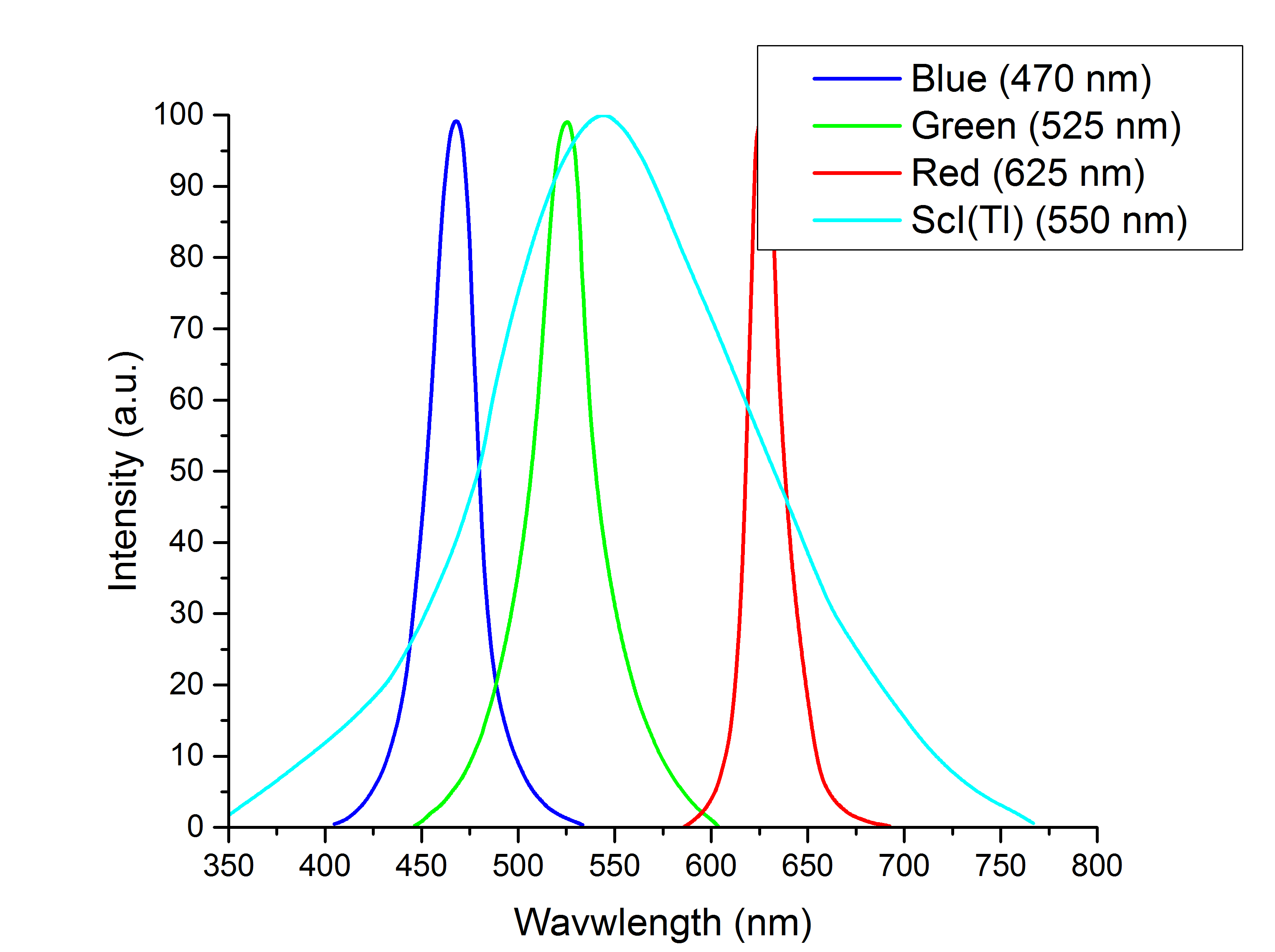}}
  \caption{\label{fig:SpectralResponses} Spectral response of VTH2090 (VTP9412 has a similar behaviour) \ref{fig:VTH} and the emission  spectra for the blue, green and red LEDs and the CsI(Tl) scintillator. \ref{fig:VTP} photodiodes. }
\end{figure}
The VTH2090 is a PIN photodiode with a 9.2x9.2 mm$^2$ active area, mounted in a black ceramic package with an epoxy window. This device is ideal for scintillation detection, spectrophotometry, or other applications requiring a fast, large area, high sensitivity device.
\begin{table}[htbp]
\centering
\caption{\label{tab:SpectralResponsesMeasures} Measured spectral responses. }
\smallskip
\begin{tabular}{|c|c|c|}
\hline
 Wavelength (nm)& VTH2090 & VTP9412 \\
\hline
    470     & 0.28$\pm$0.02 A/W & 0.33$\pm$0.02 A/W \\
    525     & 0.30$\pm$0.02 A/W & 0.34$\pm$0.02 A/W \\
    625     & 0.38$\pm$0.03 A/W & 0.42$\pm$0.03 A/W \\
 \hline
\end{tabular}
\end{table}
Its main features are: high quantum efficiency, excellent uniformity, high shunt impedance, low junction capacitance, fast response and low noise.
The VTP9412 is a fast response silicon photodiode having a 1.6 mm$^2$ active area designed for spectral response between 400 and 1150 nm. This series of photodiodes has been designed for low junction capacitance to achieve faster response time. These photodiodes are suitable for operation under reverse bias, which increases the speed of response, but can also be used in photovoltaic mode. Main features for this PD are: visible to IR spectral range, small active area, 1 to $2\%$ linearity over 7 to 9 decades, low dark current, high dark resistance, low capacitance, fast response. 
The spectral response for both VTH2090 and VTP9412 PDs are shown in figure \ref{fig:SpectralResponses}. Both are well matched to the CsI (Tl) emission spectrum also shown in the same figure, together with the emission spectra of the light sources we used for testing the PDs. We have tested both diodes  with four light sources: blue, green and red LEDs, with a wavelength of respectively 470 nm, 525 nm, 625 nm and a CsI(Tl) scintillator with a spectrum maximum at 550 nm.

% \begin{figure}[htbp]
% \centering
% \includegraphics[width=.8\textwidth]{Images/Thesis/emission_spectrum.png}
% \caption{\label{fig:emissionSpectra} Emission spectra for the blue, green and red LEDs and the CsI(Tl) scintillator. }
% \end{figure}

The main goal of these measurements was to verify the parameters provided by the manufacturer, in particular the spectral response. Comparing the given spectral response with the measured one, we have reproduced the data shown in figure \ref{fig:SpectralResponses} within 8$\%$. Our measurements are shown in table \ref{tab:SpectralResponsesMeasures}. 

\subsection{Photodiode single channel readout and calibration}

As stated previously, we have used a low noise DAQ system designed by Amptek for our single crystal/photodiode studies and measurements. The system consists of a charge sensitive preamplifier A250 and a digital pulse analyser PX5.

\begin{figure}[htb]
\centering
\includegraphics[width=0.6\linewidth]{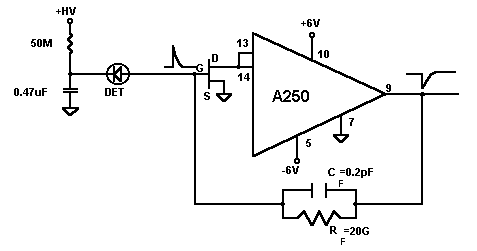}
\caption{The A250 charge preamplifier circuit used with the DC coupling for the PD. }
\label{fig:couplingCircuit}
\end{figure}

The A250  \cite{Amptek:A250} is a hybrid Charge Sensitive Preamplifier for use with a wide range of detectors having capacitance from less than one, to several thousand picofarads.
% To permit optimization for a wide range of applications, the input field effect transistor is external to the package and user selectable. This feature allows the FET to be matched to the particular detector capacitance, as well as to noise and shaping requirements. 
% The noise performance of the A250 is such that its contribution to FET and detector noise is negligible in all charge amplifier applications 
The input FET was selected and optimized to our specific application. Usually, this component is chosen with a large transconductance (\textit{$g_m$}) as possible, while matching its input capacitance (\textit{$C_{iss}$}) to the detector capacitance (\textit{$C_d$}).
\begin{table}[htbp]
\centering
\caption{\label{tab:FETcharacteristics} 2SK152 F.E.T. characteristics. }
\smallskip
\begin{tabular}{|c|c|c|c|c|c|c|c|}
  \hline
  Type      & $BV_{GSS}$ & $I_{GSS}$ & $V_{GS(off)}$ &  $I_{DSS}$    & $g_{fs}$ & $C_{iss}$ & $C_{rss}$\\
  \hline
  n-channel &    -20    &    0.1   &  -0.5/-2.0  &  50/20       &    30   &    15    &  4.0  \\
            &           &  (-10V)  &    (-10V)   &  (10V)       &  (10V)  &    (0-10V)   &  (0-10V)  \\
 %           &           &          &             &              &         &   (10V)  & (10V) \\
  \hline
  Units     &      V(min)    &     nA(max)   & V (min/max) & mA (min/max) &    mS   &     pF   &   pF \\
%            &    (min)  &   (max)  &             &              &         &          &      \\ 
  \hline
\end{tabular}
\end{table}
%IO L'HO COPIATA DALLA TESI DI OLEK MA NON CAPISCO IL SENSO DI MOLTE DELLE ENTRATE DELLA TABELLA...
Taking into account this fact and the PDs capacitances, we chose a parallel configuration of three 2SK152 FET, which has a good capacitive match with our PDs and  an optimal noise performance. The 2SK152 FET characteristics are shown in table \ref{tab:FETcharacteristics}. Moreover, we used DC coupling between PD and A250 to exclude additional noise sources. The typical coupling circuit is shown in figure \ref{fig:couplingCircuit}. 
The final configuration that has been used for the single channel measurements is characterized by 3x2SK252 as input FET, standard feedback components (R$_f$=300 M$\Omega$, C$_f$=1 pF), DC coupled photodiode and an RC bias filter (R$_b$=20 M$\Omega$, C$_b$=27 nF). 

\begin{figure}[htb]
\centering
\includegraphics[width=0.8\linewidth]{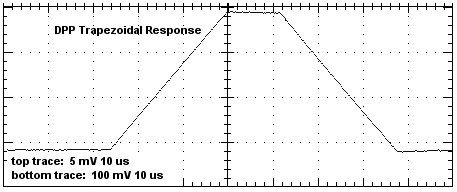}
\caption{Trapezoidal shaped pulse of DPP. }
\label{fig:trapezoidalTypicalOutput}
\end{figure}

We used the Amptek PX5 \cite{Amptek:PX5} postamplifier to interface the detector (coupled to the A250 preamplifier) to a computer running data acquisition and control software.
% The main purpose of using a postamplifier is to provide amplification and to filter out low and high frequency noise. 
% The PX5 includes a high performance digital pulse processor (replacing a conventional shaping amplifier), a multichannel analyzer and both low and high voltage power supplies. Moreover it offers several advantages over traditional systems, including improved performance, many more configuration options to optimize the system and many communications and output options . 
\begin{figure}[htb]
\centering
\includegraphics[width=0.8\linewidth]{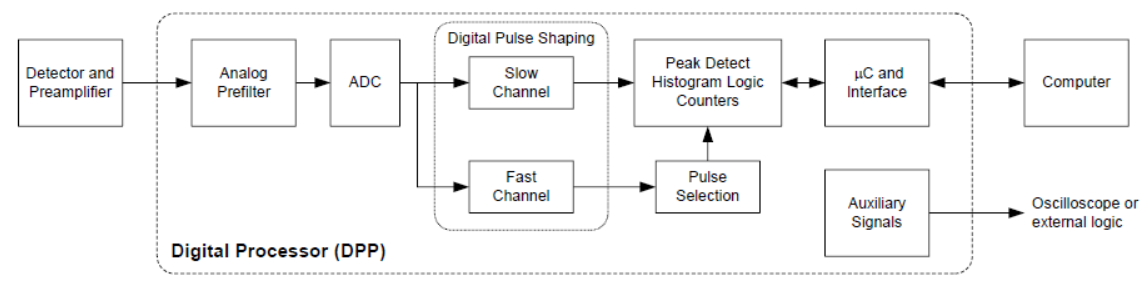}
\caption{Block diagram of the PX5  Digital Pulse Processor (DPP) in a complete system. }
\label{fig:blockDiagramPX5}
\end{figure}
The PX5 digitizes the preamplifier output, applies real-time digital processing to the signal, detects the peak amplitude (digitally) and bins this value in its histogramming memory, generating a  signal spectrum. 
% Pulse selection logic can reject pulses from the spectrum using a variety of criteria. The spectrum is then transmitted to the computer.
The Block diagram of PX5 is shown in figure \ref{fig:blockDiagramPX5}. 
% Input for the PX5 is the output of a charge sensitive preamplifier. The analog prefilter circuit prepares the signal for an accurate digitization. More precisely the appropriate gain and offset are set in order to exploit the dynamic range of the ADC and in addition some filtering and pulse shaping are carried out to optimize the digitization.  
% The 12-bit ADC digitizes the output of the analog prefilter at a 20 or 80 MHz rate. This stream of digitized values is sent in real time into the digital pulse shaper.
% The ADC output is processed continuously using a pipeline architecture to generate a real time shaped pulse. This carries out pulse shaping as in any other shaping amplifier. The shaped pulse is a purely digital entity. There are two parallel signal processing paths inside the DPP, the fast and slow channels, optimized to obtain different data about the incoming pulse train. The slow channel, which has a long shaping time constant, is optimized to obtain accurate pulse heights. 
We used a trapezoidal pulse shaping, with a typical output pulse shape shown in figure \ref{fig:trapezoidalTypicalOutput}. This shape provides a near optimum signal to noise ratio for many detectors.
%Relative to conventional analog shapers, the trapezoid provides lower electronic noise and, simultaneously, reduced pulse pile-up \cite{PulseShaping:Goulding1972}. The fast channel is optimized to obtain timing information: detecting pulses which overlap in the slow channel, measuring the incoming count rate, pulse rise-times, etc. 
The pulse selection logic rejects pulses for which an accurate measurement cannot be made. It includes pile-up rejection, rise time discrimination, logic for an external gating signal, etc.
The histogram memory operates as in a traditional multichannel analyser (MCA). When a pulse occurs with a particular peak value, a counter in a corresponding memory location is incremented. The primary output of the DPP is a histogram (an array), with a bin content that follows the signal spectrum. 

\subsubsection{Optimization and calibration of the set-up}

\begin{figure}[htbp]
\centering
\includegraphics[width=.8\textwidth]{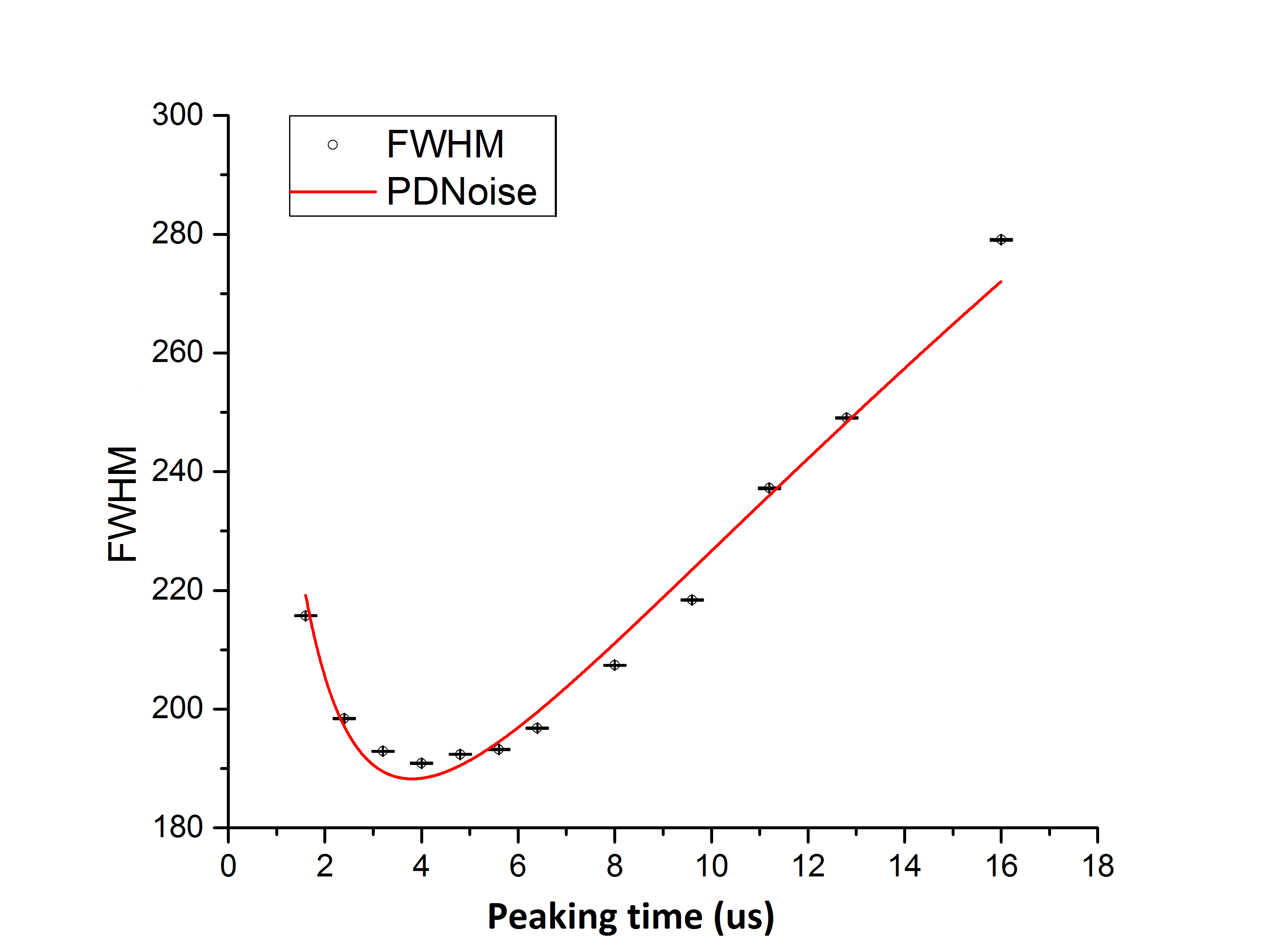}
\caption{\label{fig:noiseMeasurements} Noise dependence on shaping time. }
\end{figure}

We have optimised the shaping time by performing various noise measurements \cite{tesi_olek} for  the VTH2090  diode. The noise has been estimated using the Full Width Half Maximum of the pulse signal distributions (in ADC channels) for various shaping times.
Noise dependence on the shaping time $\tau$  \cite{NoiseCharacterization:Giacomini2011}, is described in terms of Equivalent Noise Charge  by eq. \eqref{eq:ENC_tau}, where $C_{in}$ is the input capacitance (sum of the detector capacitance plus the 3 FET in parallel), $R_s$ is the sum of the series resistance of the input FET (from source to gate), the cables arriving to the gate and the other parasitic series resistances (usually few $\Omega$), $R_f$ is the feedback resistance, $\Gamma$ is a numerical coefficient equal to 2/3 for JFET, $flattop$ is the smaller base of the trapezoid, $g_m$ is the transconduttance of the FET (in our case it is the sum of 3FETs), $e_{pink}$ is the pink noise term (1/$f$ noise term) and $I_{leak}$ is the leakage current through the detector. We have fitted the experimental data using eq. \eqref{eq:ENC_tau}. The red line present in figure \ref{fig:noiseMeasurements} shows the result of the fit,  with the value of $4kT$ fixed at 0.1 eV (room temperature) and with all other parameters free to vary within an appropriate range. 
\begin{equation}
  ENC = \frac{1}{q} \sqrt{ ( \frac{\tau}{3} + \frac{flattop}{2} ) \cdot ( 2qI_{leak} + \frac{4kT}{R_f} ) + C^2_{in} 4kT ( \frac{\Gamma}{g_m} + R_s ) \cdot \frac{1}{\tau} + C^2_{in} q e_{pink} } \label{eq:ENC_tau}
\end{equation}
 From the fit of figure \ref{fig:noiseMeasurements}, we derive an optimal shaping time 
 in the range 3.8$\div$4.2 $\mu$s. As a cross check we have compared nominal and fit derived values for the free parameters as shown in table \ref{tab:fitResults}. 

\begin{table}[htbp]
  \centering
  \caption{\label{tab:fitResults} Parameter fit values. }
  \smallskip
  \begin{tabular}{|c|c|c|}
    \hline
    % \multirow{2}{*}{Parameter name} &\multirow{2}{*}{Nominal value} & \multirow{2}{*}{Fit value} \\
   Parameter name  & Nominal value & Fit value \\
    % \multicolumn{2}{|c|}{Fit Results} \\
    % \cline{3-4}
%    & &  Value & Std. Err. \\
    \hline
    $C_{in}$ (F)        & 1.15 $\times 10^{-10}$    & 1.13 $\times 10^{-10}$ \\  
    $R_{s}$ ($\Omega$)  & few Ohms             & 1                     \\  
    $g_{m}$ (mS)        & 0.09                     & 0.1                     \\  
    $R_{f}$ ($\Omega$)  & 3 $\times 10^{8}$         & 3.05 $\times 10^{8}$    \\  
    $I_{leak}$ (A)       & $\sim$ 1 $\times 10^{-9}$ & 8.4 $\times 10^{-10}$  \\  

    % $C_{in}$ (F)        & 1.15 $\times 10^{-10}$    & 1.1286 $\times 10^{-10}$ & 7.17 $\times 10^{-5}$ \\  
    % $R_{s}$ ($\Omega$)  & few $\Omega$             & 1                       & 8.3596 $\times 10^{6}$ \\  
    % $g_{m}$ (mS)        & 0.09                     & 0.1                     & 20866 \\  
    % $e_{pink}$           & unknown                  & 38351                   & 4.87 $\times 10^{10}$ \\  
    % $R_{f}$ ($\Omega$)  & 3 $\times 10^{8}$         & 3.05 $\times 10^{-8}$    & 4.898 $\times 10^{15}$ \\  
    % $I_{leak}$ (A)       & $\sim$ 1 $\times 10^{-9}$ & 8.4 $\times 10^{-10}$    & 0.00262 \\  
    \hline
  \end{tabular}
\end{table}

\subsubsection{Absolute charge calibration procedure}
\label{subsec:charge}
To calibrate the  ADC scale versus energy or charge we used a $^{241}$Am source which decays into $^{237}$Np emitting a 5.5 MeV $\alpha$ particle. Such a source emits also X and $\gamma$ lines which originate mostly from decay products of $^{237}$Np. We have used two of these lines (8 keV and 59.6 keV) to calibrate the system. 
\begin{figure}[htbp]
\centering
\includegraphics[width=.65\textwidth]{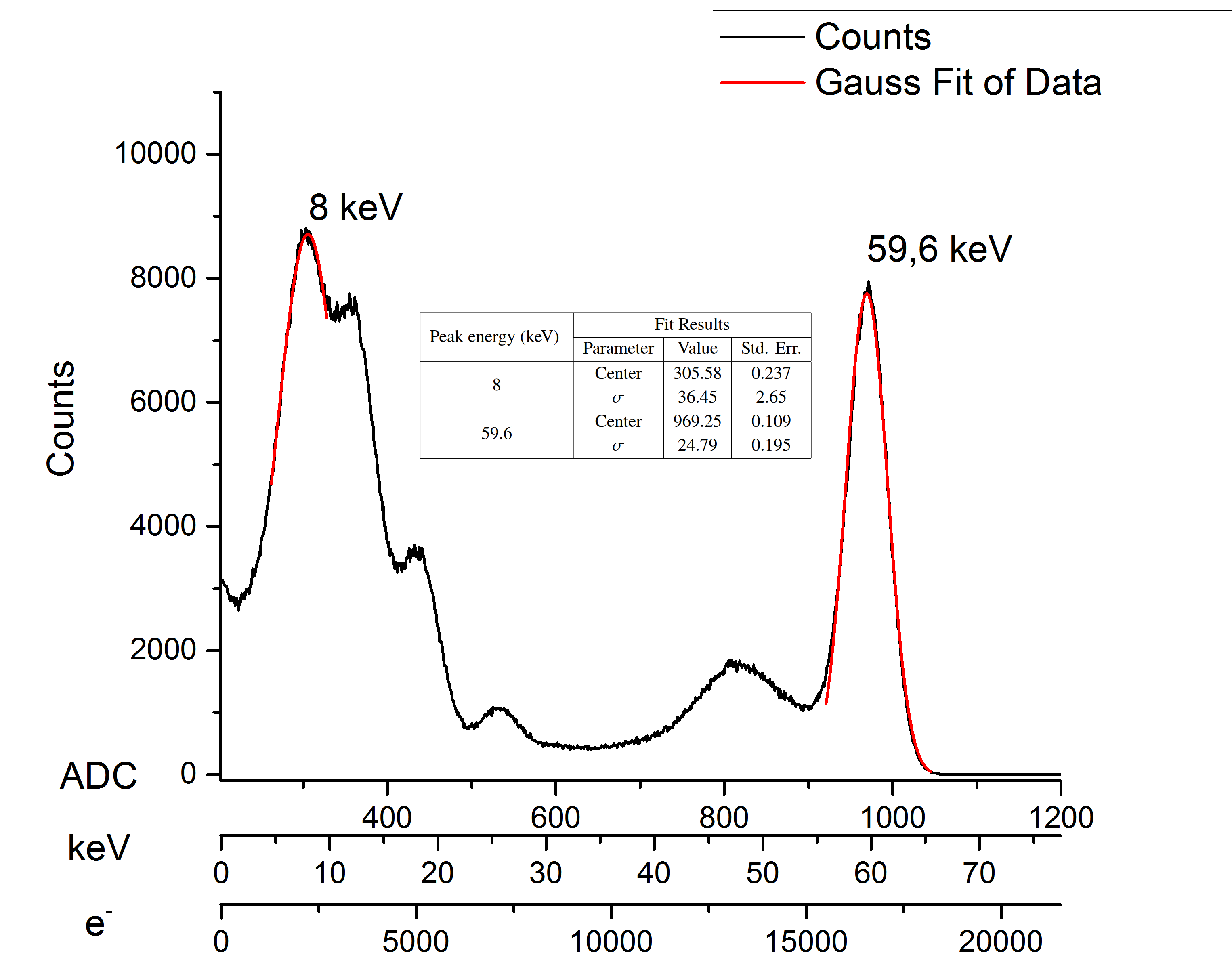}
\caption{\label{fig:alphaSourceSpectrum} $^{241}$Am source spectrum, with the fit results on the 8 keV and 59.6 keV peaks. }
\end{figure}
The $^{241}$Am  spectrum measured directly by the VTH2090 photodiode coupled to the source, is shown in figure \ref{fig:alphaSourceSpectrum}. The two calibration peaks can be easily identified and both of them have been fitted with a Gaussian function.
The resulting Gaussian fits are shown with red solid lines in the figure. 
\begin{figure}[htbp]
\centering
\subfigure{\includegraphics[width=.42\textwidth]{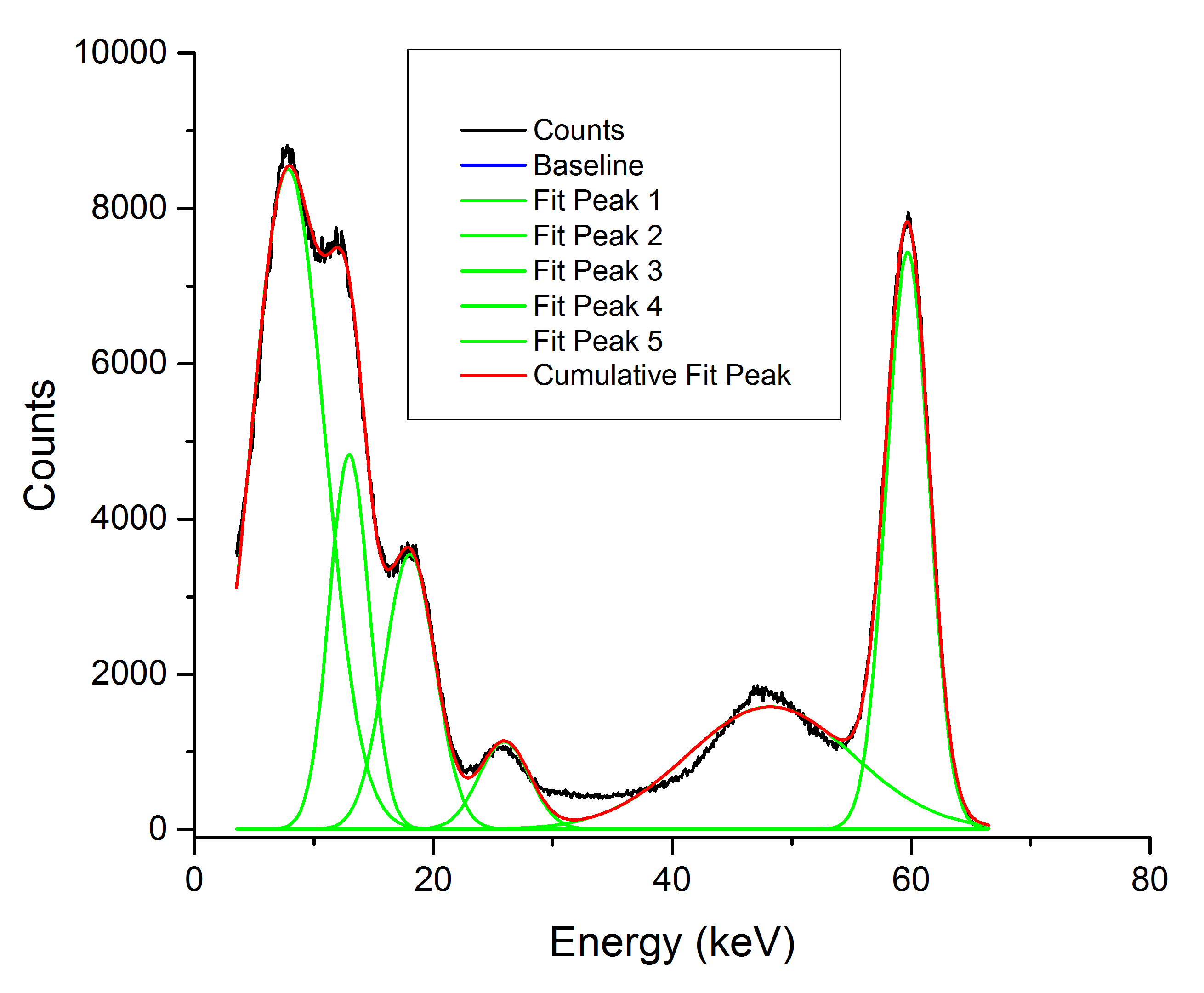}}
\subfigure{\includegraphics[width=.42\textwidth]{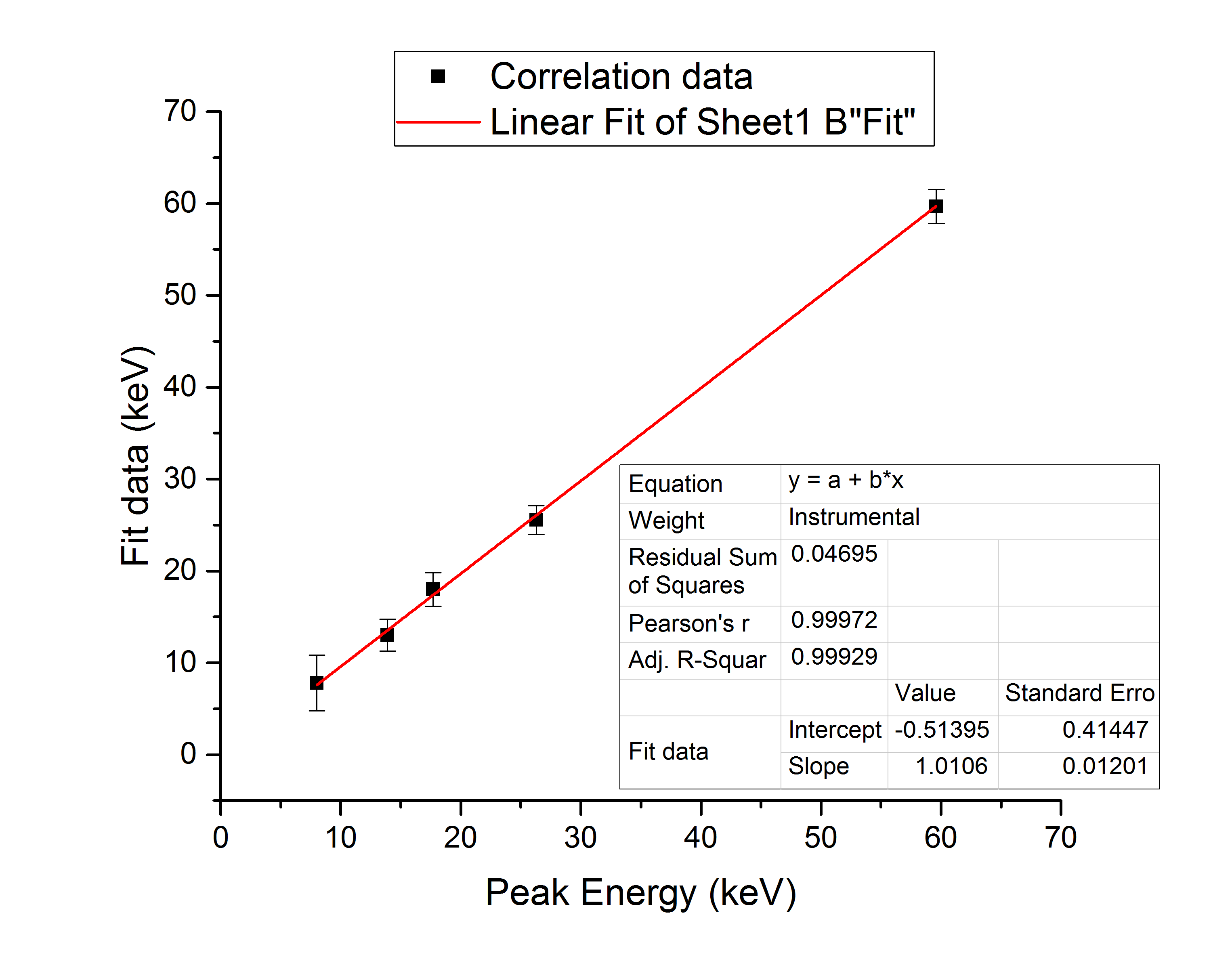}}
 \caption{\label{fig:alphaSourceSpectrumFITS} Multiple peak fits (left) and correlation between nominal and derived values of the $^{241}$Am spectrum peaks (right). }
\end{figure}The energy difference between the two calibration peaks is 51.6 keV, corresponding to 663.7 ADC channels from which we obtain a calibration factor of 77.75 eV per ADC channel. Dividing this value by the energy needed for electron-hole pair creation in silicon, which is equal to 3.67 eV at room temperature \cite{SiPairProd}, we obtain a charge calibration factor of 21.19 e$^-$ per ADC channel. 
Using the position of the two peaks both in energy and ADC channels, the offset can be evaluated to be about 202.7 ADC channels. In the same figure, both the energy and charge scale obtained are shown at the bottom. 
%calibration is 0.261 ADC channels or 0.03$\%$ for 59.6 keV.
% \begin{table}[htbp]
  % \centering
  % \caption{\label{tab:gausFitResults} Calibration peaks fit results. }
  % \smallskip
  % \begin{tabular}{|c|c|c|c|}
    % \hline
   %  \multirow{2}{*}{Peak energy (keV) } &\multicolumn{3}{|c|}{Fit Results} \\
    % \cline{2-4}
    % & Parameter & Value & Std. Err. \\
    % \hline
    % \multirow{2}{*}{8}    &   Center   &  305.58 & 0.237 \\
         %                  &  $\sigma$  &   36.45 & 2.65 \\
   %  \multirow{2}{*}{59.6} &   Center   &  969.25 & 0.109 \\
        %                   &  $\sigma$  &   24.79 & 0.195 \\
    % \hline
  % \end{tabular}
% \end{table}
We have extended this procedure to the other peaks present in the $^{241}$Am spectrum. The measured energy spectrum of $^{241}$Am with multiple fits is shown in figure \ref{fig:alphaSourceSpectrumFITS}, with the result of the correlation fit using the previous scale factors to predict the peak energies. The correlation slope coefficient is 1.01, indicating an excellent agreement when using the scale factors derived above.

\subsection{Scintillating crystal choice and characterisation}
%\label{sec:crystal}
%\subsection{Scintillating materials characterisation}
\label{sec:NonGaussiaTail}

The materials that have been considered for the Calocube detector are CsI(Tl) \cite{CsI}, BGO (Bi$_4$Ge$_3$O$_{12}$) \cite{BGO} and LYSO (Lu$_{2(1-x)}$Y$_{2x}$SiO$_{5}$(Ce)) \cite{LYSO}. \begin{table}[htbp]
  \centering
  \caption{\label{tab:scintMaterials} Scintillator properties. }
  \smallskip
  \begin{tabular}{|c|c|c|c|}
    \hline
    Scintillator                        & CsI(Tl)            & BGO                     & LYSO \\
    \hline
    Density (g/cm$^3$)                  & 4.51               & 7.13                    & 7.30 \\
    Radiation length X$_0$ (cm)         & 1.85               & 1.12                    & 1.16 \\
    Decay time (ns)                     & 1000               & 300                     & 50 \\
    Light yield (e$^-$/keV$_{\gamma}$)    & 5.5 $\cdot$ 10$^4$ & $\sim$ 8 $\cdot$ 10$^3$ & $\sim$ 2.5 $\cdot$ 10$^4$ \\
    Wavelength of maximum emission (nm) & 550                & 480                     & 428 \\
    \hline
  \end{tabular}
\end{table}
All are high density materials with a small radiation length, capable of containing showers in a relatively small volume, a critical issue as stated before, in space applications.
In particular CsI(Tl), which is a common scintillator used in nuclear
\begin{figure}[htbp]
\centering
\includegraphics[width=.65\textwidth]{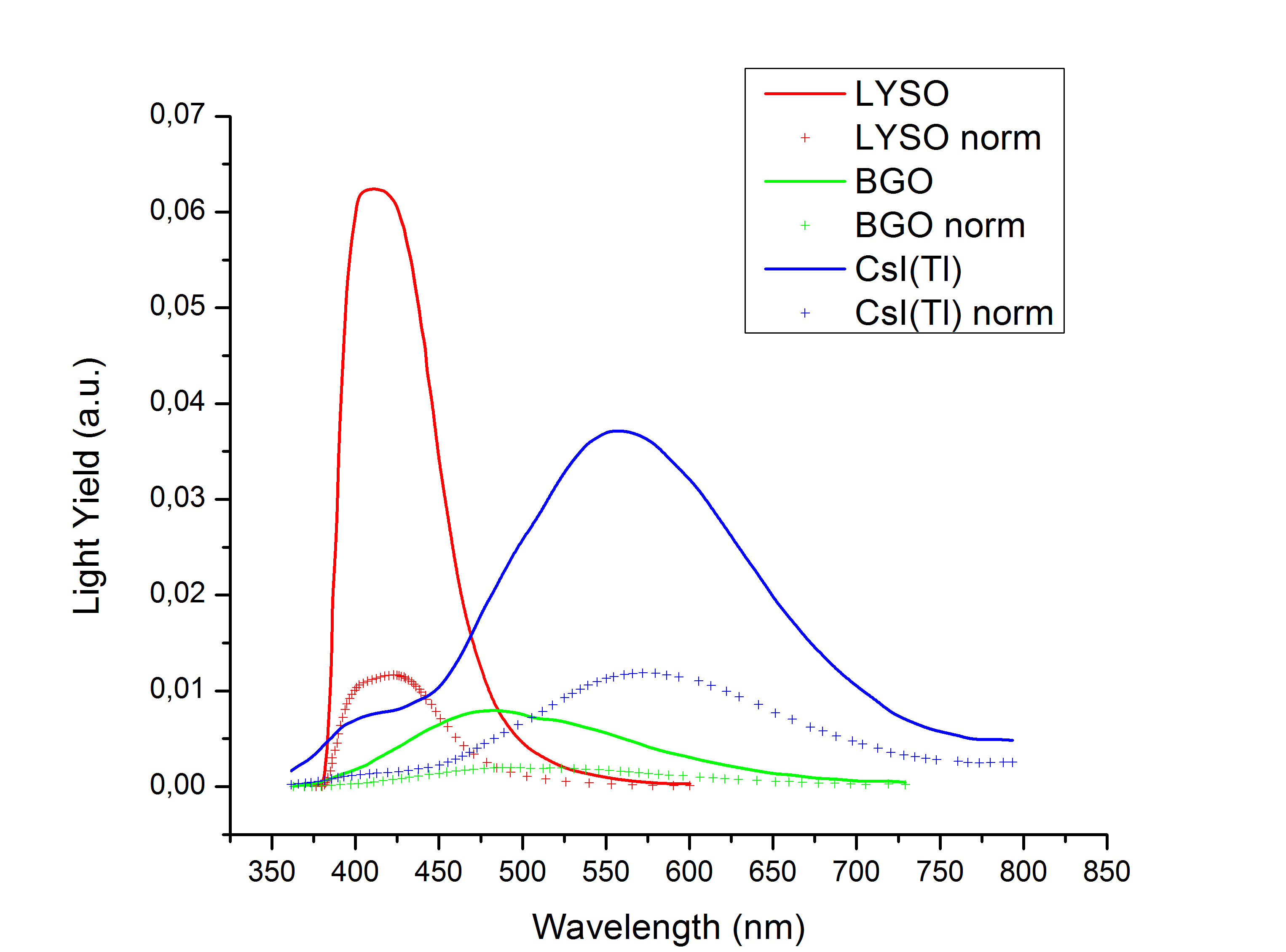}
\caption{\label{fig:scintConvEmSpectra} Scintillator emission spectra~\cite{rihuama2008} and their convolution with the VTH2090 response (norm). }
\end{figure}
 and particle physics with a very high light yield, became our final choice for the prototype given its wide availability with many producers in the market, at an affordable price which is a very important parameter since the propotype involved the use of ~1000 such crystals. Another advantage  (from a space mission point of view) is that it suffers plastic deformation when stressed but is not brittle. Also the spectral response of CsI(Tl), which peaks at around 580 nm, is very well matched to normal silicon photodiodes that are available in a wide variety of formats and electrical parameters.
We summarise the main properties of our chosen crystals  in table \ref{tab:scintMaterials}.

\begin{figure}[htbp]
\centering
\includegraphics[width=.6\textwidth]{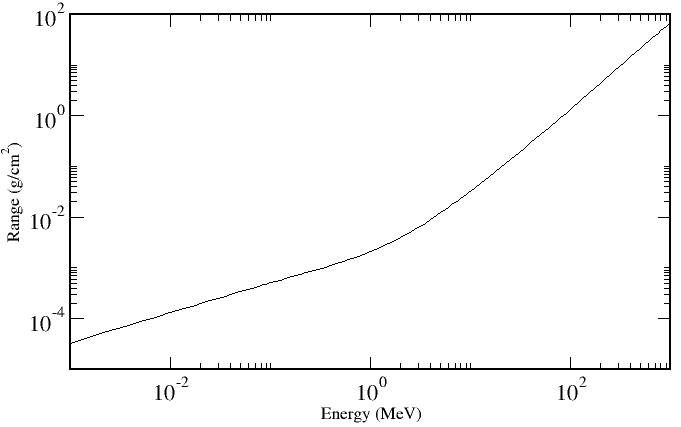}
\caption{\label{fig:AlphaCsIrange} $\alpha$ particles range in CsI(Tl) obtained with the use of the ASTAR software from NIST. }
\end{figure}

We have found consistent discrepancies in the available literature on LYSO crystals. In fact LYSO is a relatively new material and its characteristics can change depending on the sources, this can also happen for more established scintillators such as BGO (cf.~\cite{rihuama2008,scintOtherSources1,scintOtherSources2:JunweiDu2009}). 
% \begin{figure}[htbp]
% \centering
% \subfigure[\label{rihua}]{\includegraphics[width=.45\textwidth]{Images/Thesis/LYSO_BGO_emission_spectrum.png}}
% \subfigure[\label{rihuavt2090}]{\includegraphics[width=.45\textwidth]{Images/Thesis/Scint_Emiss_spectrums.png}}
 % \caption{\label{fig:scintEmSpectra} Emission spectra of CsI(Tl), BGO and LYSO\ref{rihua}  and their convolution with the VTH2090 spectral response (symbols) \ref{rihuavt2090}. }
% \end{figure}
A comparison of the emission spectra of the three scintillators was performed by the group of Mao Rihua \cite{rihuama2008} from which the emission spectra shown in figure \ref{fig:scintConvEmSpectra} are taken. We convoluted these with the spectral response of our photodiodes to obtain the curves shown in the same figure.  The VTH2090 spectral response can be approximated by a line with offset and slope respectively equal to -0.1779 A/W and 8.815 $\cdot$ 10$^4$ A/Wnm in the range up to 800 nm. These values have been used to convolute the scintillators emission spectra. 
% \begin{figure}[htbp]
% \centering
% \includegraphics[width=.5\textwidth]{Images/Thesis/LYSO_BGO_emission_spectrum.png}
% \caption{\label{fig:scintEmSpectra} Emission spectra of CsI(Tl), BGO and LYSO \cite{scintillatorsComparision:RihuaMa2008}. }
% \end{figure}
% To estimate the light yield for all three scintillators, the convolution of the emission spectra presented in figure \ref{fig:scintEmSpectra} with the PDs spectral response has been used.
% The VTH2090 spectral response can be approximated by a line with offset and slope respectively equal to -0.1779 and 8.815 $\cdot$ 10$^4$ in the range up to 800 nm. These values have been used to convolute the scintillators emission spectra and the original (lines) and convoluted (symbols) curves are presented in figure \ref{fig:scintConvEmSpectra}.
\begin{table}[htbp]
  \centering
  \caption{\label{tab:scintSignLevel} Relative signal level for the scintillators. }
  \smallskip
  \begin{tabular}{|c|c|c|}
    \hline
    Scintillator & Integral value (A.U.) & $\%$ of CsI(Tl) \\
    \hline
    CsI(Tl) & 2.44 & 100    \\
    LYSO     & 0.85 &  35  \\
    BGO    & 0.37 &  15 \\
    \hline
  \end{tabular}
\end{table}
After convolution CsI(Tl) shows the highest light yield indicating a better spectral match to the photodiodes used.  The results of the integral of the convoluted curves are presented in table \ref{tab:scintSignLevel} relative to the CsI(Tl) signal.
The result is very dependent on the manufacturers data, and suffers from the data discrepancies mentioned before. These depend on the details of crystal growth, on the exact composition, and of course on the spectral response of the photon detectors used. For example according to data from "Saint-Gobain Ceramics $\&$ Plastics, Inc." the ratio of the signals for CsI(Tl)/BGO/LYSO is 1/0.59/0.18 \cite{CsI, BGO, LYSO}. We thus performed laboratory tests of these scintillators coupled to our chosen photodiodes, to obtain the data that was used afterwards as input for the physics simulations of our calorimeter.
\begin{figure}[htbp]
\centering
\includegraphics[width=.55\textwidth]{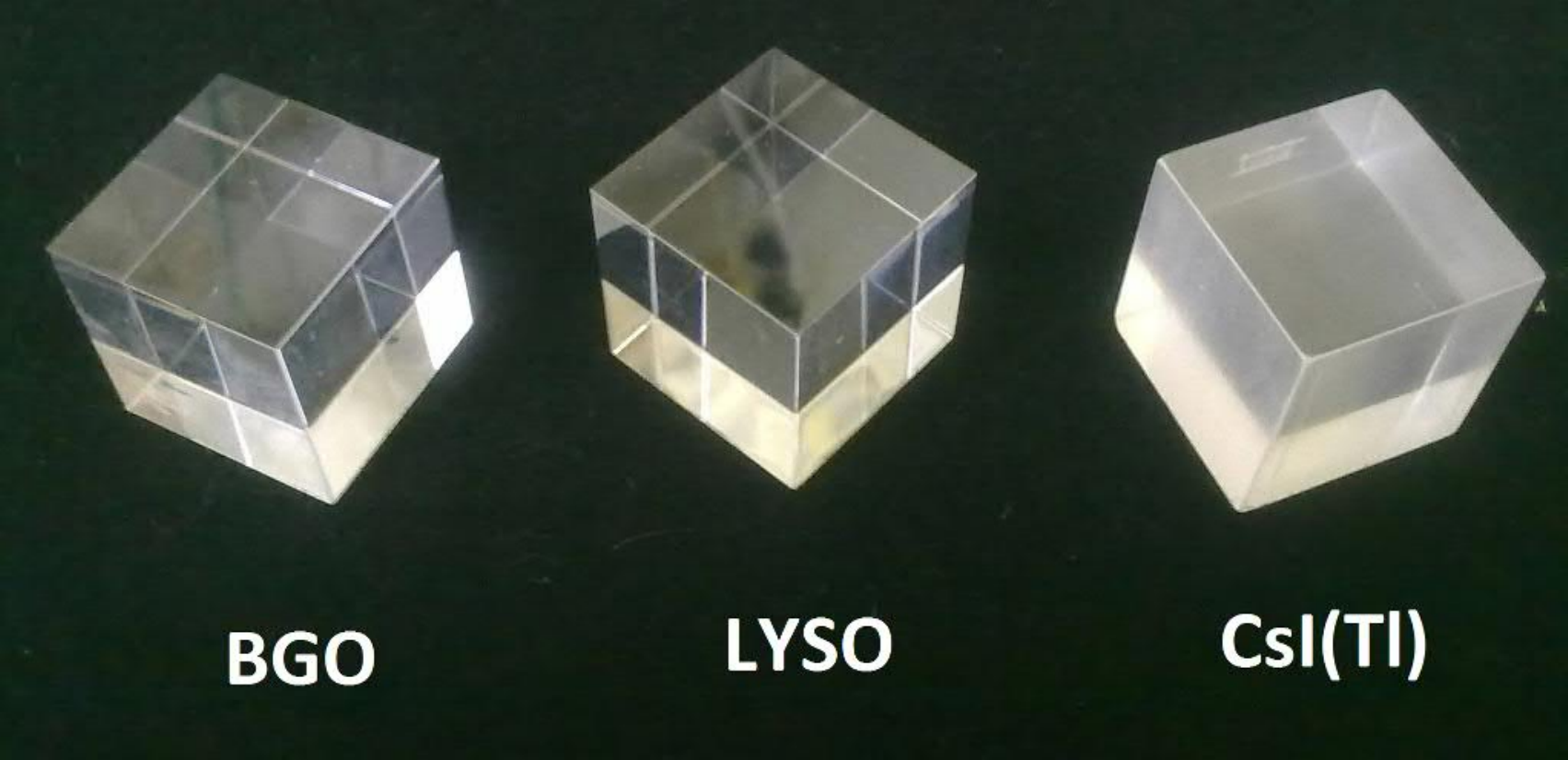}
\caption{\label{fig:scintCrystals} Cubes of CsI(Tl), BGO and LYSO with different surface finish (sanded for the CsI). }
\end{figure}
We performed these measurements on cubic crystals of 20 $\times$ 20 $\times$ 20 mm$^3$, as shown in figure \ref{fig:scintCrystals}. These are smaller than the dimensions needed for the prototypes but allowed us to keep the scintillator costs down without sacrificing the measurement siginficance. The crystals were wrapped with a 400 $\mu$m thickness of Teflon tape. All scintillators were coupled to VTH2090 diodes.
We used $\alpha$ particles from $^{241}$Am source. To minimize the external energy loss of the $\alpha$ particles in the wrapping material, a 1mm$^2$ window in the Teflon cover was opened and the source was placed in contact with the crystal surface.  The range of $\alpha$ particle in CsI(Tl) versus energy, shown in figure \ref{fig:AlphaCsIrange}, has been obtained using ASTAR software, developed by National Institute of Standards and Technology (NIST) \cite{NISTsoftware,ICRUrange}. 
From the figure, a 5 MeV $\alpha$ particle has 0.054 cm range in CsI(Tl)  (likewise for the other crystals under study), thus it is totally contained within the scintillator cubes used for the measurements and  the energy deposited depends only on the $\alpha$ particle energy independently of the scintillator under examination. 

\begin{figure}[htbp]
\centering
\includegraphics[width=.6\textwidth]{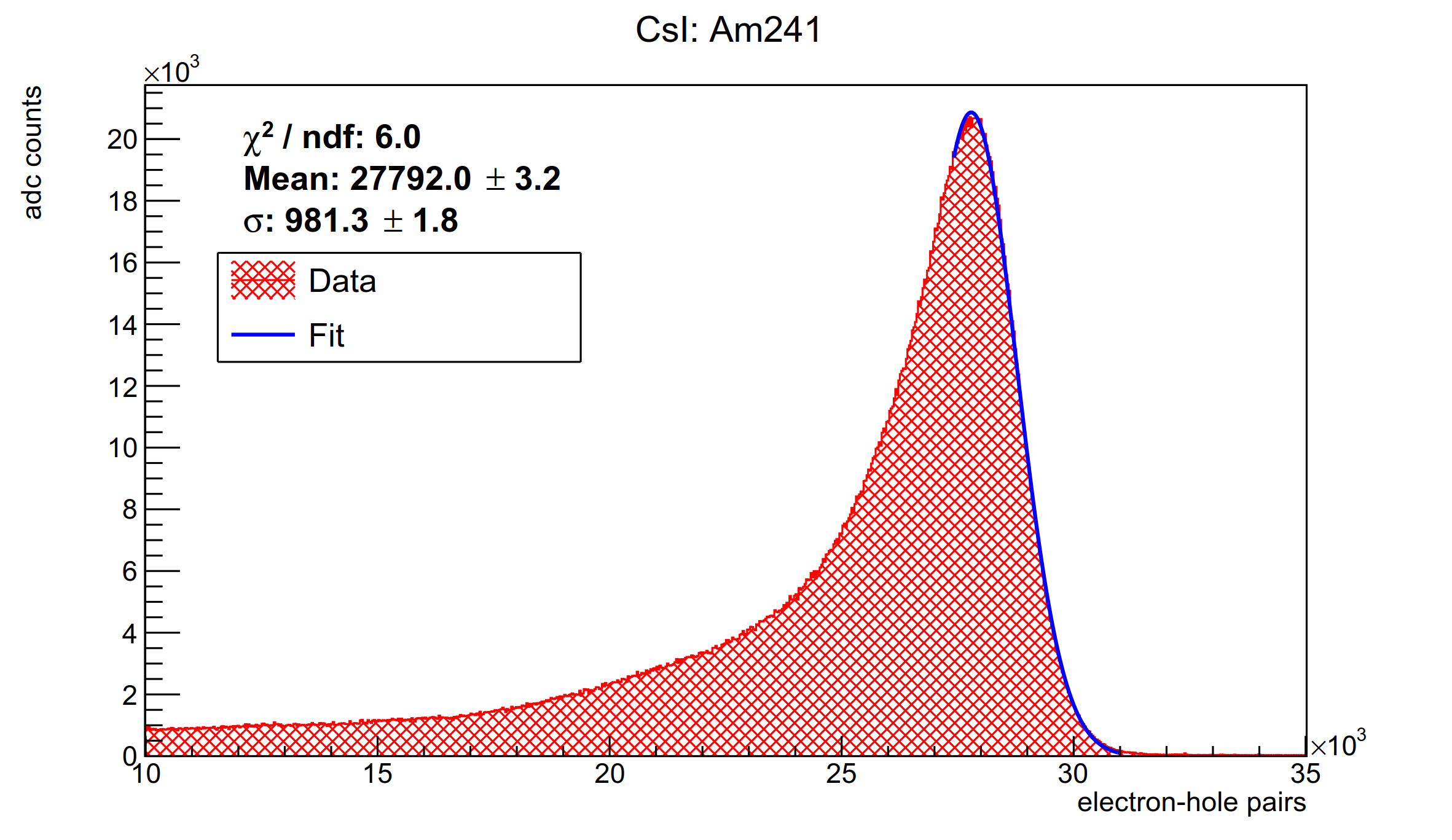}
\caption{\label{fig:AlphaCsI} Measured $\alpha$ particles energy in the CsI(Tl) crystal (red) and Gaussian fit of the peak (blue).}
\end{figure}

\textbf{CsI(Tl)}
We performed the first measurements with the   $^{241}$Am source on the CsI(Tl) crystal. The PD signal is shown in figure \ref{fig:AlphaCsI}. The low energy tail is due to the $\alpha$ particles that have interacted with the Teflon wrapping. In addition there are also interactions with the air (2mm) separating the source from the crystal. We have fitted only the right part of the peak with a Gaussian, as shown in blue in figure \ref{fig:AlphaCsI}.
From the fit we derive the mean value and the width of the full energy $\alpha$ particle, with the resolution evaluated as the ratio between the $\sigma$  and the mean value of the Gaussian fit. For CsI(Tl) the results are: mean value equal to 27792 $\pm$ 3 electrons and resolution of 3.5$\%$ .

\textbf{BGO}
We performed the same measurement for the BGO scintillator. The analysis is more complicated because the signal is lower and superimposed with the background pedestal. We thus acquired two sets of data,  one of only the background spectrum (i.e. without the source) and the other of the background and the $\alpha$ source signals. We fitted the background only spectrum with a Gaussian to find the pedestal distribution parameters. We then fitted the second spectrum with a sum of two Gaussian, in which only the $\alpha$ peak parameters were left free while the background parameters were fixed to the values obtained from the first fit.
\begin{figure}[htbp]
\centering
\includegraphics[width=.6\textwidth]{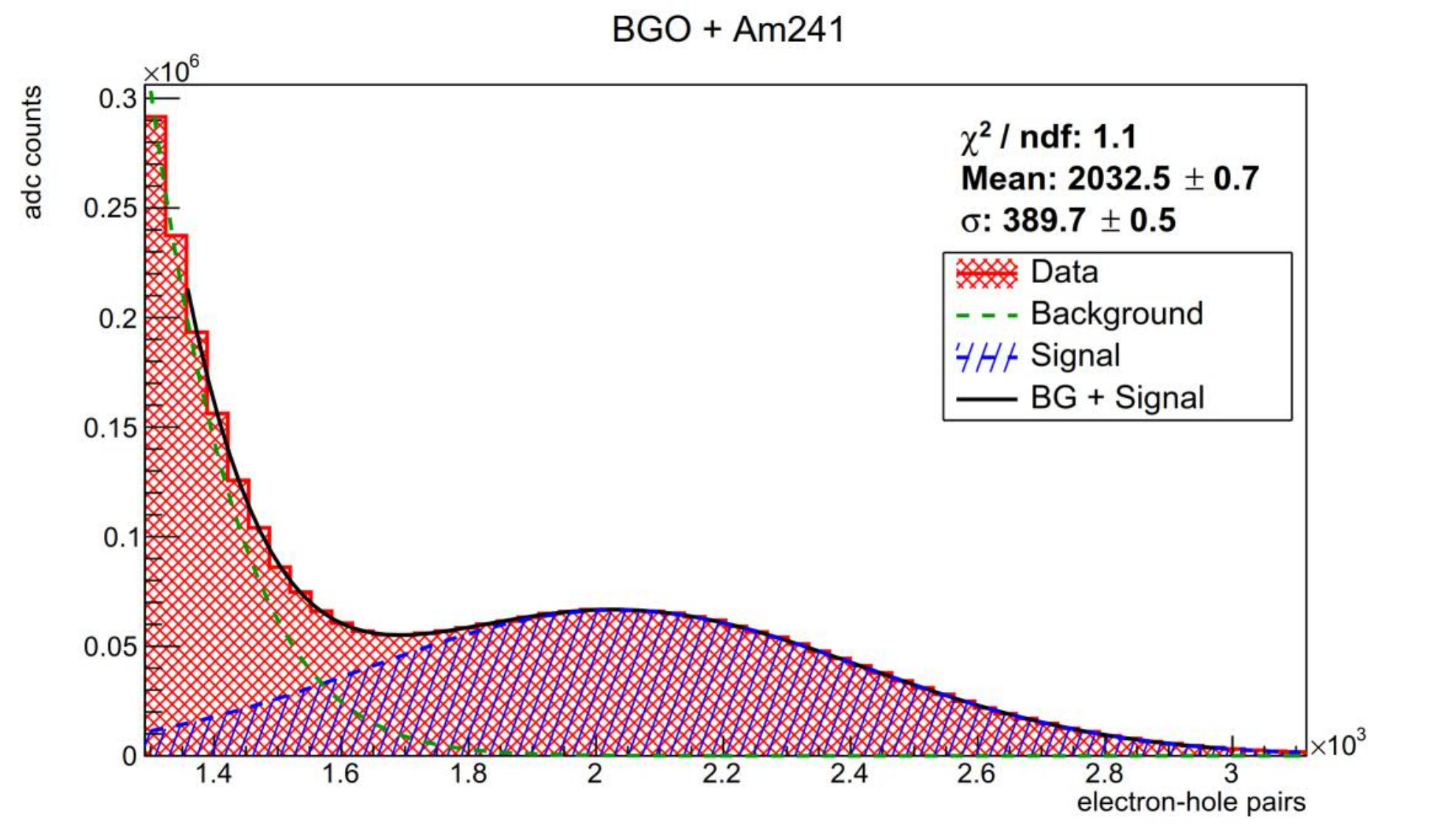}
\caption{\label{fig:AlphaBGO} Measured $\alpha$ particles energy  in the BGO crystal (red), pedestal Gaussian fit (green), $\alpha$ particle Gaussian fit (blue) and sum of both fits (black). }
\end{figure}
The results of the second fit are shown in figure \ref{fig:AlphaBGO}, in which the measured total spectrum is the red hatched plot. The green, blue and black lines show respectively the background distribution estimated with the first fit, the $\alpha$ particle fit estimated with the second fit and the sum of both fits. 
For BGO the results are: mean value equal to 2032 $\pm$ 1 electrons and resolution of 19.2$\%$.

\textbf{LYSO}
The LYSO scintillator measurement procedure is the same as the one for the BGO crystal. In addition to the low signal level, LYSO has also a natural radioactivity caused by the lutetium present in the scintillator. 
\begin{figure}[htbp]
  \centering
%  \subfigure[\label{fig:AlphaLYSObg}]{\includegraphics[width=.4\textwidth]{Images/AmLYSObg.png}}
%  \subfigure[\label{fig:AlphaLYSOsig}]{\includegraphics[width=.4\textwidth]{Images/AmLYSOsig.png}}
  \includegraphics[width=.7\textwidth]{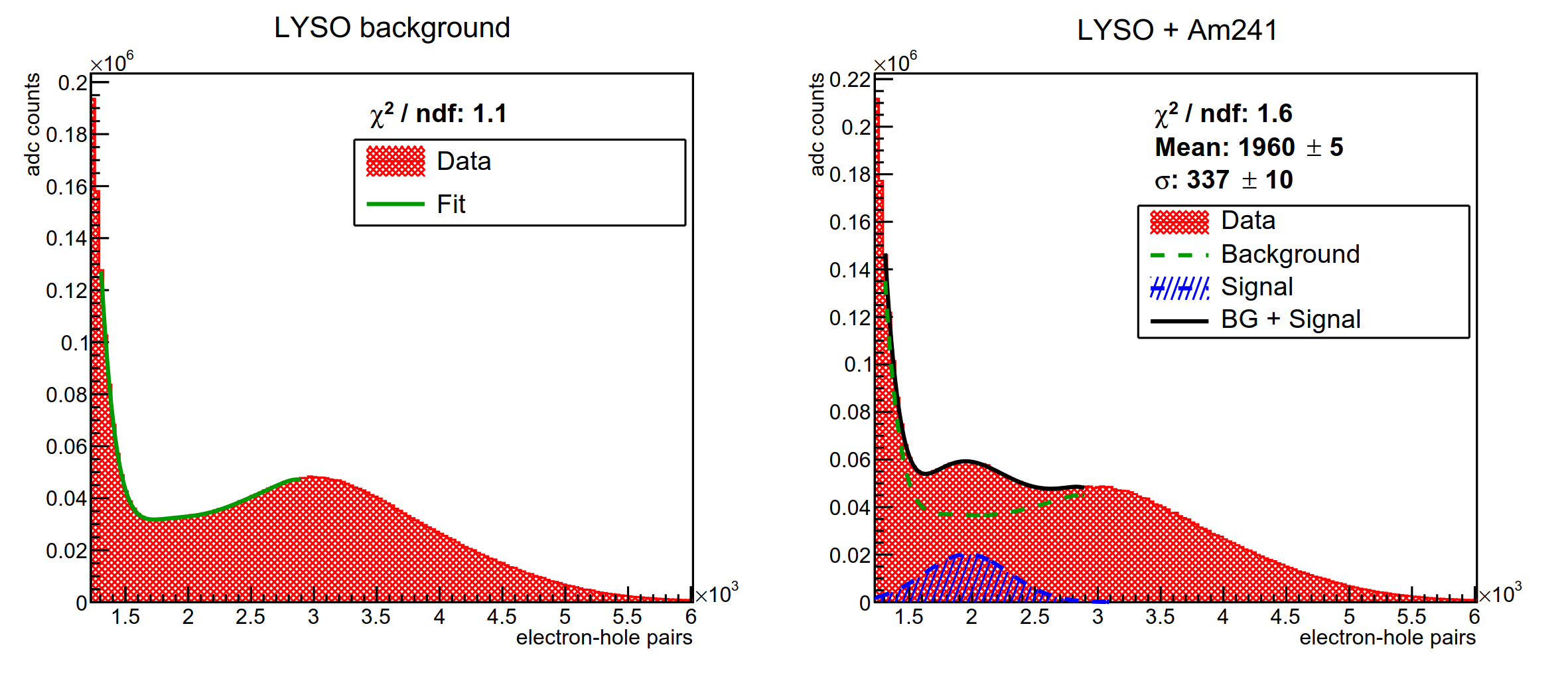}
  \caption{\label{fig:AlphaLYSO} Measured pedestal and natural radioactivity spectrum for the LYSO crystal (red) and in green the region where the source signal is expected to be present (left). Measured  $\alpha$ particles energy in the LYSO crystal with the $\alpha$ energy Gaussian fit (blue) and sum of both (black) (right). }
\end{figure}
This radioactivity generates signals even in absence of an external source and has a distribution that can not be fitted by a Gaussian. The spectrum measured in the absence of the $^{241}$Am source is shown in figure \ref{fig:AlphaLYSO} where in addition to the pedestal, there is a second peak at a higher value due to the natural radioactivity. 
The spectrum generated by the internal radioactivity of LYSO is asymmetrical and of not obvious parametrisation. We opted to fit the additional part of the spectrum with a polynomial function. The result of such fit with a six degree polynomial plus a Gaussian for the pedestal, is shown at left in figure \ref{fig:AlphaLYSO}. 
For the analysis of the $^{241}$Am spectrum, we then subtracted the green part shown in figure. In practice we fitted an additional Gaussian function in the green region of the spectrum were the $^{241}$Am signal manifests itself. The results of the fit are illustrated at right figure \ref{fig:AlphaLYSO}, in which the signal of the $\alpha$ particles (blue) falls in between the pedestal and natural  radioactivity peaks. On this figure, the green and red line represents the background fit, the blue line presents the signal fit and the black one is a sum of all functions. 
For LYSO the results are: mean value equal to 1960 $\pm$ 5 electrons and resolution of 17.2$\%$.

The summary of all measurements is reported in table \ref{tab:AmPeaks}. The measured amplitude summarised in table \ref{tab:AmPeaks} are not in good agreement with the values presented in table \ref{tab:scintSignLevel}.
\begin{table}[htbp]
  \centering
  \caption{\label{tab:AmPeaks} Amplitude and resolution of the signal generated by $\alpha$ particle 5.5 MeV of $^{241}$Am in three scintillators. }
  \smallskip
  \begin{tabular}{|c|c|c|c|}
    \hline
    Scintillator & Mean (electrons) & Mean ($\%$ of CsI(Tl)) & Resolution for a 5.5 MeV ($\alpha$) \\
    \hline
    %CsI(Tl) & 26575   $\pm$ 12    &         100             & ( 4.9 $\pm$ 0.06)$\%$  \\
    %BGO     &  2032   $\pm$  1  & (7.65 $\pm$ 0.01)$\%$ & (19.0 $\pm$ 0.06)$\%$  \\
    %LYSO    &  1907 $\pm$  2    & (7.18 $\pm$ 0.01)$\%$ & (20.3 $\pm$ 0.4)$\%$  \\
    
 CsI(Tl) & 27792   $\pm$ 3    &         100             & ( 3.5 $\pm$ 0.01)$\%$  \\
 BGO     &  2032   $\pm$  1  & (7.31 $\pm$ 0.01)$\%$ & (19.2 $\pm$ 0.1)$\%$  \\
 LYSO    &  1960 $\pm$  5    & (7.05 $\pm$ 0.02)$\%$ & (17.2 $\pm$ 0.5)$\%$  \\

    \hline
  \end{tabular}
\end{table}
As pointed out in the presentation of the scintillators characteristics, the literature has different values depending on the sources. An additional reason for the disagreement could be due to the particle source: the light yield of scintillators usually refers to $\gamma$ rays and can be different for other kind of particles. 
% For LYSO only the studies for X and  $\gamma$ rays with energies much lower than 5.5 MeV have been found \cite{scintillatorsComparision:RihuaMa2008},\cite{scintillatorsComparision:Johan2008}.
There are few works with direct comparison of these three scintillators and even in those where LYSO is mentioned, it is not possible to deduce the absolute value of the light yield \cite{scintillatorsComparision:Thiel2008}. 
%+ Misure con cristalli scintillanti: 
%  - Confronto tra più materiali scintillanti CsI(Tl), LYSO, BGO con segnali da sorgente Am-241 [ma questi sono stati fatti con i cristalli piccoli del preprototipo da 2x2x2 cm3]
%  - Studio della risposta dei cristalli 36x36x36 mm3 di CsI(Tl) con sorgente di Am-241
%  - Studio dell'efficienza della raccolta di luce per diversi materiali:
%     - Simulazioni con cristalli 36x36x36 mm3 e due materiali (Teflon e mylar alluminato) variando il trattamento delle superfici del cristallo e l'energia depositata.
%     - Misure in laboratorio: fogli di alluminio vs Teflon (non so con che cristalli è stato fatto), studio degli strati di Teflon necessari per migliorare la raccolta di luce (cristalli 36mm), Teflon vs Vikuiti (non so con che cristalli è stata fatta la misura)
%  [Olek ha fatto anche delle misure sulla variazione della luce raccolta a seconda della posizione della sorgente...volendo si può aggiungere ma non so se è funzionale per l'organicità del l'articolo...
%    anche perchè lui questa misura l'ha fatta con il Teflon anche se dai test precedenti si evince che la cosa migliore è il Vikuiti]
%  - Misura della risoluzione energetica con i raggi cosmici

%Investigation of different scintillators and wrapping materials have been performed and several measurements have been carried out on the most promising materials with $^{241}$Am source and cosmic rays to find out the most appropriate materials in terms of signal level, noise and resolution.

 The light yield measured with $\alpha$ particles can be useful for comparisons, but measurements with MIPs are more pertinent to our calorimetric application. In fact an alpha releases a large amount of energy in a very small volume at the surface of the crystal, in a way that is substantially different from that of a MIP or a particle shower. Cosmic rays provide an easy to use source, but given the crystal dimensions, such measurements are limited by a low rate of the order of 0.01 Hz. For this reason we performed cosmic ray measurements only on full sized final crystals (36x36x36mm$^3$)(see section \ref{sec:CRcalibration}). Results on the many high energy particle test beams we performed during these years on the various prototypes will be reported in a dedicated paper that is in preparation.

\subsection{Full sized crystal calibration}\label{sec:FScrystalCalib}

\begin{figure}[htbp]
  \centering
  \subfigure[\label{fig:testedal}]{\includegraphics[width=.4\textwidth]{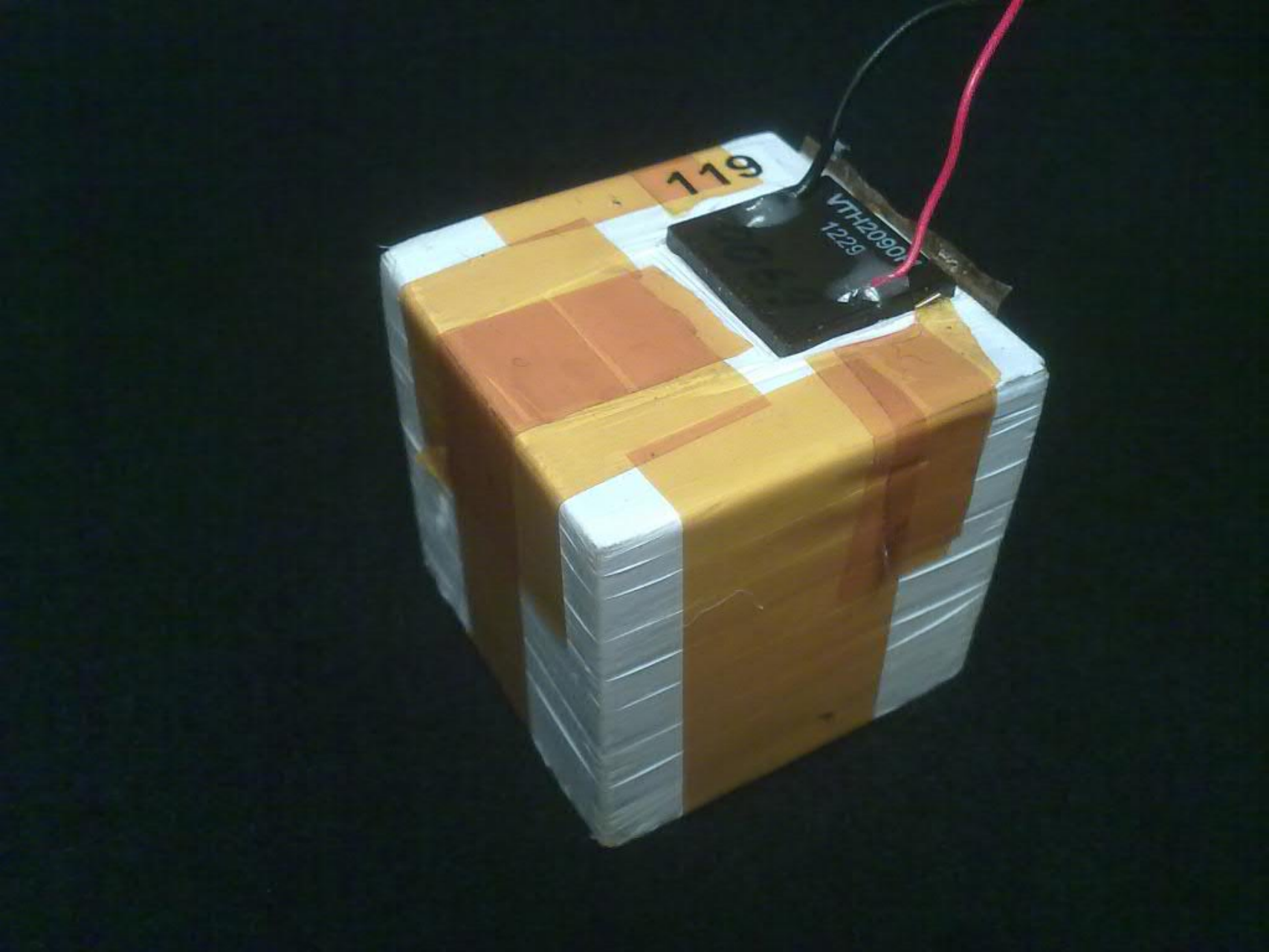}}
  \subfigure[\label{fig:testedSpectrum}]{\includegraphics[width=.48\textwidth]{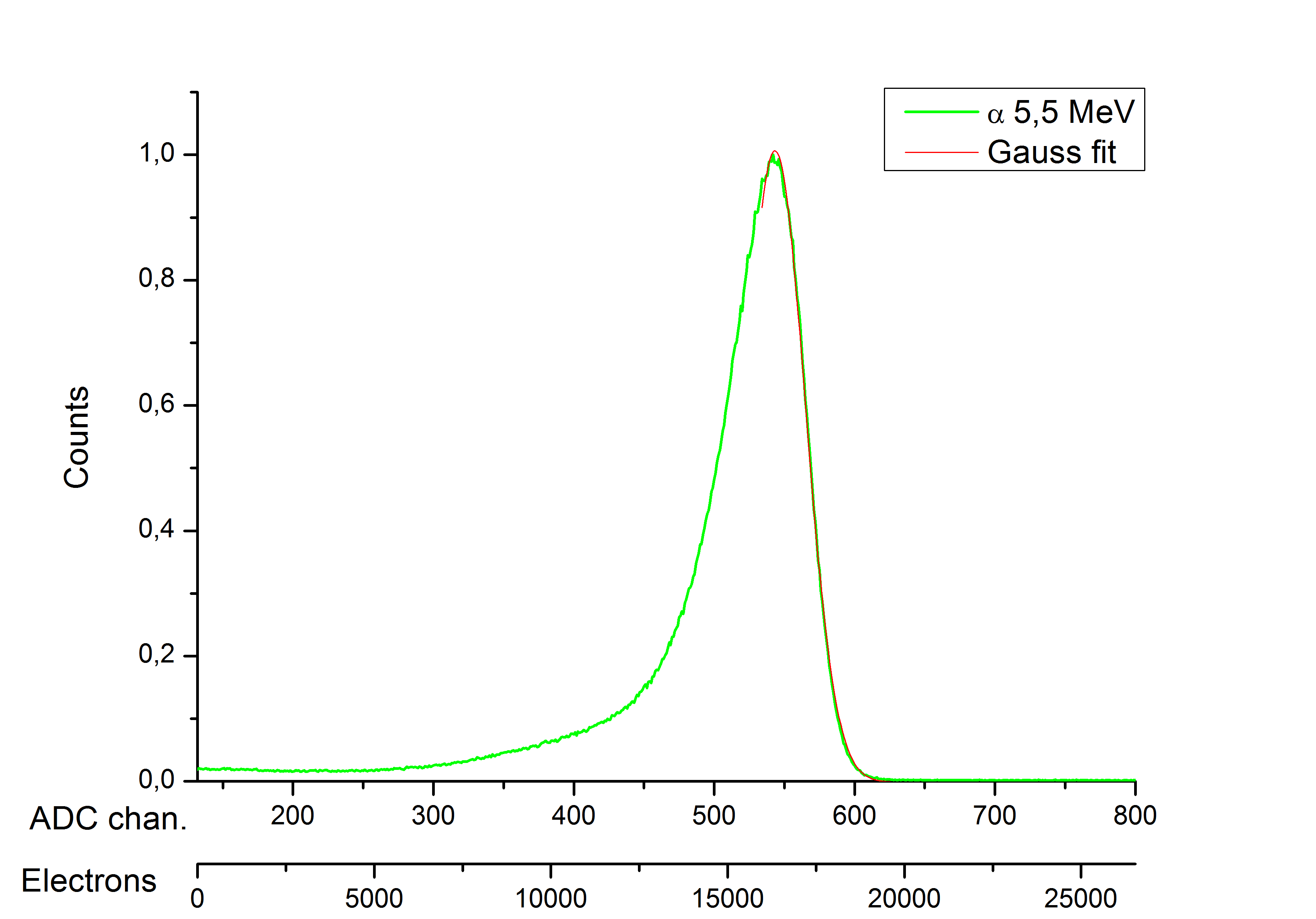}}
  \caption{\label{fig:testedCrystal} Full sized CsI(Tl) cubic crystal with the 100 $\mu$m Teflon wrapping and the VTH2090 photodiode (a).  $^{241}$Am $\alpha$ particle spectrum measured with a full sized CsI(Tl) crystal (b).}
\end{figure}

The light yield results presented in the previous section, together with the considerations on costs and availability, have led us to choose  CsI(Tl) as the scintillator for the full sized CaloCube crystals (36$\times$36$\times$36 mm$^3$). We procured roughly 700 of these crystals in order to build the prototype to be used on test beams.  We performed some preliminary tests on the light yield of a few of these full sized crystals with the same $^{241}$Am source used in the previous tests. The crystals were covered with two layers of Teflon for a total thickness of 100 $\mu$m and the source was placed on the opposite side with respect to the position of the PD. The crystal under test with attached PD is shown in figure \ref{fig:testedCrystal}.

The measured spectrum is shown in the right part of figure \ref{fig:testedCrystal}: the green line shows measured data and the red line is the gaussian fit. The fit has been done for the right part of the peak only, because of the non-Gaussian tail on the left, which is due to the energy losses in the wrapping material (cf. section \ref{sec:NonGaussiaTail}). Accordingly we estimate a signal of about 16300 photoelectrons with an energy resolution of 5.2$\%$.

We derive the number of expected photoelectrons from the following equation:
\begin{equation}
S_e = L_{ph} \cdot E \cdot L \cdot F_g \cdot Q_{PD} \label{eq:signalElectrons}
\end{equation}
where S$_e$ is the signal level in photoelectrons, L$_{ph}$ is the light yield of the scintillator, E is the energy deposited in scintillator (5.5 MeV $\alpha$ particle), L is the light collection efficiency in the scintillator, F$_g$ is the geometric factor (i.e. the ratio of the scintillator side area and the photodiode active area), and Q$_{PD}$ is the quantum efficiency of the photodiode. 

We obtained an estimate of the order of 15000 photoelectrons, indicating a qualitative agreement with the measured value. In fact there are many uncertainties in this estimate, not least the light yield of the scintillator and the light collection efficiency, that make it very difficult if not impossible to achieve a reliable quantitative agreement. While the light yield depends on the chosen manufacturer, the light collection efficiency value and reproducibility depend on the crystal surface finish and chosen wrapping material. We thus peerformed an extensive study starting with simulations that were then compared with actual measurements.

\subsection{Light collection efficiency}

We have performed a systematic study of light collection efficiency for various wrapping materials on our full sized crystal. We have performed various simulations, following previous studies \cite{LightCollEff:Onyshchenko2005,GammaVsChargedParticles:Gupta2009,Larochelle1994,Aharonian2008}, and focusing our efforts on Teflon, aluminized mylar, and an enhanced specular reflector (Vikuiti  \cite{vikuiti}). We have also varied the crystal surface finish from matt to fully polished. The actual measurements were made with a full sized CsI(Tl) cube coupled to a VTH2090 PD using an $^{241}$Am $\alpha$ source to generate the scintillation signal.

\subsubsection{Simulation}

\begin{figure}[htbp]
  \centering 
  \subfigure[\label{subfig:aepsilon1}]{\includegraphics[width=.4\textwidth]{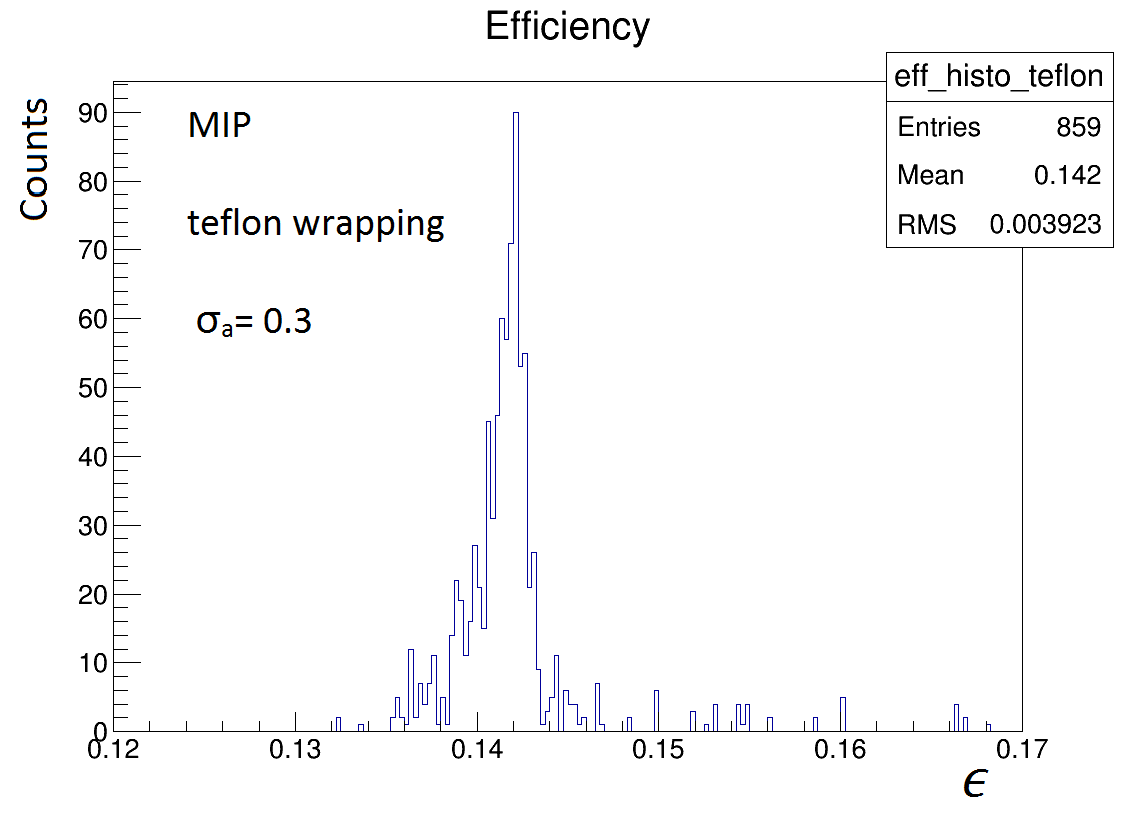}}
  \subfigure[\label{subfig:bepsilon2}]{\includegraphics[width=.4\textwidth]{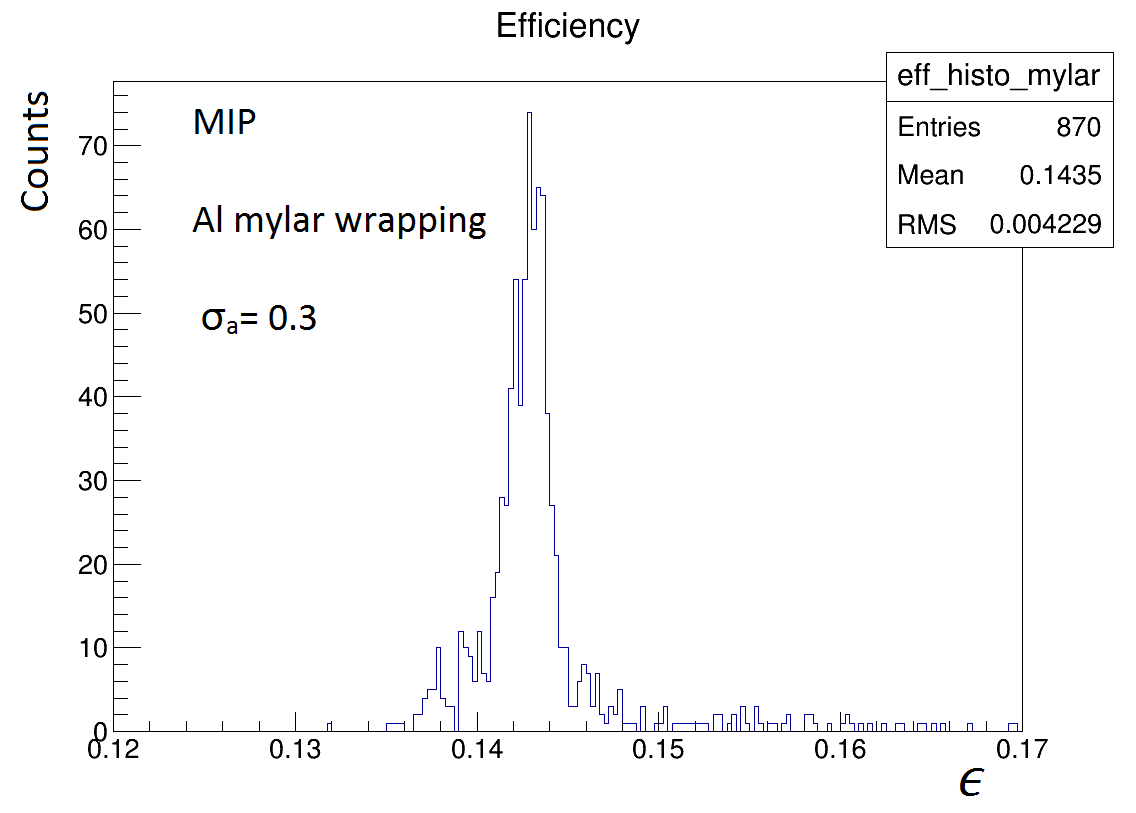}}
  \hfill  
  \subfigure[\label{subfig:aepsilon3}]{\includegraphics[width=.4\textwidth]{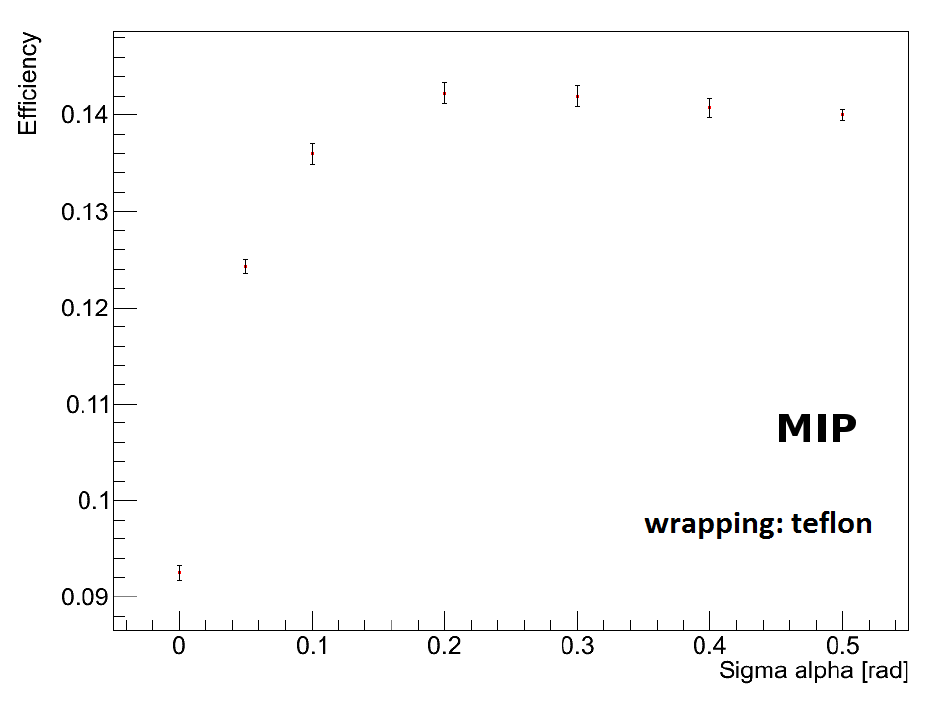}}
  \subfigure[\label{subfig:bepsilon4}]{\includegraphics[width=.4\textwidth]{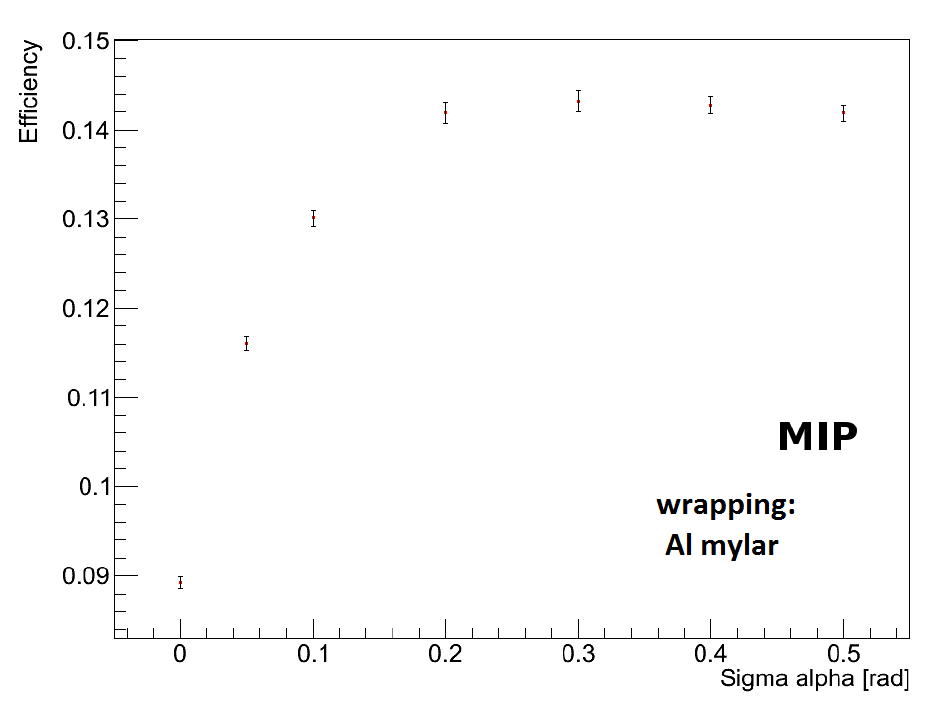}}
  \caption{\label{fig:TeflonWrappingSimMIP} Efficiency  distributions generated by MIPs in the crystal with Teflon and Aluminised Mylar wrapping and with roughness $\sigma_{\alpha}$ = 0.3 (Top).  Efficiency as a function of roughness  ($\sigma_{\alpha}$) for Teflon and Aluminised Mylar (Bottom). }
\end{figure}

To acquire a feeling on the issue at hand, we  performed optical ray tracing simulations using GEANT.  The initial photons were produced in the crystal following either a MIP interaction from any arbitray direction or an electromagnetic shower from 100 GeV electrons incident on a non-scintillating absorber situated in front of the crystal. The following parameters were varied in the simulations: surface finish (roughness $\sigma_{\alpha}$ in the range 0.0 $\div$ 0.5 radiant) and wrapping material (no wrapping, Teflon, aluminized mylar). 
The light detection efficiency $\epsilon$ seen in the plots of figure~\ref{fig:TeflonWrappingSimMIP} is defined as the ratio of the number of photons detected by the PD to the number of photons produced in the scintillation process (thus it includes also the quantum efficiency of the PD). 
%  and it is obtained by the product of the light collection efficiency $\epsilon_r$ and the quantum efficiency of the PD $\epsilon_{QE}$.
% Similarly $\epsilon_r$ is defined as the number of photons interacting in the PD over the number of photons generated in CsI, while $\epsilon_r$ is the number of interacting photons divided by the number of detected photons. 
In figure \ref{fig:TeflonWrappingSimMIP} a few examples of the results obtained are shown. The top two plots in the figure, are typical efficiency distributions obtained by considering the photons collected in the PD area from all simulated particle interactions. These distributions were then fitted with Gaussians to obtain the respective central values. The values obtained are shown in the bottom part of figure  \ref{fig:TeflonWrappingSimMIP} as a function of roughness $\sigma_{\alpha}$ for two choices of wrapping material.\\
For Teflon or aluminized mylar wrapping, the light collection efficiency starts from 9$\%$ for a polished surface and grows rapidly with $\sigma_{\alpha}$ up to 14$\%$. This behaviour is consistent with the fact that a rougher surface diffuses the photons and thus reduces the amount of reflections needed to arrive at the PD entrance window \cite{roughness:Ishibashi}. This effectively reduces the probability that the photon will be absorbed by the wrapping (because of the lower number of reflections) or in the volume of the crystal (because of the shorter path travelled).\\
We have found, that in the case of a crystal without wrapping, the efficiency descends from roughly 6$\%$ for a polished crystal to a stable value of less than 2$\%$ with increasing surface roughness. This is expected, since an internal photon will be more likely to undergo total reflection if the crystal surfaces are polished than if they are sanded (causing diffusion to the outside to occur). 

The results of the many simulations we performed, are in agreement with previous data \cite{roughness:Ishibashi} available for the crystals with comparable dimensions.  Without wrapping, crystals with polished surfaces (low $\sigma_{\alpha}$) show the highest efficiencies, whereas crystals with rough (diffusive) surfaces show better performance when using both reflective and diffusive wrappings. 
% Table \ref{tab:X0sim} shows the light collection efficiency in the case of an electromagnetic shower generated by an electron with energy 100 GeV interacting in the absorber in front of the crystal and MIPs. The same configuration (aluminized mylar and $\sigma_{\alpha}$ = 0.3) has been used in all cases and the result is that the light collection efficiency values are very similar in all the cases. 
% \begin{table}[htbp]
%   \centering
 %  \caption{\label{tab:X0sim} Light collection efficiency for electromagnetic showers for different depth of absorber in units of radiation length (X$_0$ =1.86 cm for CsI) and for MIPs. }
 %  \smallskip
 %  \begin{tabular}{|c|c|}
  %   \hline
 %    X$_0$ & $\epsilon_r$ \\
  %   \hline
  %   0 & 0.14340 $\pm$ 0.00017 \\
  %   1 & 0.14356 $\pm$ 0.00013 \\
  %   2 & 0.14363 $\pm$ 0.00010 \\
  %   \hline
  %   MIPs & 0.1433 $\pm$ 0.0012 \\
   %  \hline
 %  \end{tabular}
% \end{table}

\subsubsection{Measurements}

Using our simulations as a guiding input, we chose to have a matted crystal surface finish. Thus all the full sized crystals used for the Calocube project were sanded on five sides  and polished only on the side where the PDs are glued.
On some of the CsI(Tl) crystals we have performed various  light collection efficiency measurements with different wrapping materials (Teflon, aluminised mylar, Vikuiti~\cite{vikuiti}). These were then repeated for various Teflon thicknesses. We also measured the uniformity of response as a function of the  $^{241}$ Am $\alpha$ source position.

\textbf{Aluminised Mylar foil and Teflon}\\
For this measurement we reverted to small sized crystals used in the scintillator comparison. The CsI(Tl) cubes were wrapped with aluminised Mylar and with Teflon wrapping. Each crystal was coupled to VTH2090 PDs and tested with an $^{241}$ Am $\alpha$ source. An aperture of 2 mm$^2$ was opened in the wrapping to let the $\alpha$ particles penetrate inside the crystal.
\begin{figure}[htbp]
  \centering
  \subfigure[\label{fig:AlWrappingMeas}]{\includegraphics[width=.44\textwidth]{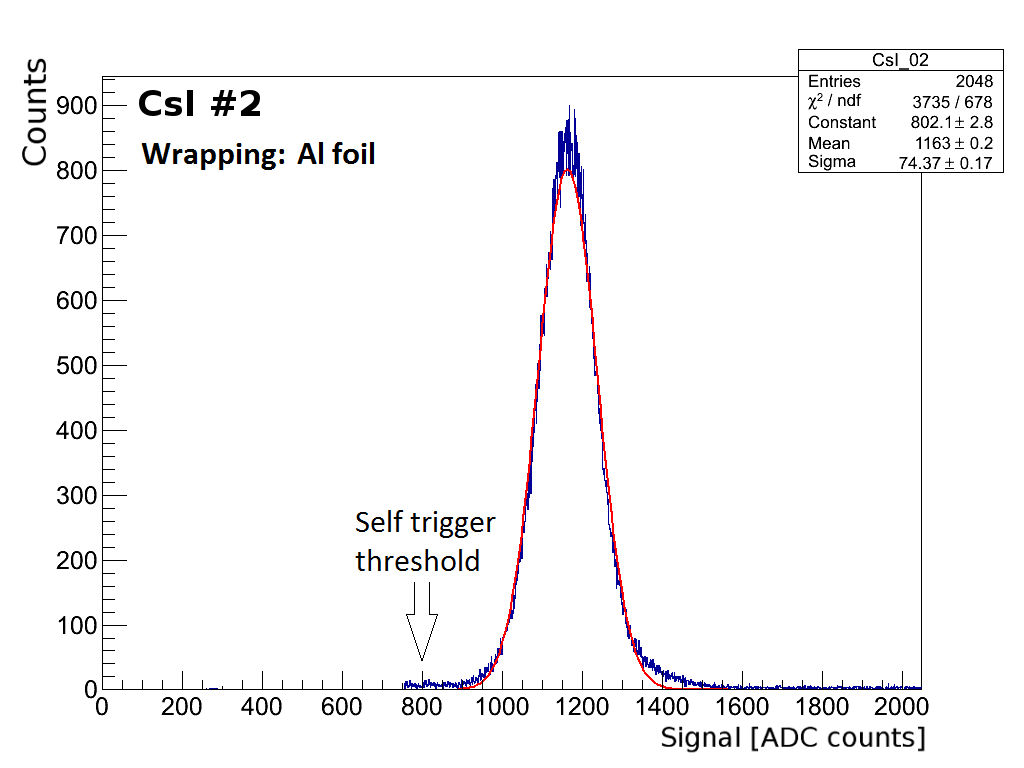}}
  \subfigure[\label{fig:TeflonWrappingMeas}]{\includegraphics[width=.44\textwidth]{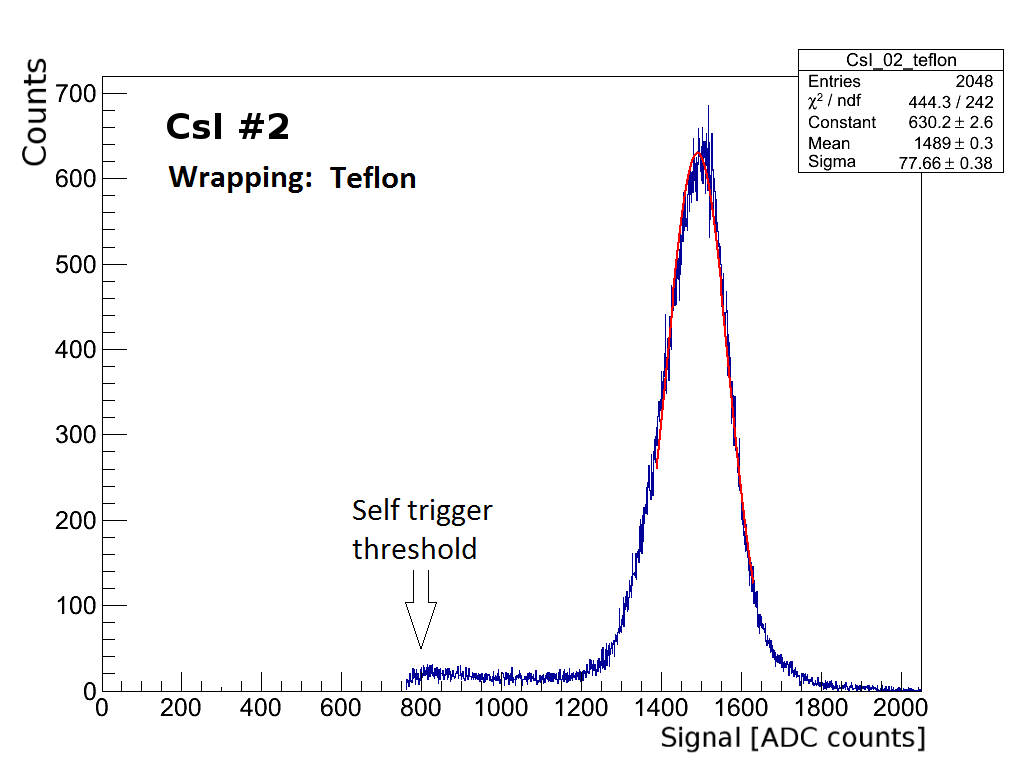}}
  \caption{\label{fig:wrappingMeas} $^{241}$Am $\alpha$ particle spectrum from CsI(Tl) crystal with aluminum foil (a) and 50 $\mu$m Teflon (b) wrapping (blue line) and Gaussian fit of the peak (red line). A relatively high threshold was used in these plots to eliminate random self triggered events due to external pickup.  The pedestal value of 203 ADC channels was taken from figure\ref{fig:alphaSourceSpectrumFITS}.}
\end{figure}
Typical spectra measured and their Gaussian fits, are shown in figure \ref{fig:AlWrappingMeas} and \ref{fig:TeflonWrappingMeas} for aluminised Mylar and Teflon wrapping respectively. 
\begin{table}[htbp]
  \centering
  \caption{\label{tab:wrappingMeas} Gaussian fit results (amplitude and resolution) for two similar crystals, using the 5.5 MeV emission peak of of $^{241}$Am $\alpha$ particle. Results are consistent within 10\% . }
  \smallskip
  \begin{tabular}{|c|c|c|c|c|}
    \hline
    CsI(Tl)+PD & Wrapping & Centroid [ADC ch.] & $\sigma$ [ADC ch.] & Resolution $\%$ \\ 
    \hline
    2 & Al     & 960 & 74 & 7.7 \\
         & Teflon & 1286 & 78 & 6.1 \\
    3 & Al     & 1068 & 90 & 8.4 \\
         & Teflon & 1435 & 90 & 6.2 \\
    \hline
  \end{tabular}
\end{table}
We found that a Teflon wrapping gives a signal increase of about of 35$\%$, while the energy resolution is improved up to roughly $\sim$6$\%$ for both crystals. The results for two crystals with both wrappings are presented in table \ref{tab:wrappingMeas}.  Both show a marked improvement from aluminised Mylar to Teflon.

\textbf{Light collection efficiency for different Teflon thicknesses}\\
We used full sized CsI(Tl) cubic crystals to study the influence of Teflon wrapping thickness. 
We started with two layers of Teflon tape ($\sim$100 \textmu m) which was a standard configuration used on a small (roughly 100 crystals) Calocube protoype used to collect preliminary electromagnetic shower data. 
\begin{figure}[htbp]
  \centering
  \subfigure[\label{fig:multiteflonMeas}]{\includegraphics[width=.45\textwidth]{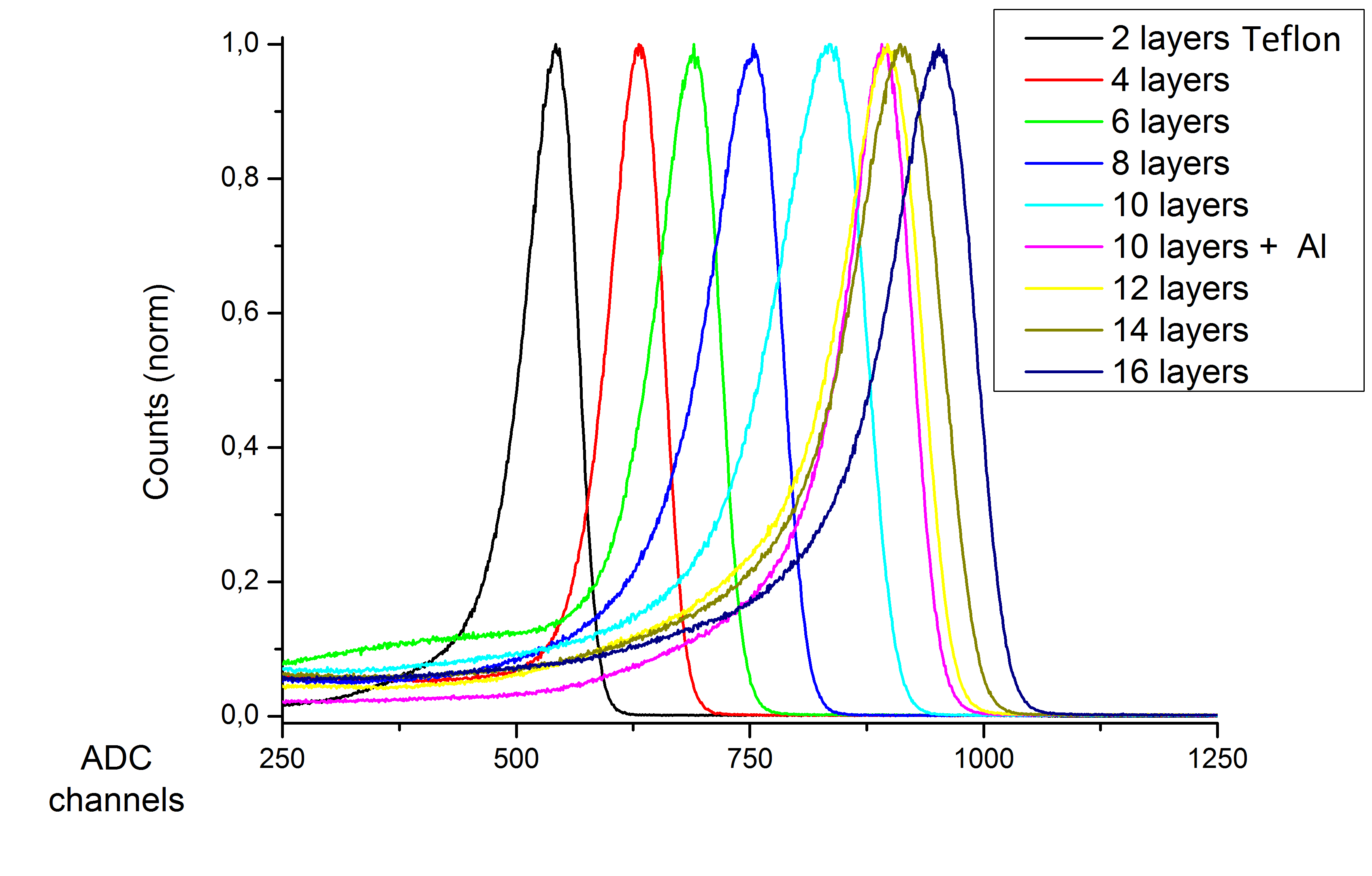}}
  \subfigure[\label{fig:TeflonFit}]{\includegraphics[width=.45\textwidth]{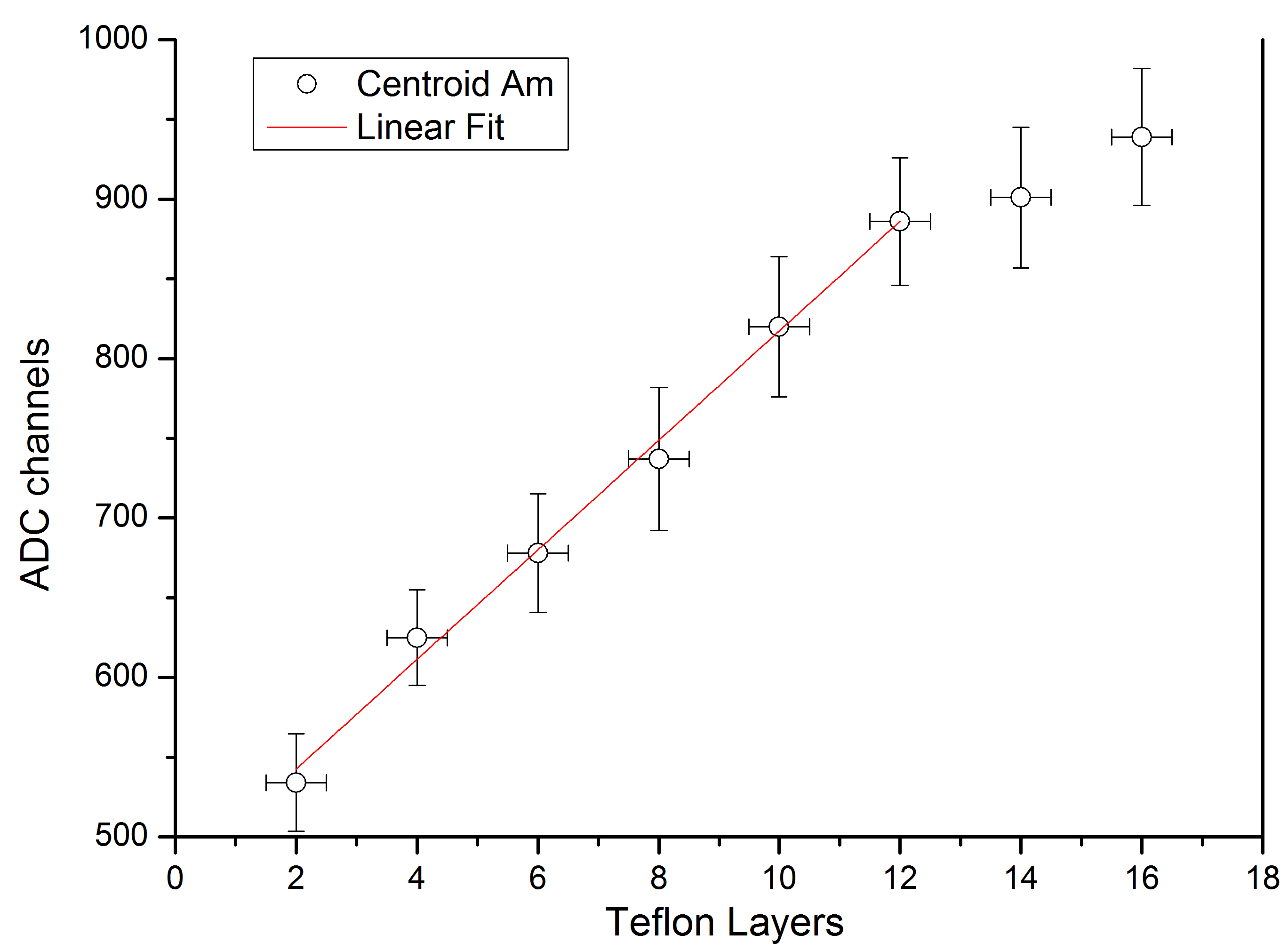}}
  \caption{\label{fig:TeflonThickness}  $^{241}$Am $\alpha$ particle spectra measured by PD+CsI(Tl) for different thicknesses of Teflon wrapping (a). Signal amplitude (centroid of the previous spectra) dependence on Teflon thickness (b). }
\end{figure}
The $\alpha$ spectrum was measured every time after adding two additional layers of Teflon, till we reached a 16 layers (800 \textmu m ) thickness. After 10 layers we also added an aluminised Mylar external wrapping which further increased the signal measured. All the spectra are shown in figure \ref{fig:TeflonThickness}. Each peak has been fitted with a Gaussian function and plotted as a function of Teflon thickness. The peak and width values are listed in table \ref{tab:TeflonThicknessFit}. 
\begin{table}[htbp]
  \centering
  \caption{\label{tab:TeflonThicknessFit} Gaussian fits (amplitude and width in ADC channels) for different thicknesses of Teflon wrapping. Peak values are different from what previously shown due to different crystal size and different scintillator manufacturer. }
  \smallskip
  \begin{tabular}{|c|c|c||c|c|c|}
    \hline
    Teflon layers & Peak (ADC ch.) & $\sigma$ (ADC ch.)&  Teflon layers & Peak & Width \\ 
    \hline
     2 & 534 & 30.6 & 10 & 820 & 43.9 \\
     4 & 625 & 30.1 & 12 & 886 & 40   \\
     6 & 678 & 37.3 & 14 & 901 & 44.1 \\
     8 & 737 & 44.8 & 16 & 939 & 43   \\
    \hline
  \end{tabular}
\end{table}

We have found that the signal with 16 Teflon layers wrapping increases by a factor $\sim$ 1.76 respect to the signal with only 2 Teflon layers. The signal amplitude rises linearly with the number of Teflon layers up to 12 layers (600 \textmu m). After 12 layers, the signal increase becomes less pronounced and begins to flatten out even  14-16 layers. The measurements indicate that thicker layers of Teflon wrapping result in better performance. Also the resolution increases reaching values of the order of 5\%. In practice though, this is difficult to implement as the wrapping with so many layers (plus the aluminised Mylar) is a time consuming procedure that also increases dead space in the actual calorimeter.

\begin{figure}[htbp]
  \centering
  \subfigure[\label{fig:TeflonStdMeas}]{\includegraphics[width=.4\textwidth]{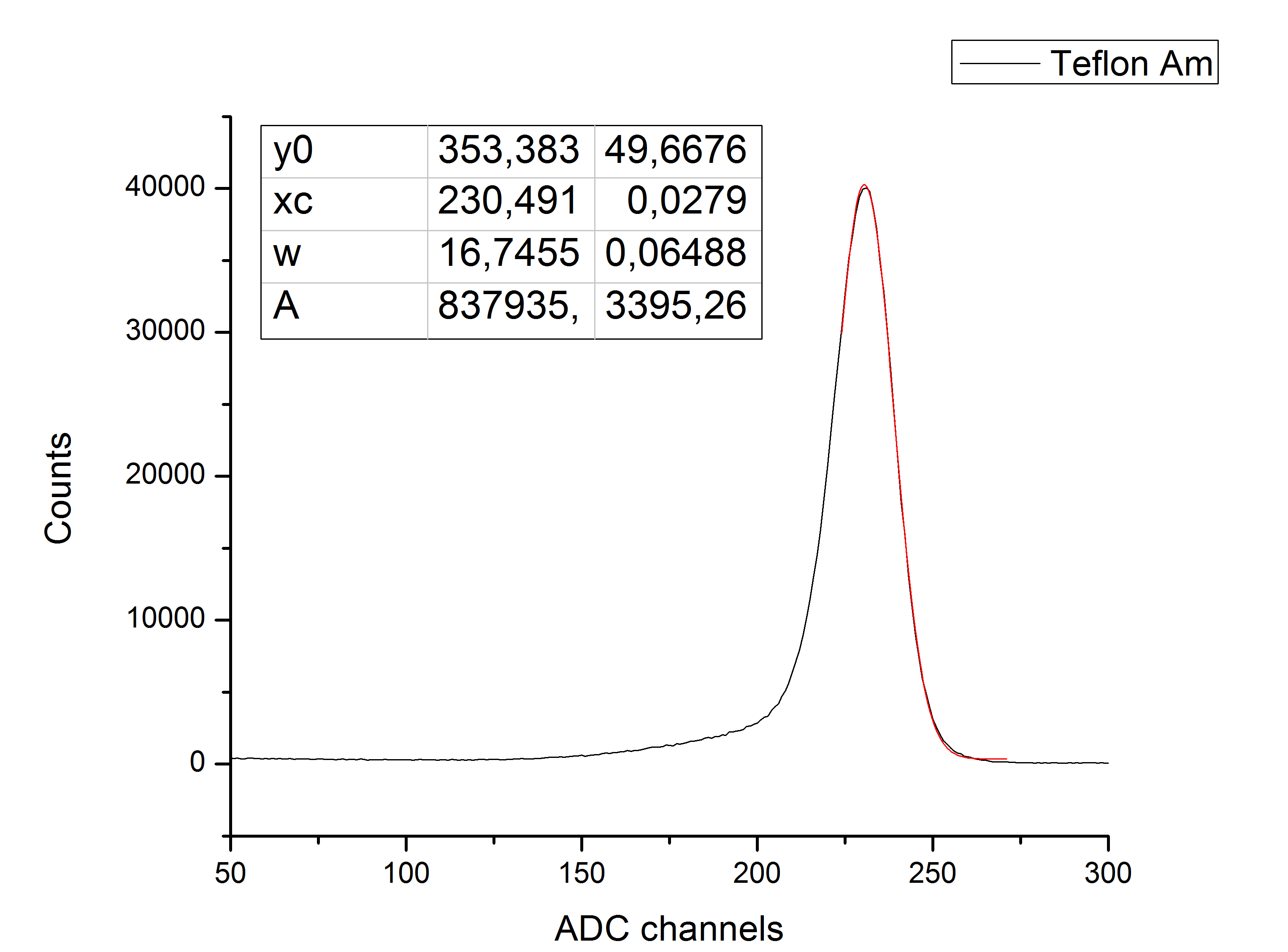}}
  \subfigure[\label{fig:VikuitiMeas}]{\includegraphics[width=.4\textwidth]{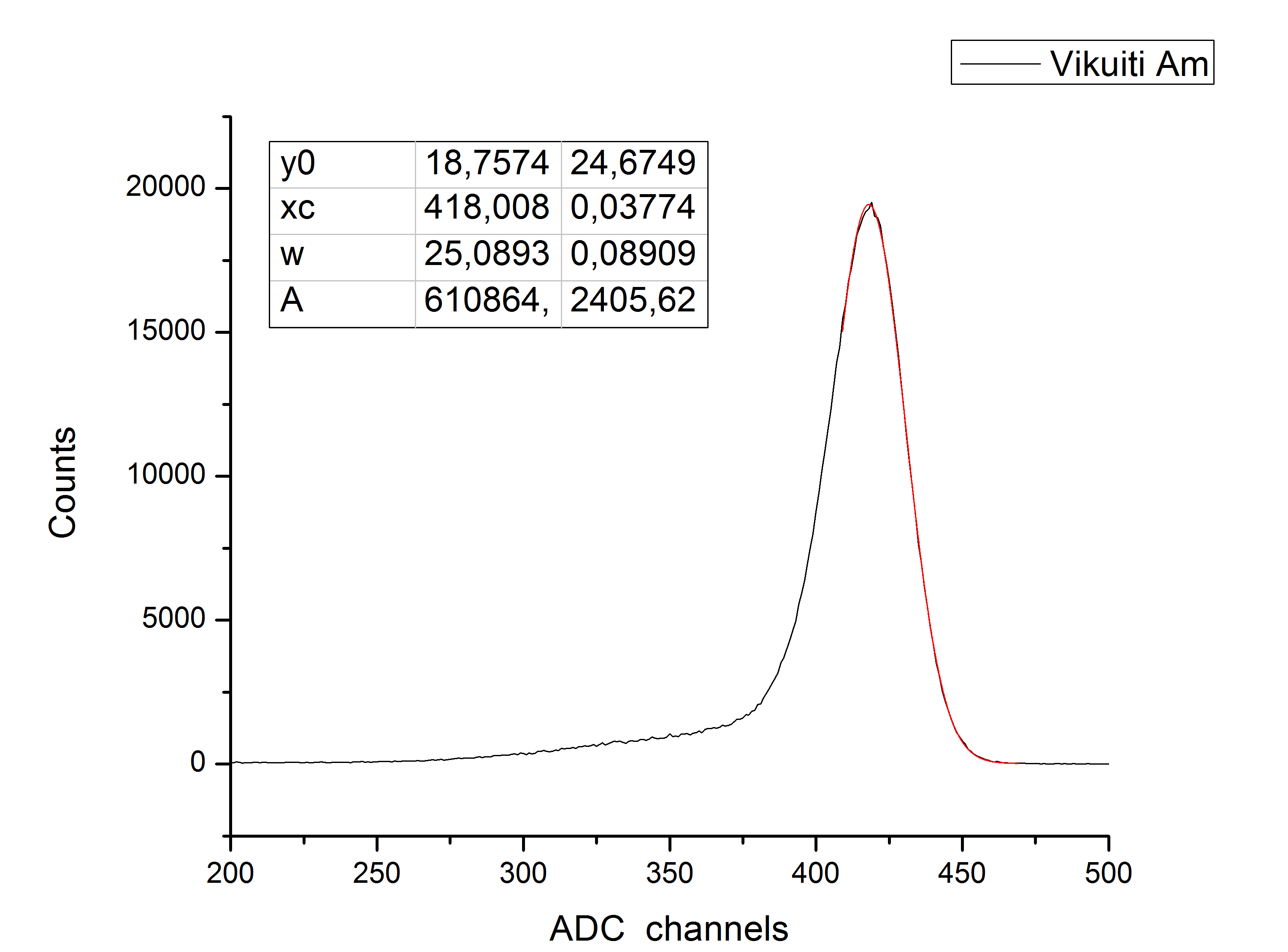}}
  \caption{\label{fig:TeflonVSvikuiti} $^{241}$Am $\alpha$ particle spectrum from CsI(Tl) crystal with $\sim$100$\mu$m Teflon (a) and Vikuiti (b) wrapping (black line) and Gaussian fit of the peak (red line). }
\end{figure}

\textbf{Comparison between Teflon and Vikuiti}\\
Vikuiti Enhanced Specular Reflector (ESR) is an ultra-high reflectivity, mirror-like optical enhancement film that is widely used in the industry \cite{vikuiti}. The thin, non-metallic, 100$\%$ polymer film, built with multi-layer polymer technology, has a 98$\%$ reflectance across the visible spectrum. Vikuiti ESR is an excellent specular surface reflector. We decided to test this type of wrapping on one of the full sized crystals to see how it performed respect to the Teflon wrapping. The signal distribution measured with the crystal wrapped with two Teflon layers and then with Vikuiti is shown in figure \ref{fig:TeflonStdMeas} and \ref{fig:VikuitiMeas} respectively. The Gaussian fits to the spectra give a peak value of 230 ADC channels for the 2 layer Teflon wrapping and 418 ADC channels for Vikuiti. While the absolute ADC values are different from the other measurements presented in this section (534 \ref{tab:TeflonThicknessFit}) because of a different gain setting on our AMPTEK amplifier, the increase by a factor of 1.8 is the same as that obtained previously with a 16 layer Teflon wrapping.
This result was confirmed with more measurements on more full sized crystals, we thus chose to use Vikuiti wrapping for all the CsI(Tl) crystals (roughly 700) used for the Calocube protoype.

\textbf{Light collection uniformity for the different crystal faces}\\
We also performed various measurements regarding the uniformity of light collection as a function of the $\alpha$ source placement, in particular the crystal face used. We measured the signal amplitude, with the source placed in front of each different face of the crystal, using a full sized CsI(Tl) crystal with a 100 micron Teflon wrapping.
\begin{figure}[htbp]
  \centering
  \subfigure[\label{fig:sourcepos}]{\includegraphics[width=.45\textwidth]{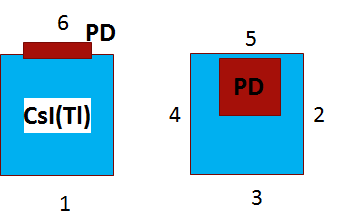}}
  \subfigure[\label{fig:posmeas}]{\includegraphics[width=.45\textwidth]{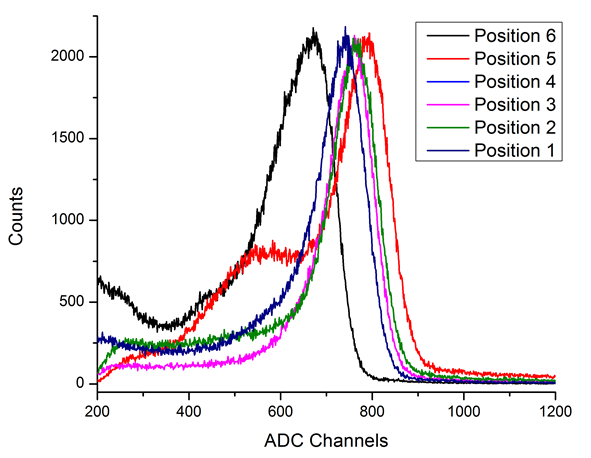}}
  \caption{\label{fig:sourcePos} Schematic view of PD and source positions with respect to the scintillating cubic crystal (a). Signal amplitude dependence on source position. Position 5 has a significant shoulder which is probably due to a slight misalignement of the source respect to the aperture in the wrapping material. The ADC scale changes from the one of table  \ref{tab:TeflonThicknessFit} due to a different F.E. gain (b).}
\end{figure}
A schematic view of the PD-crystal setup and the six positions of the $^{241}$Am source used for the measurements is shown in figure \ref{fig:sourcepos}. For each data acquisition the source was placed in the centre of the corresponding face (with an aperture of 2 mm$^2$). The acquired spectra are shown in figure \ref{fig:posmeas}. Position 5, which is the lateral face closest to the PD, gives the highest signal (810 ADC channels). The other faces anyway show similar peak values (order of 780 ADC channels) except for position 6 which shows the lowest signal (650 ADC channels). In fact this is the same face where the PD is glued and photons from the alpha particle have to travel much further in order to reach the PD. These results demonstrate a very good uniformity of response especially when considering that they were obtained with a two layer Teflon wrapping. With the Vikuiti wrapping chosen for the prototype the uniformity improves further.

\subsection{CsI(Tl) crystal response with cosmic rays} \label{sec:CRcalibration}

\begin{figure}[htbp]
\centering
\includegraphics[width=.75\textwidth]{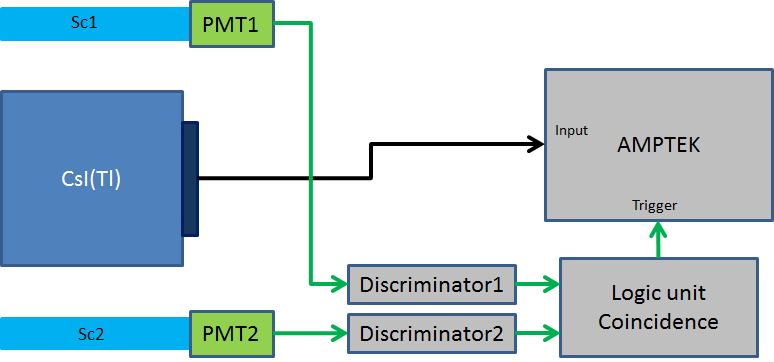}
\caption{\label{fig:CRblock}  Block diagram of the Cosmic rays measurement bench. }
\end{figure}

All the measurements described so far have used  a $^{241}$Am $\alpha$ particle source. This has obvious advantages in terms of particle rate and ease of measurements in general but needs to be compared to the crystal response to MIPs since the crystals will be used to build a calorimeter for highly energetic particle showers. We have thus performed some comparisons using a cosmic ray bench whose schematic diagram is shown in figure  \ref{fig:CRblock}. An external trigger system consisting of two fast plastic scintillators 30 $\times$ 30  mm$^2$ , coupled to PMTs and placed above and below the CsI(Tl) scintillator cube selects the cosmic ray events. This trigger is used to drive the Amptek data acquisition. The same system is used in self triggering mode with the $\alpha$ source.
% \begin{figure}[htbp]
%   \centering
%   \subfigure[\label{fig:schema}]{\includegraphics[width=.3\textwidth]{Images/Thesis/CRmeasBlockDiagr.png}}
%  \subfigure[\label{fig:crspectrum}]{\includegraphics[width=.5\textwidth]{Images/Thesis/CsI_CR.png}}
%  \caption{\label{fig:CRblock}  Cosmic rays measurement. Block diagram \ref{fig:schema} and Cosmic rays (green) and 5.5 MeV $\alpha$ particles (red) spectra \ref{fig:crspectrum}. }
% \end{figure}

\begin{figure}[htbp]
\centering
\includegraphics[width=.75\textwidth]{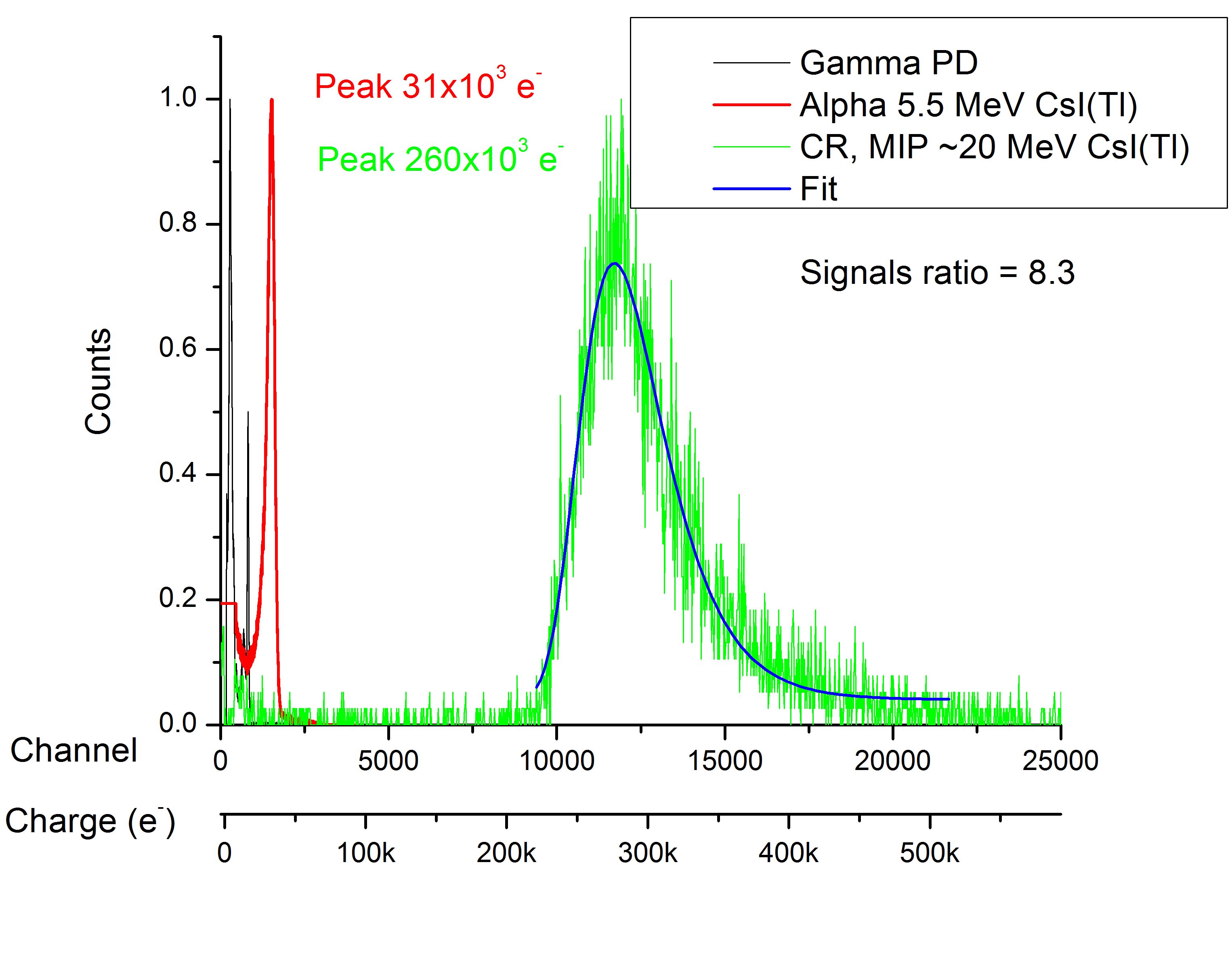}
\caption{\label{fig:CRspectrum}   Cosmic rays spectrum (green), compard to the 5.5 MeV $\alpha$ particle (red) response. See text for an explanation of the differences observed. Also plotted is the PD response to the $^{237}$Np lines used for calibration of the electronic chain (see Subsec.~\ref{subsec:charge}).}
\end{figure}

A MIP in the cube with 36 mm side releases $\sim$20 MeV of energy\footnote{CsI(Tl) provides the following energy loss for a unit of distance: 1 MIP/cm = 1.25 MeV/(g/cm$^2$) $\cdot$ 4.5 g/cm$^3$ = 5.62 MeV/cm. For a CsI cubic crystal of 36 mm side, 1 MIP has energy loss equals to 5.62 MeV/cm $\cdot$ 3.6cm $\approx$ 20 MeV (for orthogonal tracks). }. This is roughly a factor 4  higher than the energy of our $\alpha$ particle source. From the spectra shown in figure  \ref{fig:CRspectrum} the ratio is a factor two higher than what expected. The Landau fit to the cosmic ray spectrum gives a apeak value of 260000 electrons, while with the same electronic chain and Amptek gain settings the $\alpha$ spectrum peaks at 31000 electrons.

Regarding the disagreement between signals for MIP and $\alpha$ particles, one should consider that a MIP produces scintillation within the whole crystal volume, whereas an $\alpha$ particle only in the small local area near the interaction point at the surface. For a MIP this difference can increase the number of the photons detected per unity of deposited energy because of a higher average light collection efficiency. 
Other factors that could contribute are: at low energies the light response of a CsI(Tl) is known to show a non-linear correlation with deposited energy especially for heavy ions, and a dependence of such correlation on both the charge Z and mass A of the detected particle \cite{Larochelle1994,Knoll4th}. These effects also depend on the details of the Tallium doping of the crystal. We have used this last measurement as the reference value for the typical signal level and light collection efficiency to be expected for all the Calocube crystals.

\section{Calorimeter mechanical design studies}
\label{sec:concept}
%\subsection{Introduction}
%\label{sec:concept_intro}
We have investigated the mechanical properties of a carbon fibre tray structure large enough to hold a 28 x 28 crystal array. This was an initial foray for a space qualified design that could be brought into orbit by a rocket. We also designed and realised a smaller tray structure using common polyoxymethilene that was used to build a prototype for particle test beam measurements.

\subsection{Carbon fibre tray prototypes and development}
\label{sec:mechanicsCF}
The final Calocube mechanics must use a lightweight support material, structurally compatible with integration in a standard payload for a space mission, so the decision was taken to use trays to support the crystals, made out of space qualified, high modulus carbon fibre (CF) and epoxy resin.
\begin{figure}[htbp]
\centering 
% \begin{center}/\end{center} takes some additional vertical space
\includegraphics[width=.44\textwidth,trim=15 20 20 10,clip]{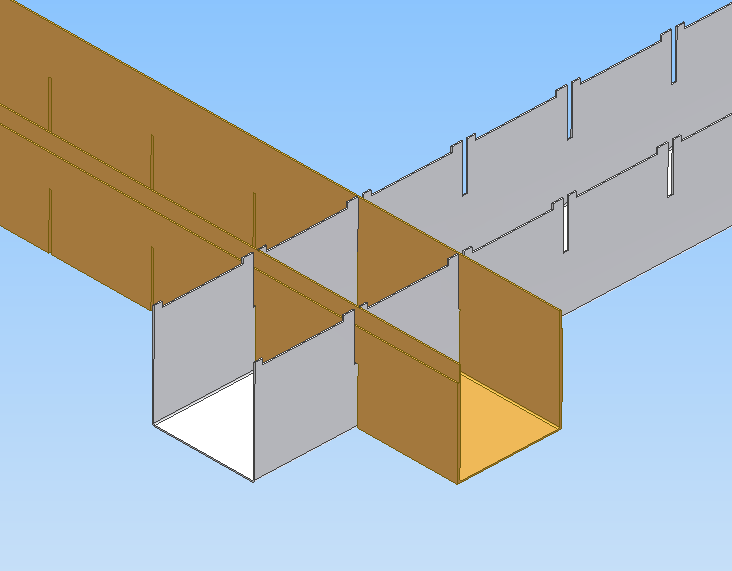}~
\includegraphics[width=.55\textwidth,trim=10 10 10 10,clip]{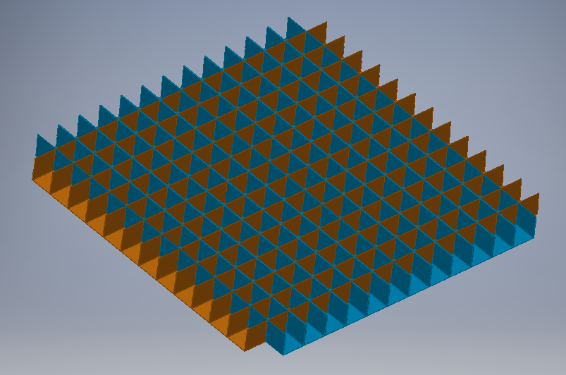}
%\qquad
%\includegraphics[width=.4\textwidth,origin=c,angle=180]{Images/derlin_tray}
% "\includegraphics" from the "graphicx" permits to crop (trim+clip)
% and rotate (angle) and image (and much more)
\caption{\label{fig:carbon_uprof1} Two U carbon fibre profiles intersecting each other (left), a tray consisting of many intercrossing U profiles (right).}
\end{figure}
The design of these fibre trays was made in collaboration with the LOSON Company\cite{Loson} (an Italian company that produces CF components for aerospace and that produced for us the first 6x6 prototype and the last 28x28 CF trays). A suitable design was developed that can be used in a production run using a glueing process between basic carbon fibre U-profiles (see figure \ref{fig:carbon_uprof1}).
Each U profile is obtained by laminating the carbon fibre using custom designed metal moulds. In case the trays require a high number of cells one can group together in a single mould many U profiles reducing the number of steps in the work flow. After lamination but before glueing, the profiles must be reworked (cut-outs, grooves, etc.) using specific custom made tools that not only support the piece, but prevent the eventual de-lamination of the carbon fibre.
\begin{figure}[htbp]
\centering 
% \begin{center}/\end{center} takes some additional vertical space
\includegraphics[width=.8\textwidth,trim=2 2 2 2,clip]{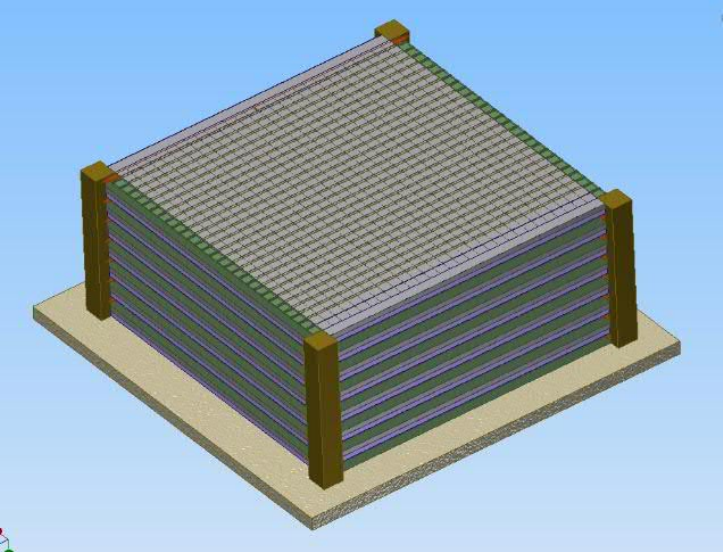}
%\includegraphics[width=.5\textwidth,trim=10 10 10 10,clip]{Images/carbon_uprof2}
%\qquad
%\includegraphics[width=.4\textwidth,origin=c,angle=180]{Images/derlin_tray}
% "\includegraphics" from the "graphicx" permits to crop (trim+clip)
% and rotate (angle) and image (and much more)
\caption{\label{fig:carbon_uprof3} A rendering showing how the carbon fibre trays would be stacked on top of each other in a calorimeter assembly.}
\end{figure}

In the final calorimeter assembly all these trays are stacked on top of each other and above a support base plate. Four lateral columns run along the four edges to further rigidify the structure (see figure \ref{fig:carbon_uprof3}). Finally, each tray is shifted, in the X and Y directions, by half a crystal width respect to each other, to eliminate unintended "escape corridors" for the impinging particles. 

\begin{figure}[htbp]
\centering 
% \begin{center}/\end{center} takes some additional vertical space
\includegraphics[width=.3\textwidth,trim=2 2 2 2,clip]{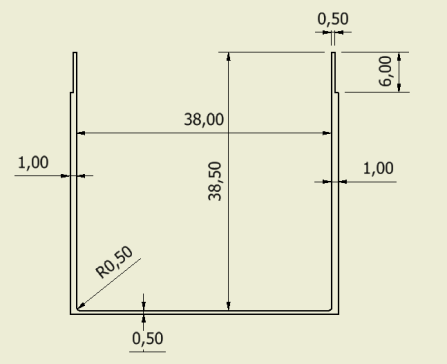}
\includegraphics[width=.25\textwidth,trim=2 2 2 2,clip]{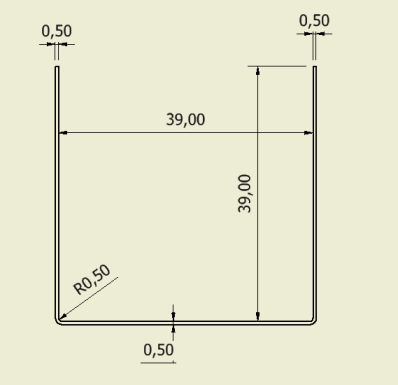}
\includegraphics[width=.37\textwidth,trim=2 2 2 2,clip]{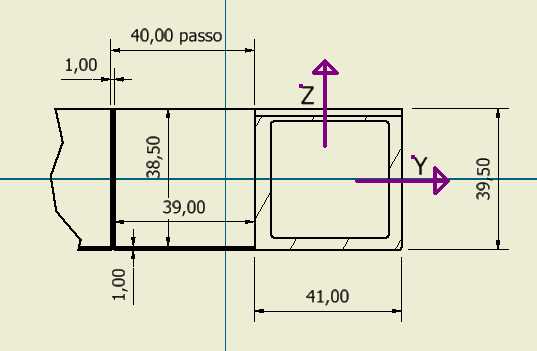}
\includegraphics[width=.46\textwidth,trim=2 2 2 2,clip]{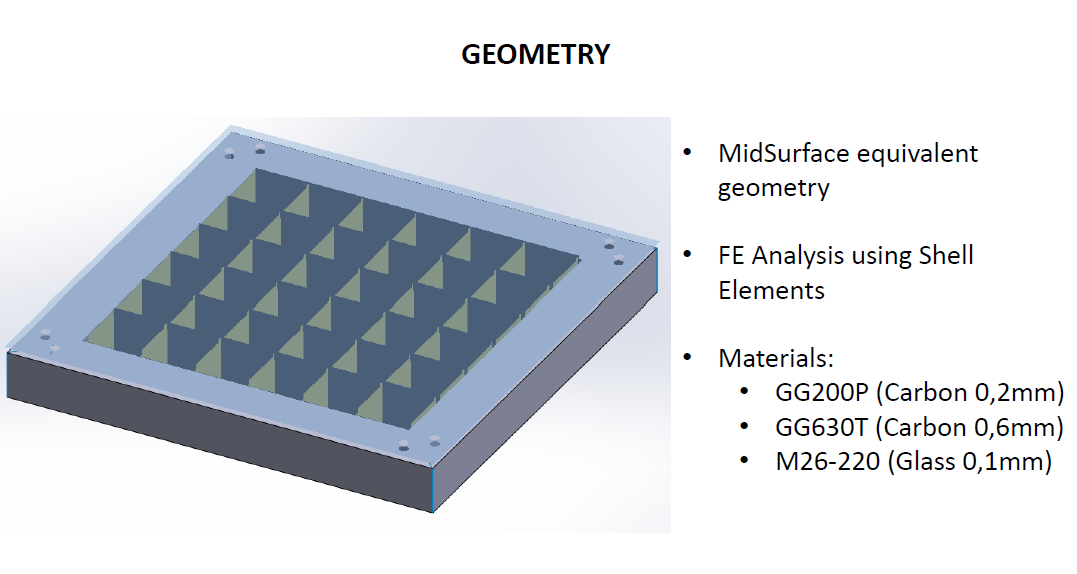}
\includegraphics[width=.46\textwidth,trim=2 2 2 2,clip]{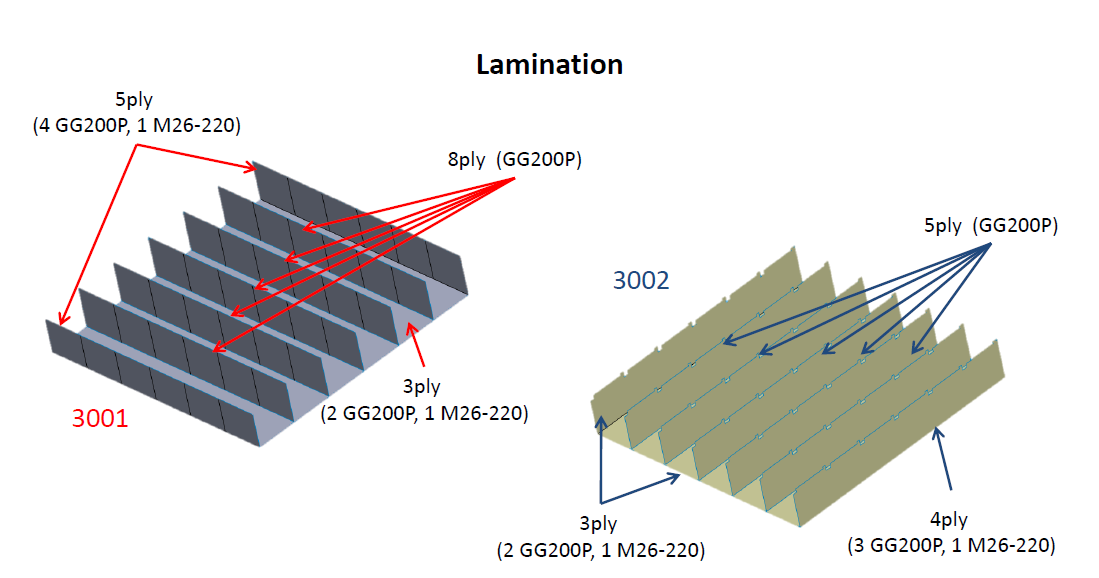}
%\qquad
%\includegraphics[width=.4\textwidth,origin=c,angle=180]{Images/derlin_tray}
% "\includegraphics" from the "graphicx" permits to crop (trim+clip)
% and rotate (angle) and image (and much more)
\caption{\label{fig:drawing_uprof3} The drawings (TOP), with the relevant dimensions (mm) of the two U profiles and of the stiff border rim used for the 6x6 carbon fibre  cell prototype. Detailed simulation drawings (BOTTOM) used for verification. }
\end{figure}

We realised a scaled down 6x6 carbon fibre tray, following the above described construction procedure and drawings (see figure \ref{fig:drawing_uprof3}), with each cell measuring  38 x 39 mm$^2$ (x and y), and with a depth of 38.5 mm (z).  From these drawings, we also derived a detailed simulation model used for verification (see following sections). 

\begin{figure}[htbp]
\centering 
% \begin{center}/\end{center} takes some additional vertical space
\includegraphics[width=.8\textwidth,trim=20 20 2 30,clip]{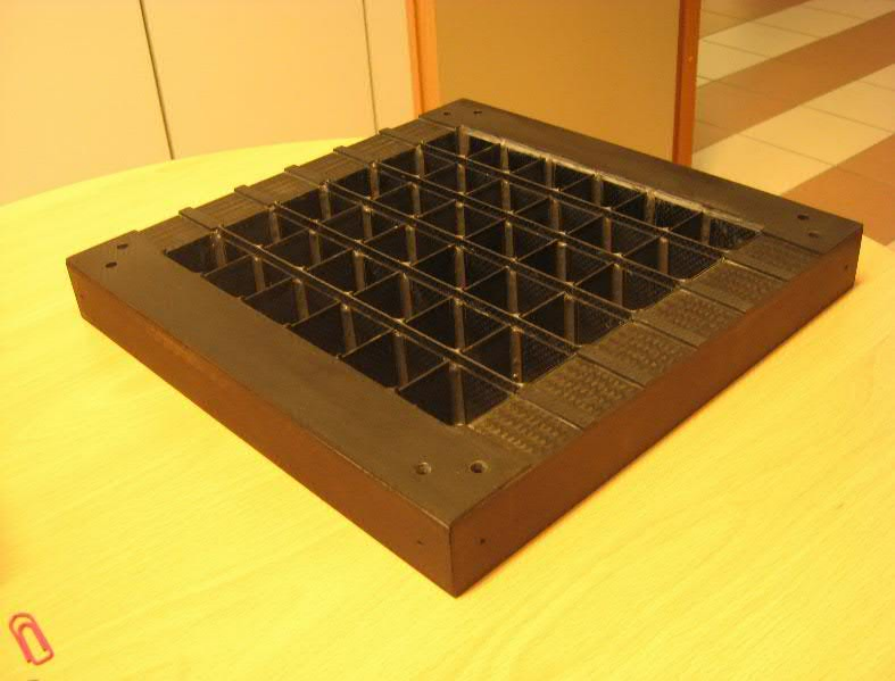}
%\includegraphics[width=.5\textwidth,trim=10 10 10 10,clip]{Images/carbon_uprof2}
%\qquad
%\includegraphics[width=.4\textwidth,origin=c,angle=180]{Images/derlin_tray}
% "\includegraphics" from the "graphicx" permits to crop (trim+clip)
% and rotate (angle) and image (and much more)
\caption{\label{fig:carbon_6x6} Photo of the 6x6 carbon fibre  prototype used for the verification tests.}
\end{figure}

The border was stiffened  with a 40 x 40 mm$^2$ empty square section (see figure \ref{fig:carbon_6x6}) of 3mm thickness. The prototype built by LOSON  was precisely measured at the INFN laboratories in Pisa and the results compared to the original drawing parameters.

\subsection{Carbon fibre tray measurements results}
\label{sec:mechanicsCFres}
\begin{figure}[htbp]
\centering 
% \begin{center}/\end{center} takes some additional vertical space
\includegraphics[width=.50\textwidth,trim=2 2 2 2,clip]{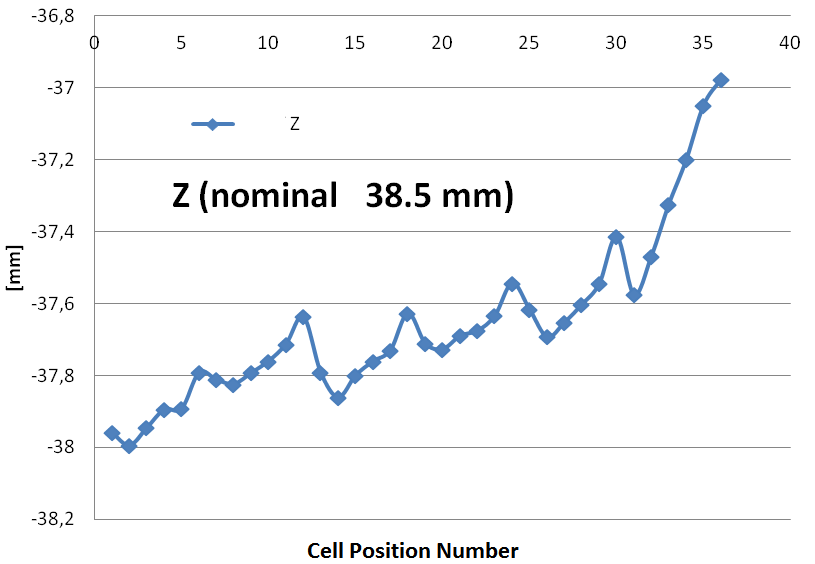}
\includegraphics[width=.48\textwidth,trim=2 2 2 2,clip]{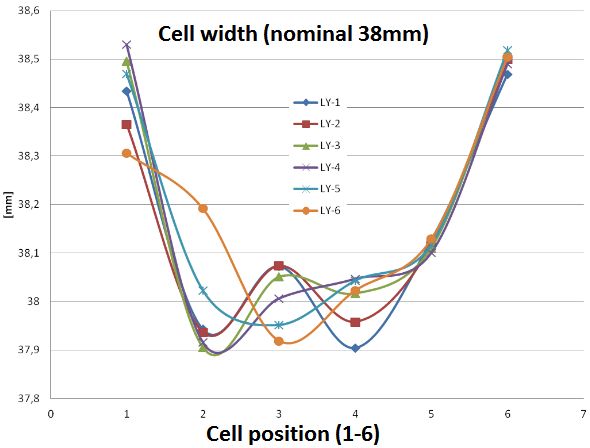}
%\includegraphics[width=.5\textwidth,trim=10 10 10 10,clip]{Images/carbon_uprof2}
%\qquad
%\includegraphics[width=.4\textwidth,origin=c,angle=180]{Images/derlin_tray}
% "\includegraphics" from the "graphicx" permits to crop (trim+clip)
% and rotate (angle) and image (and much more)
\caption{\label{fig:measuredepthwidth}Uniformity measurements: Plots showing the 36 cells depths (left), and their widths (6x6 matrix, cell position along the row for six different columns).}
\end{figure}

\begin{figure}[htbp]
\centering 
% \begin{center}/\end{center} takes some additional vertical space
\includegraphics[width=.49\textwidth,trim=2 2 2 2,clip]{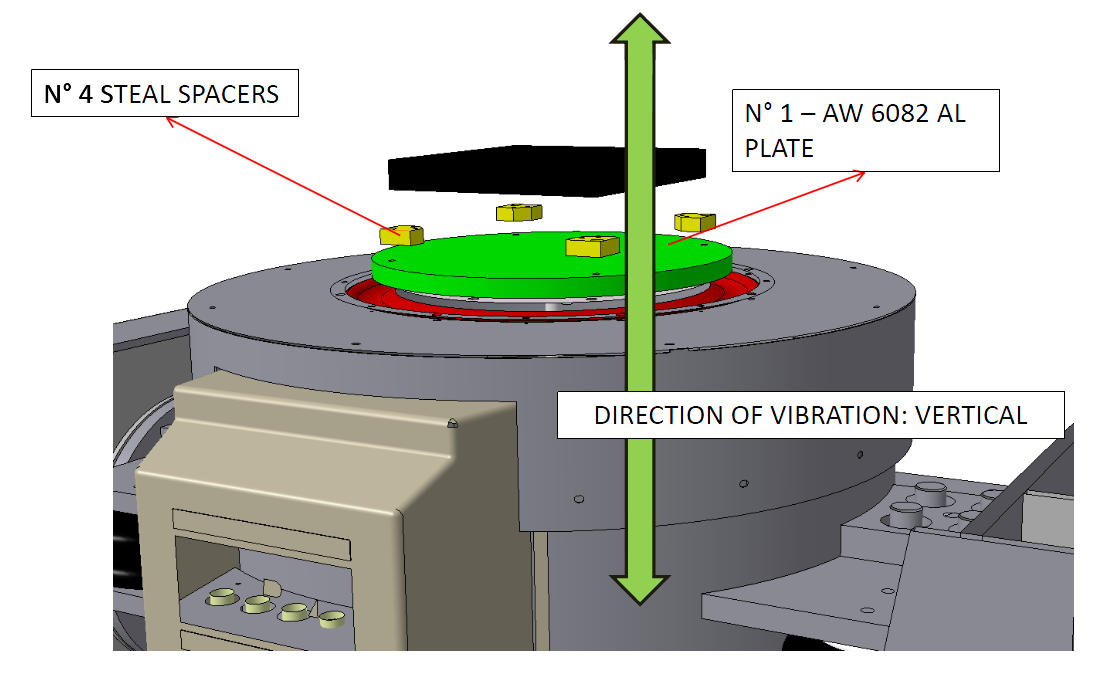}
\includegraphics[width=.49\textwidth,trim=2 20 2 2,clip]{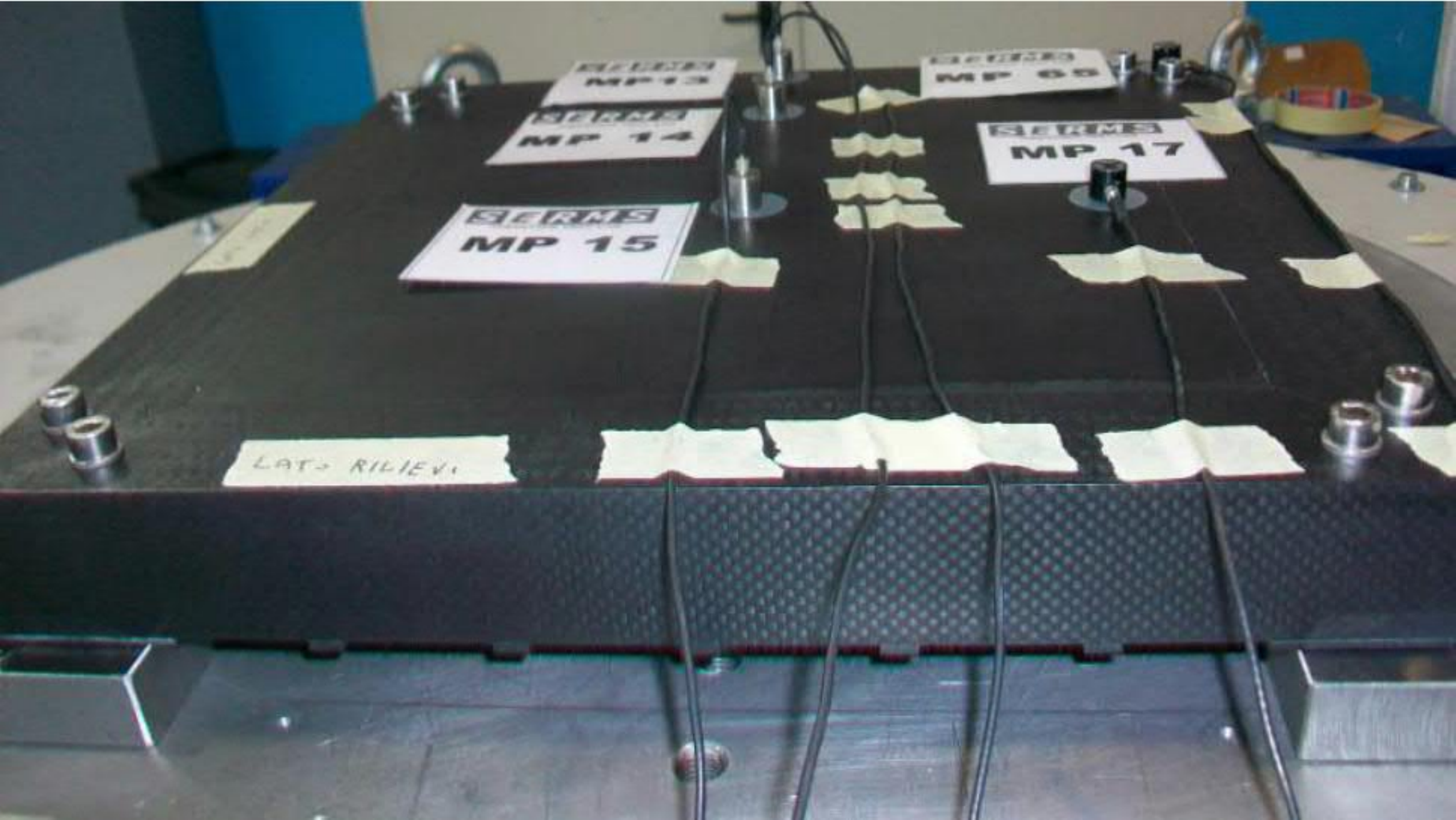}
%\includegraphics[width=.5\textwidth,trim=10 10 10 10,clip]{Images/carbon_uprof2}
%\qquad
%\includegraphics[width=.4\textwidth,origin=c,angle=180]{Images/derlin_tray}
% "\includegraphics" from the "graphicx" permits to crop (trim+clip)
% and rotate (angle) and image (and much more)
\caption{\label{fig:vibra12}Drawing of the vibrating fixture (left), and photo of the fixation plate and carbon fibre tray with the accelerometers.}
\end{figure}

\begin{figure}[htbp]
\centering 
% \begin{center}/\end{center} takes some additional vertical space
\includegraphics[width=.70\textwidth,trim=2 2 2 2,clip]{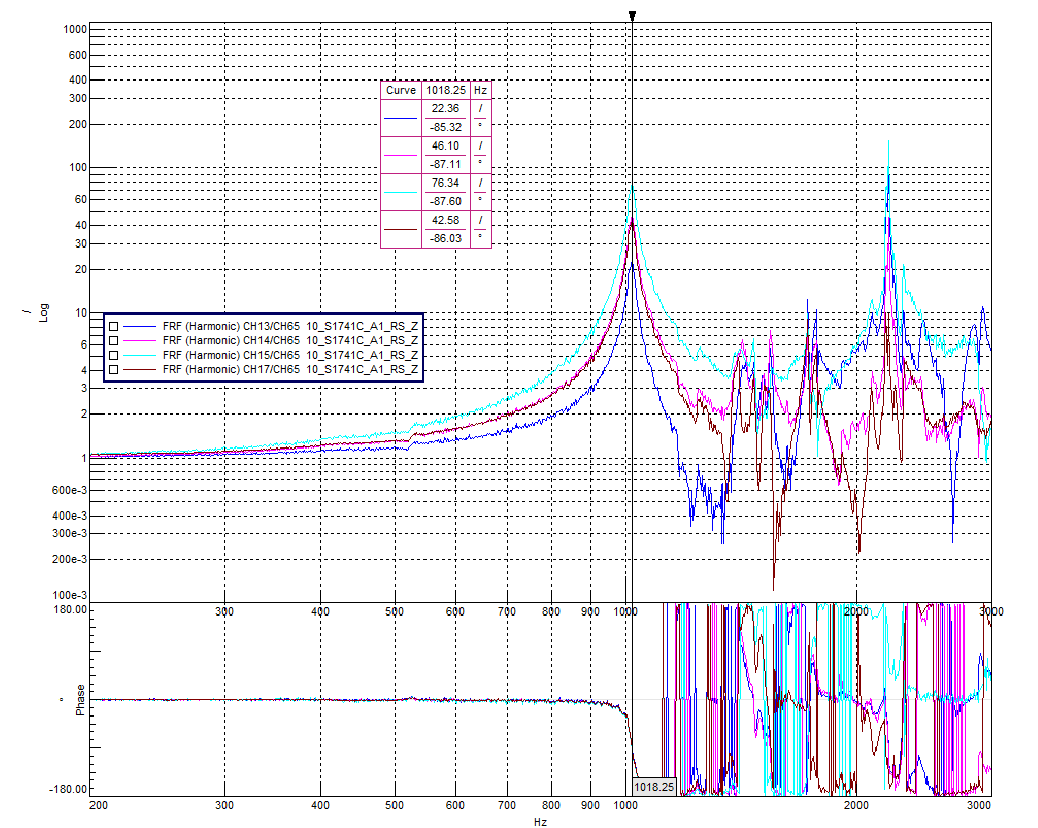}
% and rotate (angle) and image (and much more)
\caption{\label{fig:vibrares1}Graph of the resonant frequencies obtained from the 6x6 cell CF prototype.}
\end{figure}

The measurement results on cell size uniformity and depth of the 6x6 tray are shown in the following plots. From these measurements, obtained with a precision coordinate-measuring machine (CMM) in our clean rooms (see figure \ref{fig:measuredepthwidth}), it appears that the depth of the cells is less than the design value (37.5 mm $\pm$0.5 mm instead of 38.5 mm) which is probably due to the glueing procedure followed during assembly.
Also a border effect in the X-Y plane is manifest both in the cell sizes and their pitch (not shown).
This prototype was then used for vibration tests in order to find the resonant frequencies, in particular those concerning the z axis orthogonal to the tray.
These tests were performed at the SERMS~\cite{SERMS} facility. A drawings and photo of the installation are shown below  (see figure \ref{fig:vibra12}). 
The tests performed were first used to check the validity of our Finite Element Modelling (FEM). The frequency graph (see figure \ref{fig:vibrares1}) shows a fundamental resonance at 1018Hz evident also from the phase change. This frequency was measured with more or less the same intensity  on all accelerometers. The test was repeated many times, checking the reproducibility of the results. Sometimes we also exchanged the accelerometers. All measurements gave consistent results.
These results are to be compared with the first simulations performed on the 6x6 cell protoype. Using the drawings shown in figure \ref{fig:drawing_uprof3}, the structure was simulated implementing the detailed knowledge of the carbon fibre types and lamination processes used. 
\begin{figure}[htbp]
\centering 
% \begin{center}/\end{center} takes some additional vertical space
\includegraphics[width=.44\textwidth,trim=2 2 2 2,clip]{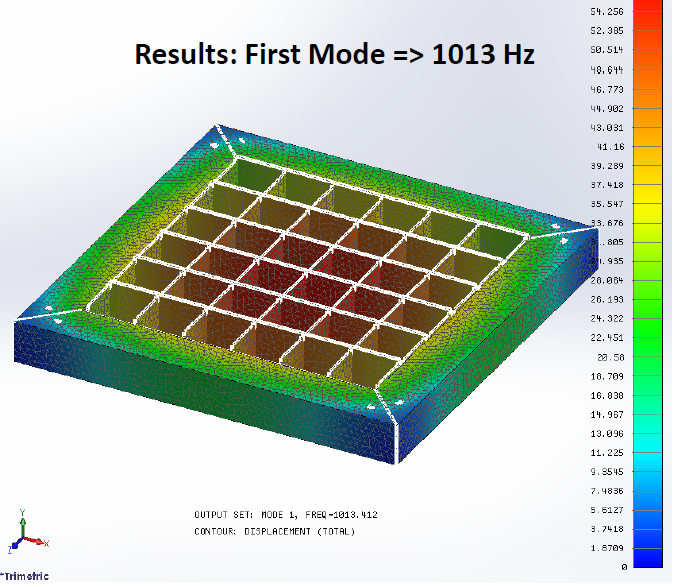}
\includegraphics[width=.52\textwidth,trim=2 2 2 2,clip]{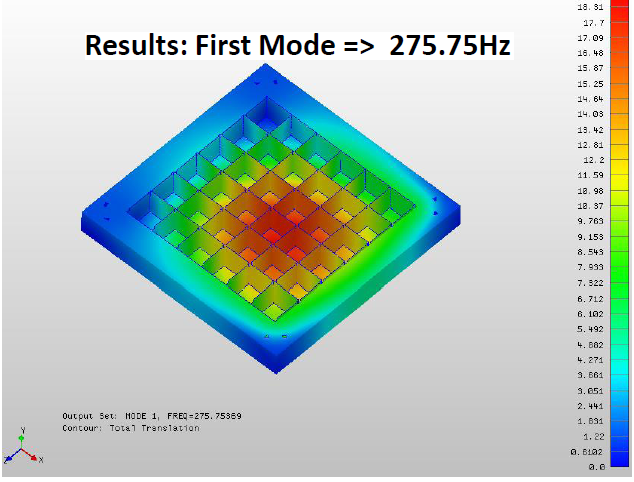}
%\includegraphics[width=.5\textwidth,trim=10 10 10 10,clip]{Images/carbon_uprof2}
%\qquad
%\includegraphics[width=.4\textwidth,origin=c,angle=180]{Images/derlin_tray}
% "\includegraphics" from the "graphicx" permits to crop (trim+clip)
% and rotate (angle) and image (and much more)
\caption{\label{fig:6x6simu34}First mode resonant frequency of the simulated tray: with no crystals (left), and loaded with crystals (right). The weight of the 36 crystals is roughly 8 Kg. Both configurations are without cover lid.}
\end{figure}
The tray was simulated with and without a cover lid, which anyway has little effect on the resonant frequency being very thin ($\sim$1mm). The results indicate, as expected, a strong vertical displacement at the centre of the tray with the resonant frequency increasing from 882 Hz to 1013 Hz depending on the boundary conditions implemented (see figure \ref{fig:6x6simu34}) in the simulations. The results match very well what was found in the tests at SERMS in Terni (figure \ref{fig:vibra12}) where the tray borders were fixed to the supporting plate.
% \begin{figure}[htbp]
% \centering 
% % \begin{center}/\end{center} takes some additional vertical space
% \includegraphics[width=.45\textwidth,trim=2 2 2 2,clip]{Images/6x6simu1}
% \includegraphics[width=.45\textwidth,trim=2 2 2 2,clip]{Images/6x6simu2}
% \caption{\label{fig:6x6simu12}Drawing of the simulated tray (left), and details of the carbon fibre ply used.}
% \end{figure}
In simulations, the tray was also loaded with the crystals which provide an increase in mass of the structure with no additional structural support or added rigidity. As shown in the right picture of figure \ref{fig:6x6simu34} the resonant frequency, in this case, drops dramatically to less than 300 Hz with a significant doubling of the vertical displacement at the centre. Given the typical  power-spectral-density vibration levels of a rocket, such a displacement would translate into certain destruction of the tray as it is way beyond the deformation that the CF structure can tolerate.
\begin{figure}[htbp]
\centering 
% \begin{center}/\end{center} takes some additional vertical space
\includegraphics[width=.53\textwidth,trim=122 2 2 80,clip]{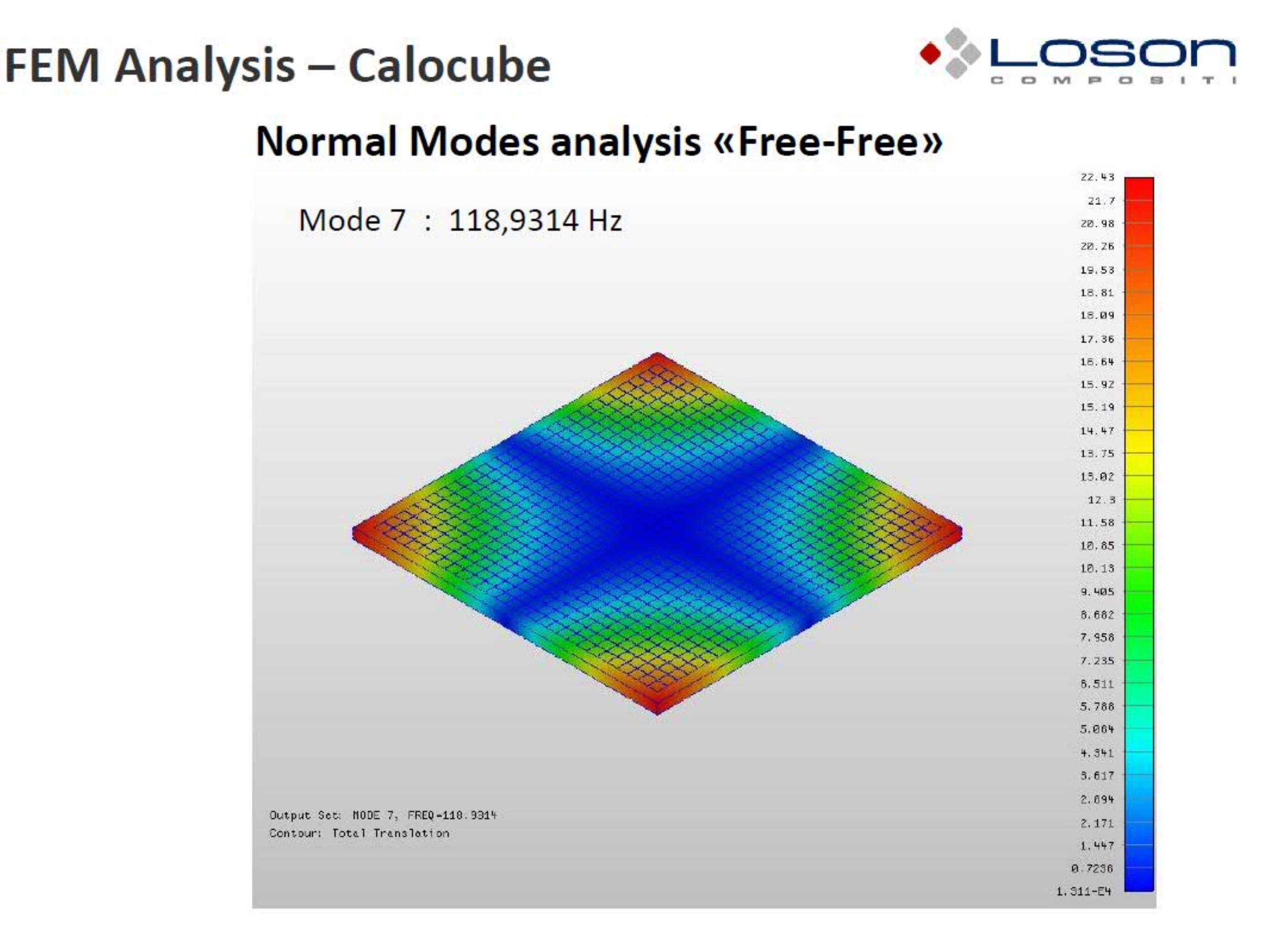}~
\includegraphics[width=.53\textwidth,trim=122 2 2 80,clip]{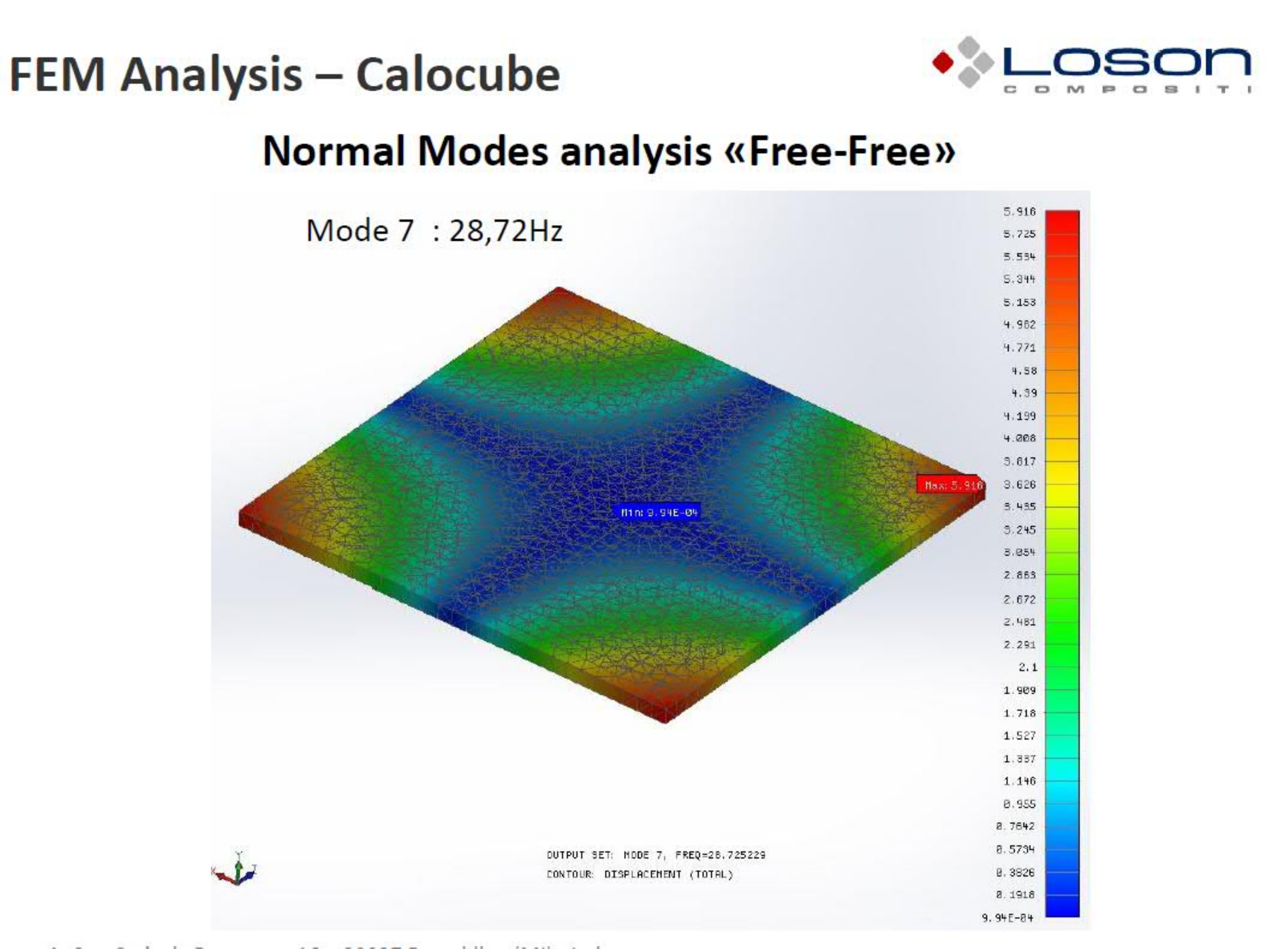}
%\includegraphics[width=.5\textwidth,trim=10 10 10 10,clip]{Images/carbon_uprof2}
%\qquad
%\includegraphics[width=.4\textwidth,origin=c,angle=180]{Images/derlin_tray}
% "\includegraphics" from the "graphicx" permits to crop (trim+clip)
% and rotate (angle) and image (and much more)
\caption{\label{fig:28x28simu12}First mode resonant frequency of a simulated full sized 28x28 cell tray: with no crystals (left), and with crystals (right). The weight of the crystals in the simulation is 165 Kg.}
\end{figure}
These results were expanded to the simulation of a full scale 28x28 cell CF tray, where the first mode resonant frequency dropped to around 100 Hz, (figure \ref{fig:28x28simu12}, left). Once loaded with crystals the frequency dropped further to 29 Hertz  (figure \ref{fig:28x28simu12}, right). Vertical displacements were reduced by lowering the input stimulus.
Such a low frequency poses a serious risk during the launch phase of the experiment. The trays suffer considerable deformation and will break and impact one on top of the other. The finite element  simulations performed by LOSON have confirmed that the stiffness of the 28x28 tray is too low to cope with the load of the crystals, totaling about 165 kg.
\begin{figure}[htbp]
\centering 
% \begin{center}/\end{center} takes some additional vertical space
\subfigure[\label{tie01}]{\includegraphics[width=0.46\textwidth,trim=2 2 2 10,clip]{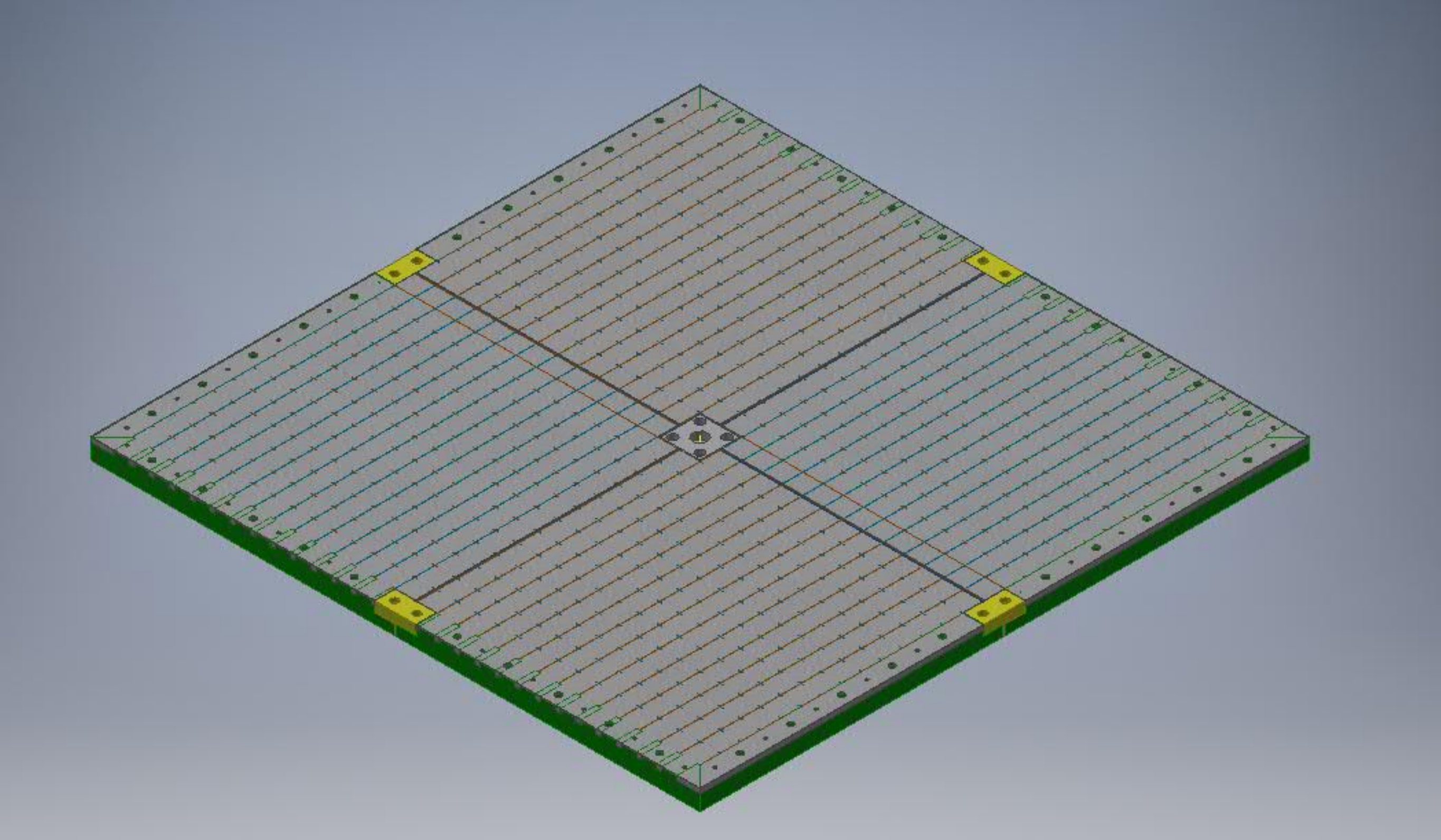}}~
\subfigure[\label{tie02}]{\includegraphics[width=0.46\textwidth,trim=2 2 2 10,clip]{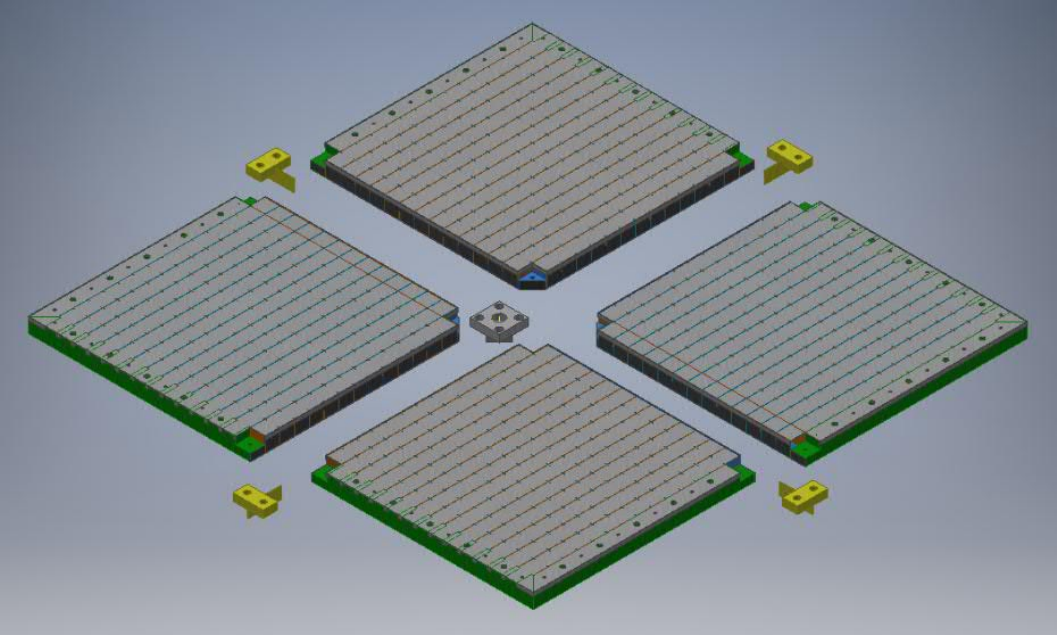}}
\subfigure[\label{tie03}]{\includegraphics[width=0.46\textwidth,trim=2 2 2 40,clip]{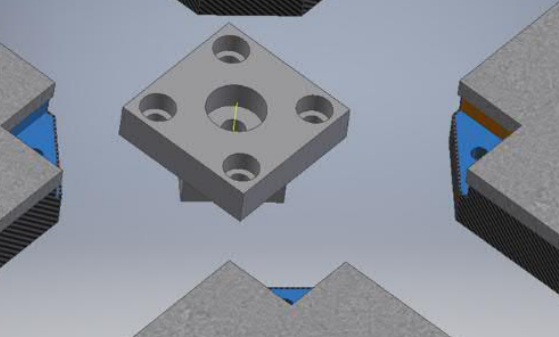}}~
\subfigure[\label{tie04}]{\includegraphics[width=0.46\textwidth,trim=2 2 2 37,clip]{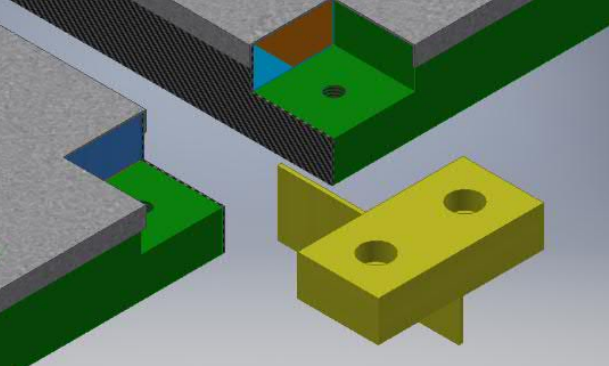}}
%\includegraphics[width=.5\textwidth,trim=10 10 10 10,clip]{Images/carbon_uprof2}
%\qquad
%\includegraphics[width=.4\textwidth,origin=c,angle=180]{Images/derlin_tray}
% "\includegraphics" from the "graphicx" permits to crop (trim+clip)
% and rotate (angle) and image (and much more)
\caption{\label{fig:CFtie} Improved version of the 28x28 cell CF tray. The tray is now subdivided in four smaller trays, and there is a further anchor point in the middle of each lateral section. From top left: the assembled tray (a), an exploded view (b),  detail of the central tie (c), and detail of the lateral binding (d).}
\end{figure}

\begin{figure}[htbp]
\centering 
% \begin{center}/\end{center} takes some additional vertical space
\includegraphics[width=.42\textwidth,trim=2 2 2 8,clip]{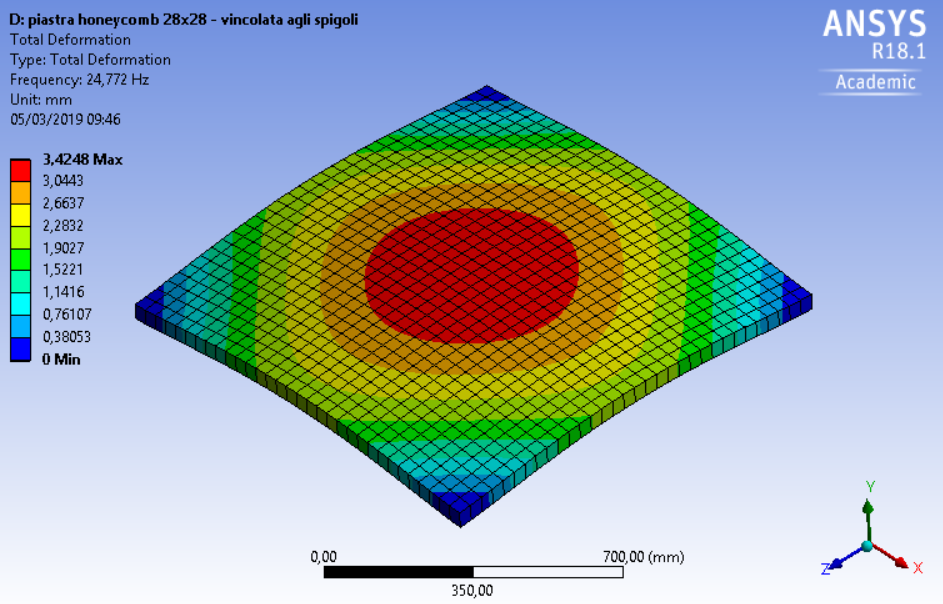}~
\includegraphics[width=.42\textwidth,trim=2 2 2 2,clip]{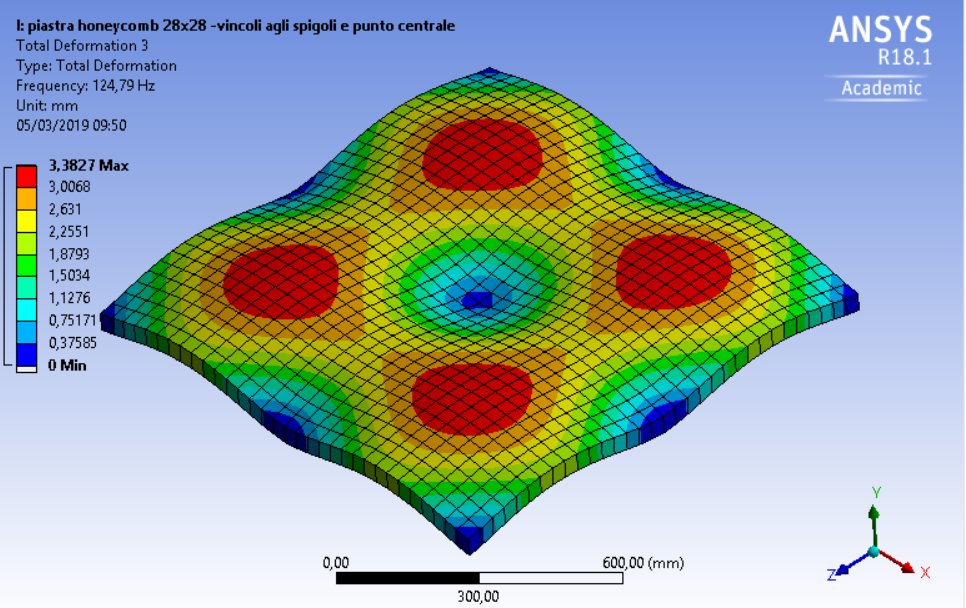}
%\includegraphics[width=.5\textwidth,trim=10 10 10 10,clip]{Images/carbon_uprof2}
%\qquad
%\includegraphics[width=.4\textwidth,origin=c,angle=180]{Images/derlin_tray}
% "\includegraphics" from the "graphicx" permits to crop (trim+clip)
% and rotate (angle) and image (and much more)
\caption{\label{fig:CFtiesimu}First mode resonant frequency of a simulated full sized 28x28 cell tray loaded with crystals. Before subdividing it into 4 sections (25 Hz resonance), after subdividision (resonance 125 Hz).}
\end{figure}

\begin{figure}[htbp]
\centering 
% \begin{center}/\end{center} takes some additional vertical space
\subfigure[\label{CFtiephoto01}]{\includegraphics[width=.42\textwidth,trim=2 2 2 8,clip]{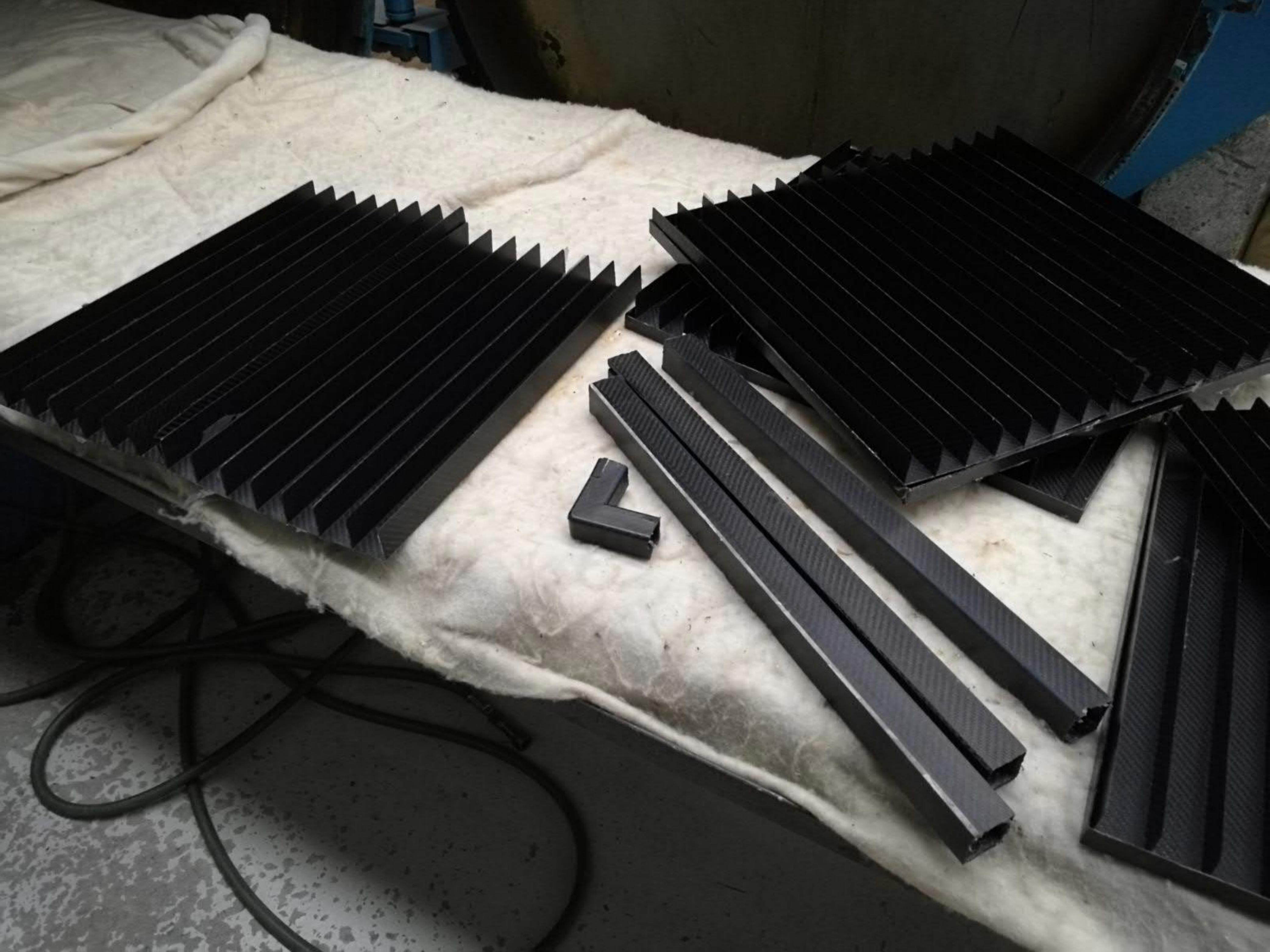}}~
\subfigure[\label{CFtiephoto02}]{\includegraphics[width=.42\textwidth,trim=2 2 2 8,clip]{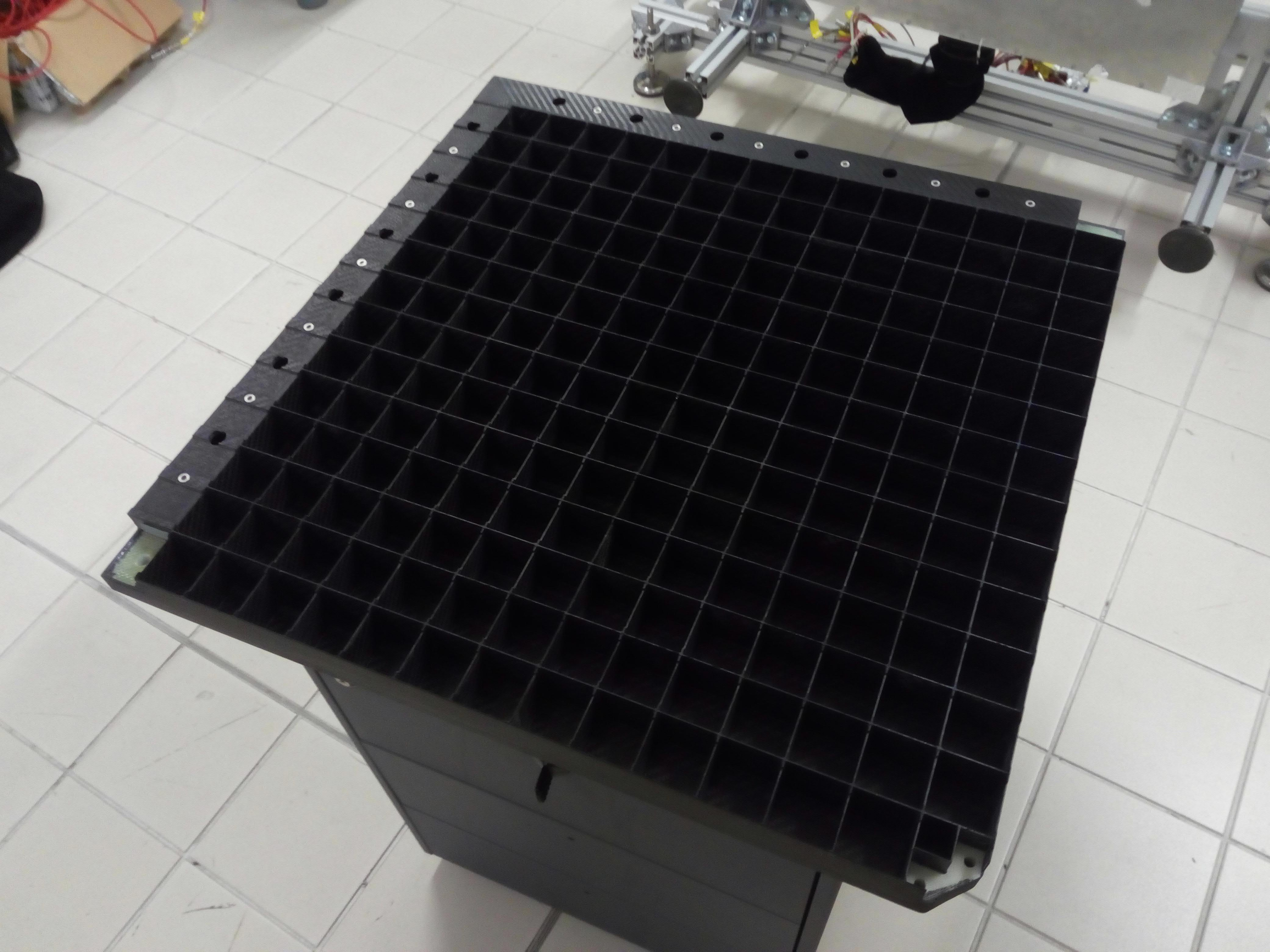}}
\subfigure[\label{CFtiephoto03}]{\includegraphics[width=.42\textwidth,trim=2 2 2 8,clip]{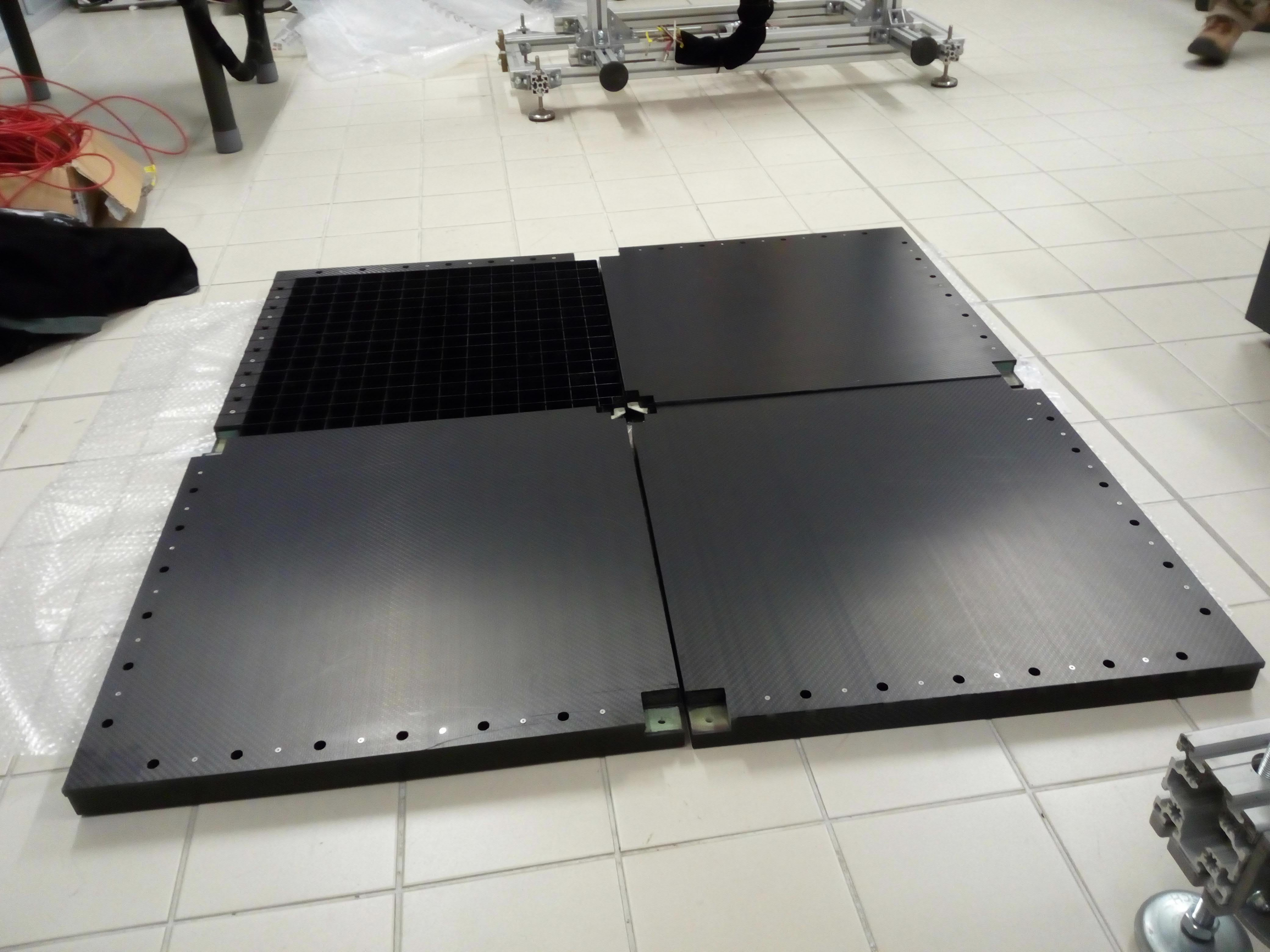}}~
\subfigure[\label{CFtiephoto04}]{\includegraphics[width=.42\textwidth,trim=2 2 2 8,clip]{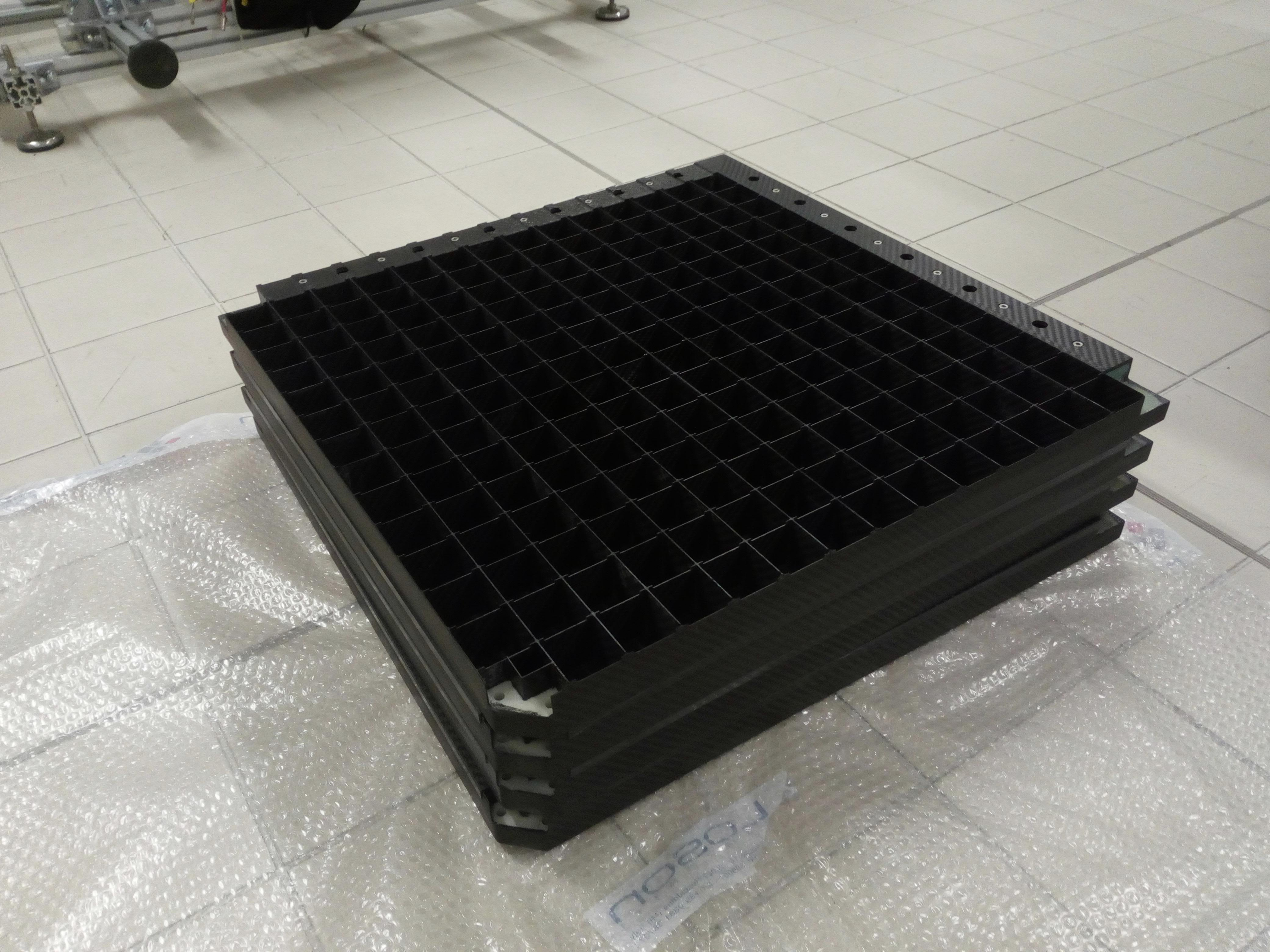}}
%\includegraphics[width=.5\textwidth,trim=10 10 10 10,clip]{Images/carbon_uprof2}
%\qquad
%\includegraphics[width=.4\textwidth,origin=c,angle=180]{Images/derlin_tray}
% "\includegraphics" from the "graphicx" permits to crop (trim+clip)
% and rotate (angle) and image (and much more)
\caption{\label{fig:CFtiephoto}From top left: CF elements before the glueing procedure (a), a single sub-tray (b), full size tray with lid (c), and four stacked sub-trays (d).}
\end{figure}

We have made various modifications, in order to increase the rigidity of the tray in the simulation. We have investigated the possibility of increasing the thickness of the base, of the lid (up to 1.5 mm), we have also toyed with the idea of having two rigidly coupled trays, but even in this case, once the crystals are loaded, the resonant frequencies drops to roughly 30 Hz. All these models were made extrapolating the parameters  from the FEM model of the 6x6 tray that was tested experimentally, as described in the previous sections.
Given the results, we decided to change the design and to split the 28x28 tray into four 14x14 sub-trays, in an effort to increase the first mode resonance frequency. We further improved the CF tray structure robustness, by implementing an anchor at the centre with a thick CF tie (see figure \ref{fig:CFtie}). 
To study the stiffness of this consame figuration we implented using Ansys a FEM model of a panel the size of the 28x28 tray with the same overall thickness, with two skins of 0.5 mm thick carbon on the outside of an internal honeycomb core with the pattern of the crystal housings; to simulate the crystals we considered a distributed load of 165 kg total over the entire surface. 
We then divided this model into the 4 sub-parts and bound it at the centre and at the corners. With this solution, the first resonant frequency obtained (loaded tray) was = 125 Hz (see figure \ref{fig:CFtiesimu}), so these simulations were then used as an initial input for the much more detailed simulations performed by LOSON that confirmed our results. The full sized CF trays were then built, following this updated design  (see figure \ref{fig:CFtiephoto}). 

There are four sub-trays with a pattern of 14x14 cells that can be joined in the centre with a CF tie and on the outer border of each frame.
In the centre there is enough space to house  4 crystals with half the standard dimensions, minimizing the impact of the hole in the middle.
The sub-trays are independent and can even be stacked together. We have built a total of eight sub-trays with 14x14 cells that can also be stacked to form an eight layer calorimeter prototype that can be used at test beams.

\subsection{Calorimeter test beam prototype}
\label{sec:mechanics}
\begin{figure}[htbp]
  \centering
  \subfigure[\label{subfig:smalltray}]{\includegraphics[width=.57\textwidth]{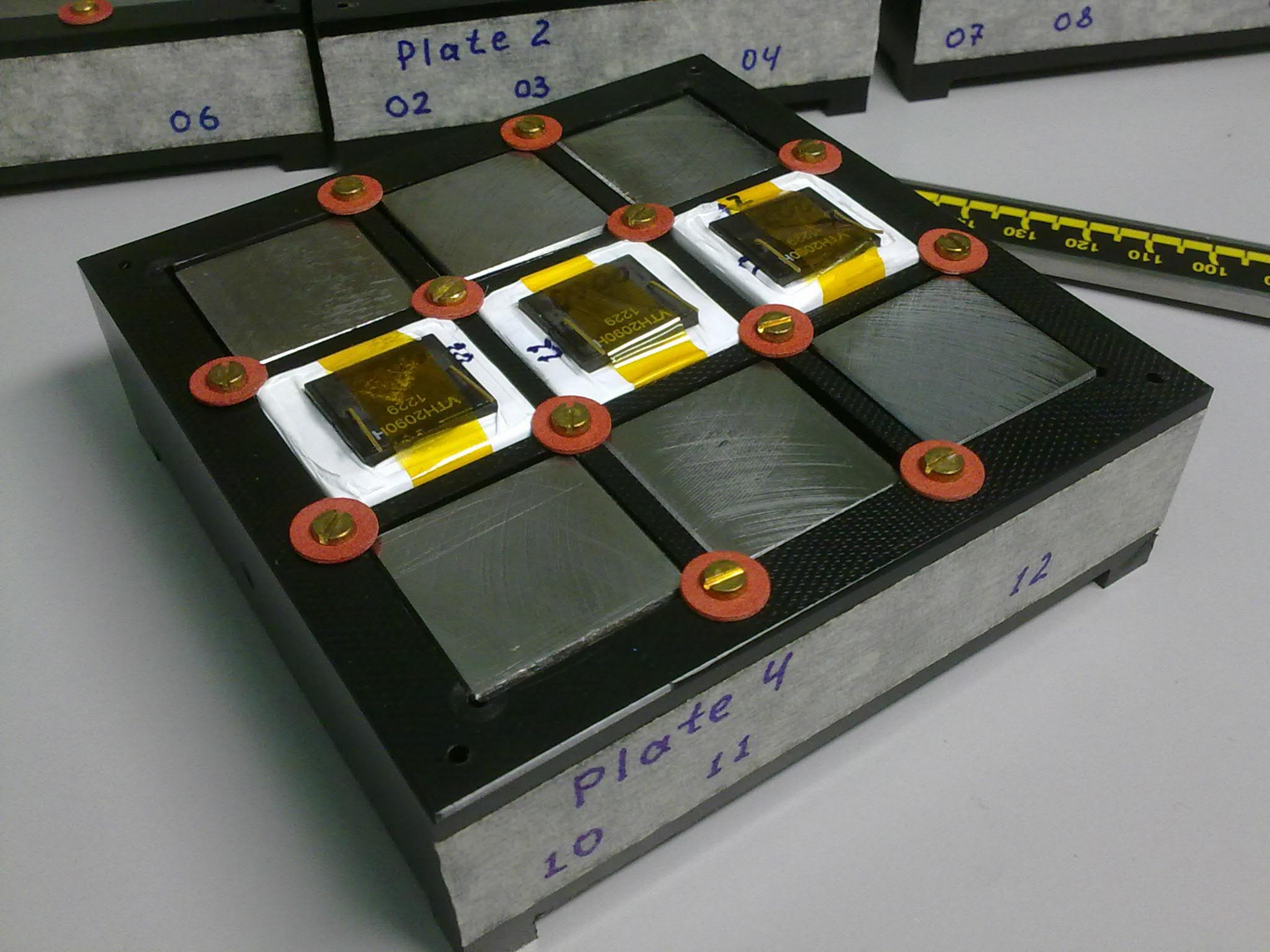}}
  \subfigure[\label{subfig:smalltower}]{\includegraphics[width=.32\textwidth]{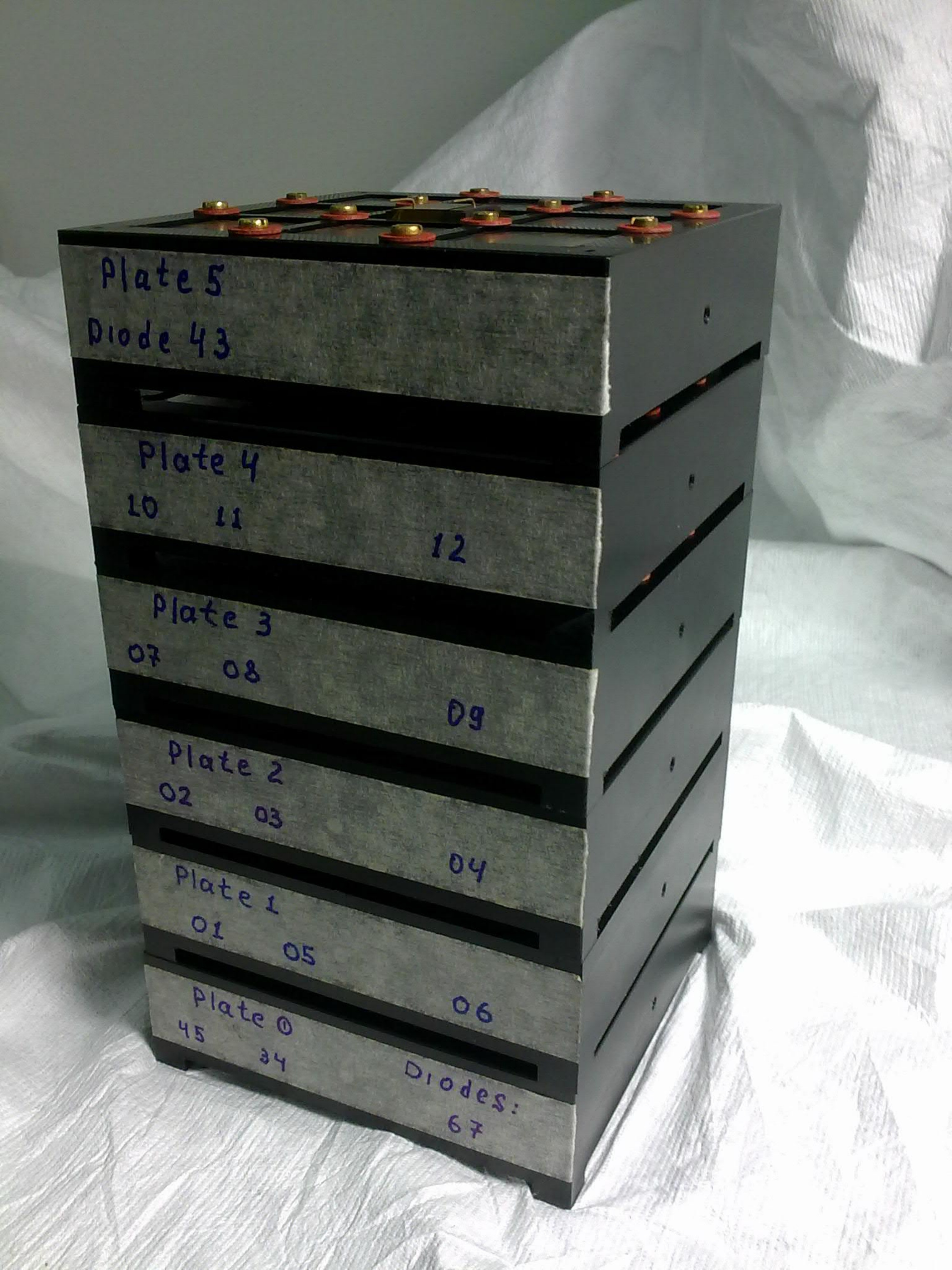}}
  \caption{\label{fig:preproto} Photo of a small tray with 3 CsI(Tl) crystals and iron cubes used as filler (a). Photo a the first small size calorimeter prototype used for our preliminary tests on elcromagnetic showers (b).}
\end{figure}

We have used the knowledge gained with all the measurments and studies performed on single crystals to build various calorimeter protoypes. The first prototype was built with only a few crystals of CsI(Tl) ($\sim$20), coupled to large area VTH2090 PDs, with iron cubes used to fill the empty spaces without crystals, as shown in figure  \ref{subfig:smalltray}.
 These small trays were then piled together to form small calorimetric towers that we tested with high energy electron beams to study shower profiles  (figure  \ref{subfig:smalltower}). This first small size prototype was build to test the main ideas and basic characteristics of the calorimeter. The prototype was tested with 50 GeV electrons and muons at Super Proton Synchrotron (CERN).
Its 12 CsI(Tl) crystals of 25×25×25 mm$^3$ were the first crystals acquired by our group. The size was chosen because we had some CsI(Tl) bars left from a previous experiment and they could be recycled using these dimensions. In fact, these first scintillators were cut and polished at our workshop in Florence (INFN). The mechanical support structure consisted of 6 plates with 9 elements in 3×3 matrix, 54 elements in total. The gap between crystals was 3 mm in all directions. The PDs were read out by specially designed Front End Boards developed by us for this project.
This paved the way for the full size prototype which we built once we optimised the scintillator material, size, wrapping, and PD layout.

\begin{figure}[htbp]
\centering % \begin{center}/\end{center} takes some additional vertical space
\subfigure[\label{delr01}]{\includegraphics[width=.42\textwidth,trim=2 20 2 1,clip]{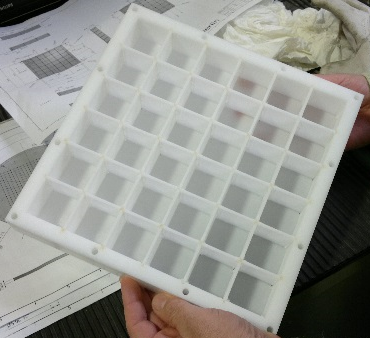}}~
\subfigure[\label{delr02}]{\includegraphics[width=.42\textwidth,trim=2 70 2 20,clip]{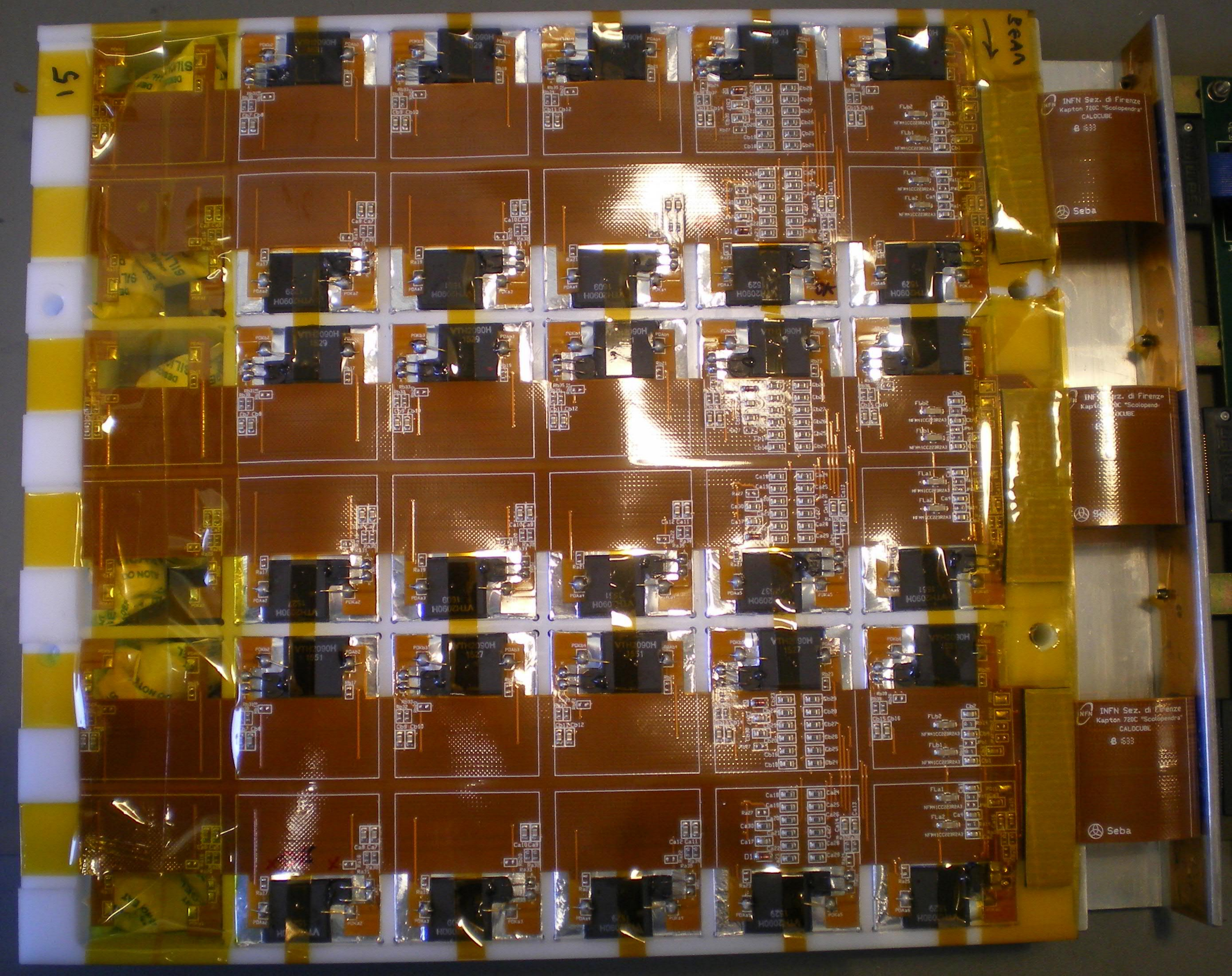}}
\subfigure[\label{delr03}]{\includegraphics[width=.42\textwidth,trim=2 255 2 1,clip]{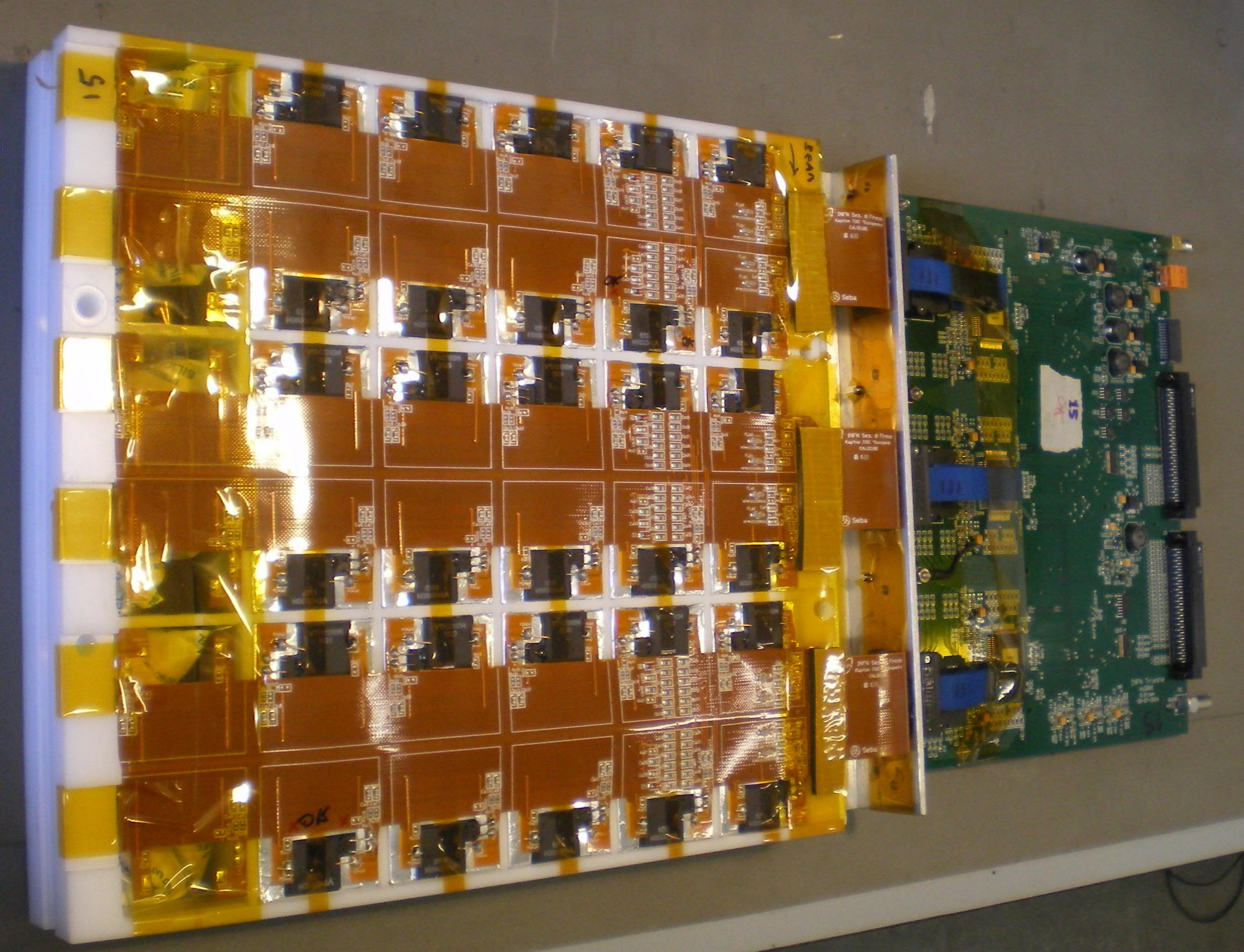}}~
\subfigure[\label{delr04}]{\includegraphics[width=.42\textwidth,trim=2 0 2 0,clip]{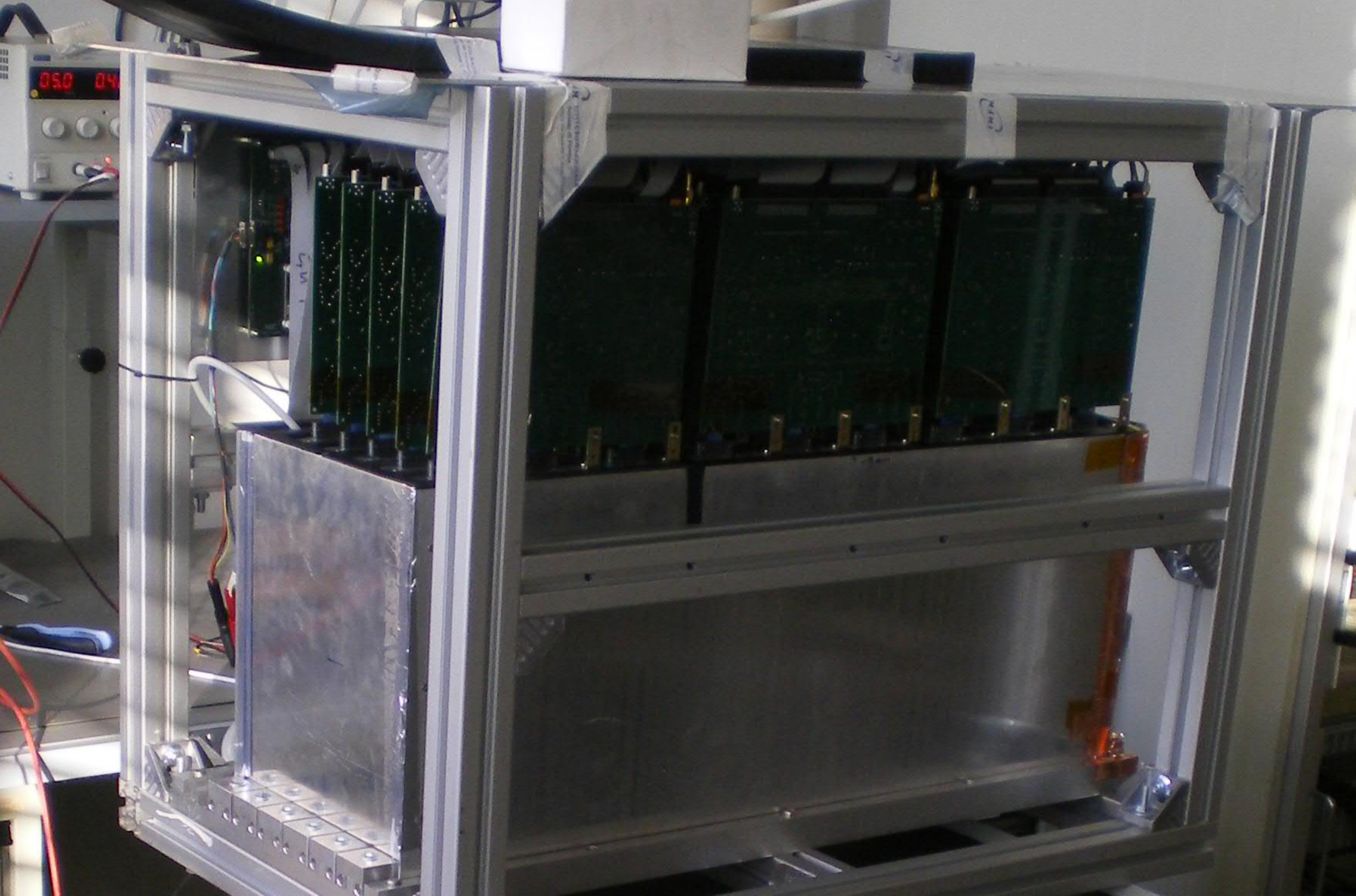}}
%\qquad
%\includegraphics[width=.4\textwidth,origin=c,angle=180]{Images/derlin_tray}
% "\includegraphics" from the "graphicx" permits to crop (trim+clip)
% and rotate (angle) and image (and much more)
\caption{\label{fig:delrintray} From top left: One of the first polyoxymethylene trays developed for the Calocube prototype (a), loaded with CsI (Tl) crystals and with the aluminium C section bolted at the end (b), with the  F.E. boards fixed to the C section (c), fully assembled prototype calorimeter (d). Also visible are the kapton cables used to route the photodiodes signals to the F.E. boards}
\end{figure}

 The development of the Calocube mechanics continued then with a design used for a test beam prototype in 2012. To support the CsI(Tl) scintillator cubes  (36x36x36 mm$^3$) and the electronic read-out boards we chose a structure based on polyoxymethylene trays with a pattern of 6 x 6 cells, each cell with a transverse dimensions of 37x37 mm$^2$ (x and y), and a depth of 36 mm (z). The cell walls are 3 mm thick, resulting in a 40 mm pitch in the x,y directions  for the crystals (see figure \ref{fig:delrintray}).
These trays have been used for all tests performed by our collaboration both at test beam facilities and in our laboratories. Each tray has a C section in alumium attached to the end that allows us to interconnect the crystals inside with the prototype Front End electronic boards (F.E.) developed by our collaboration in Trieste \cite{Bonvicini:2007, Bonvicini:2010}. Fifteen loaded trays have been stacked together, with a 4 mm spacing inbetween trays in order to have a 40 mm pitch also in the z direction. This assembly is what constitutes our Calocube calorimeter prototype that has been subject to many test at various beam facilities, (see bottom of  figure \ref{fig:delrintray} where a photo of a fully assembled tray with its F.E. and of the complete calorimeter protype are shown). A total of 25 trays were produced at the INFN Pisa workshop using a CNC milling machine from a single 40 mm thick plate. 

\begin{figure}[htbp]
\centering % \begin{center}/\end{center} takes some additional vertical space
\includegraphics[width=.35\textwidth,trim=2 2 2 1,clip]{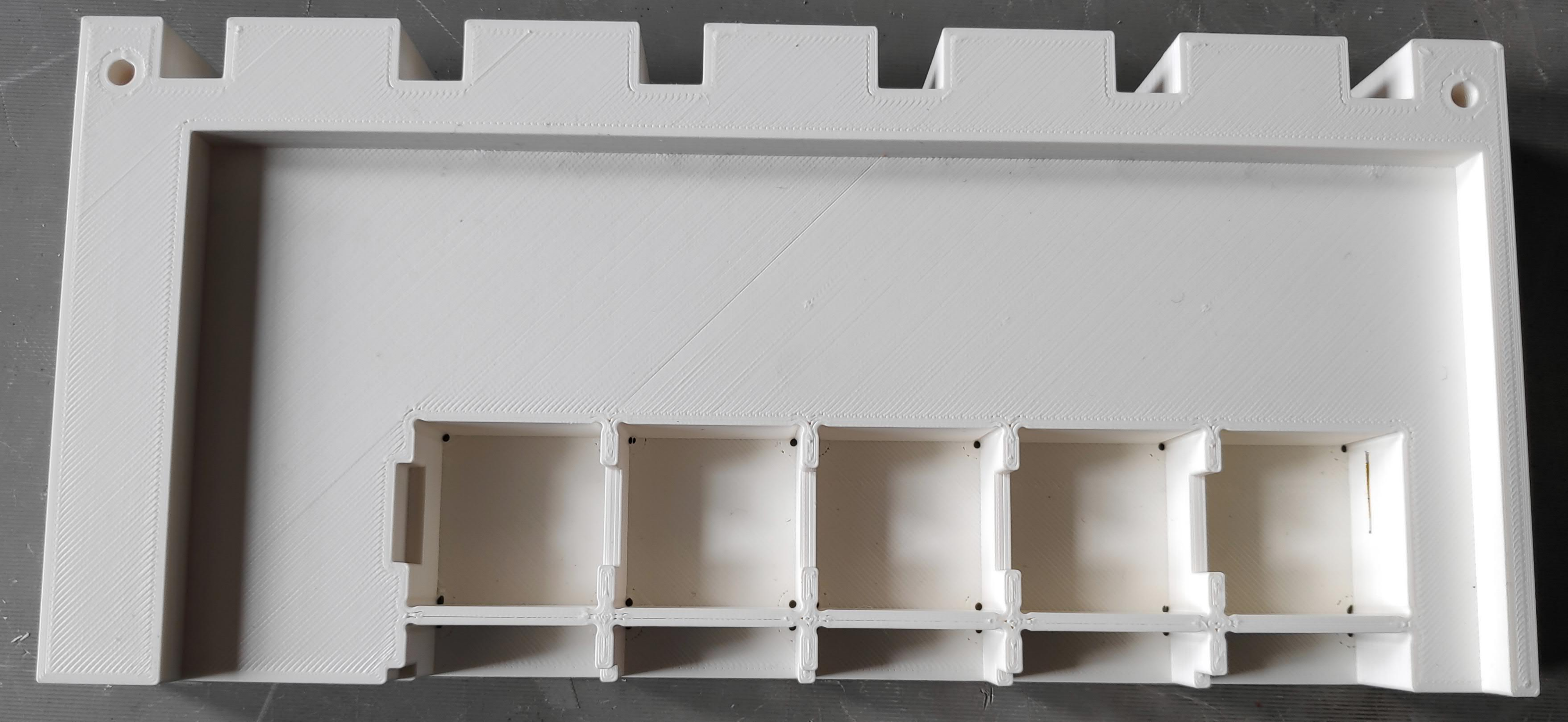}
\includegraphics[width=.60\textwidth,trim=2 2 2 1,clip]{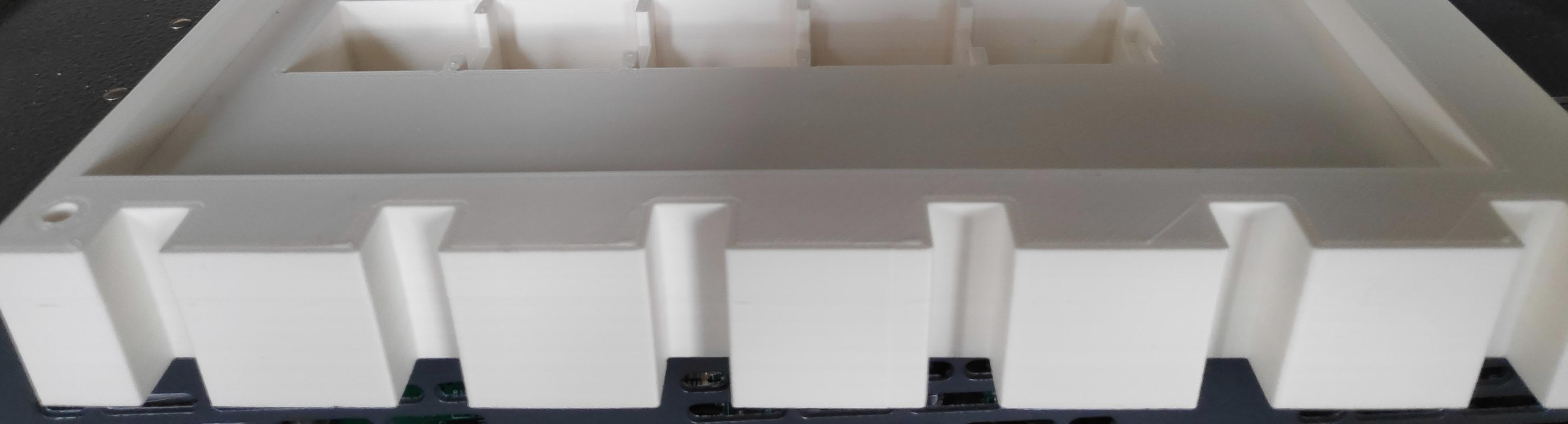}
%\qquad
%\includegraphics[width=.4\textwidth,origin=c,angle=180]{Images/derlin_tray}
% "\includegraphics" from the "graphicx" permits to crop (trim+clip)
% and rotate (angle) and image (and much more)
\caption{\label{fig:3Dtray}The first 3D printed tray protoype developed for future Calocube crystal tests with different scintillator materials (i.e. LYSO). The tray is nearly 40mm thick.}
\end{figure}

In order to experiment with different crystal assemblies of varying sizes (i.e. CsI(Tl) 36x36mm$^2$, LYSO 30x30mm$^2$), we have developed a new design based on a honeycomb plastic matrix suitable for 3D printing (figure \ref{fig:3Dtray}). This will be our baseline for future tests of new calorimeter designs. The honeycomb structure allows us to achieve a very high rigidity while keeping the weight of the tray and plastic consumption at a minimum. The printing takes only a couple of hours and it is relatively easy to change cell dimensions to accomodate different crystal sizes.

\section{ Front End electronics and final Test Beam prototype}
\label{sec:CASIS}
While the A250 by Amptek provides exceptional performance in terms of signal to noise ratio, it remains a very expensive single channel device with a limited dynamic range. For our calorimeter R\&D we pursued the development of a dedicated very low power consumption Front End amplifier.
\subsection{The CASIS Front End Electronics}
The CASIS (CAlorimetry in SIlicon for the Space ) chip \cite{Bonvicini:2007,Casis01} is an Application Specific Integrated Circuit (ASIC) specifically designed for space calorimetry by the INFN section of Trieste. For this scope the chip has been designed with a high dynamic range ($\sim$50pC maximum input charge with a ENC of 2000-3000 electrons), that allows to study the interactions of high energy particles, keeping  a low power consumption (2-3 mW per channel).
\begin{figure}[htbp]
  \centering
  \subfigure[\label{subfig:casisFE}]{\includegraphics[width=.55\textwidth]{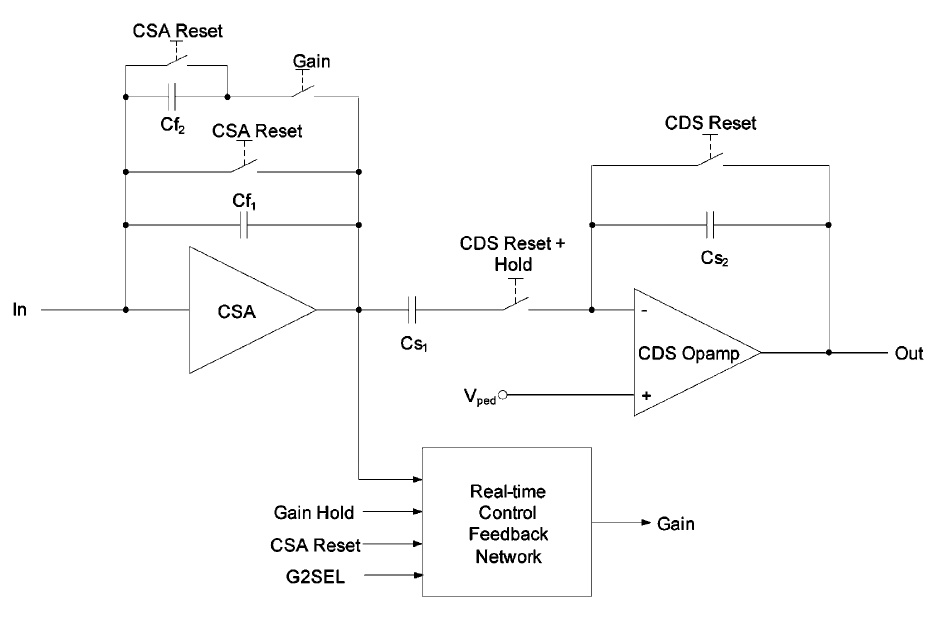}}
  \subfigure[\label{subfig:casisSIL}]{\includegraphics[width=.38\textwidth]{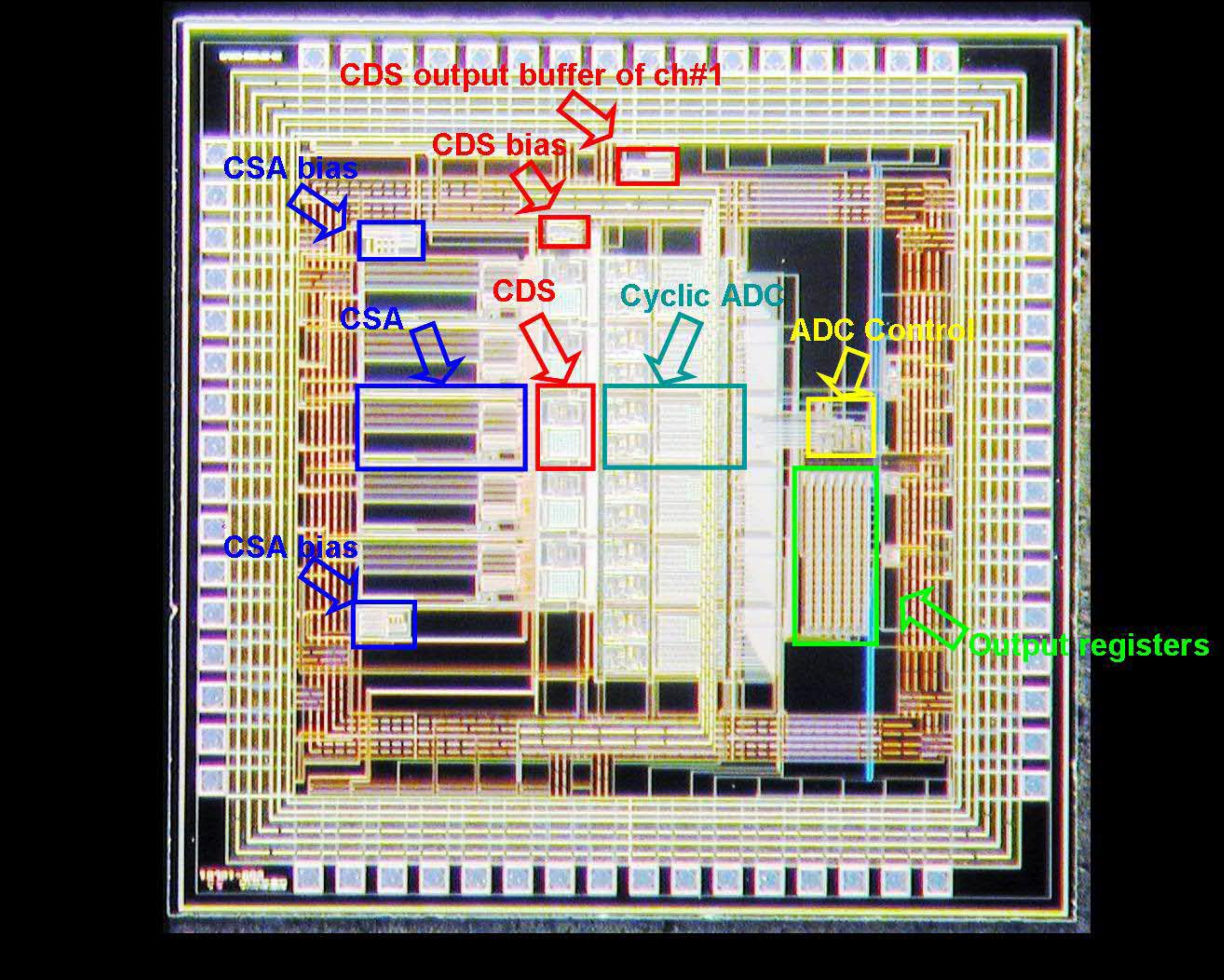}}
  \caption{\label{fig:CASIS01} Circuit block diagram of the Front End amplifier (a) and silicon layout of a complete 16 channel chip (b).}
\end{figure}
The chip consists of 16  front-end channels with double correlated sampling and one multiplexed ADC. The control circuitry uses external clock, convert and reset lines to generate all the required signals to operate the sampler and then the ADC. The chip has been designed and realized with the 0.35 µm C35B4 CMOS technology of Austria Micro Systems \cite{Casis02}, which provides two polysilicon and four metal layers, with a power supply voltage of 3.3 V .

The schematic block diagram of the front end section is shown in figure  \ref{subfig:casisFE}. A charge sensitive amplifier (CSA) is followed by a correlated double sampling filter (CDS). To achieve the very large required dynamic rage, a new architecture of the CSA feedback was implemented. It consists of a double gain loop with a real-time selection circuitry: a 1.6 pF capacitor is permanently connected to a folded cascode amplifier which sets the high gain, while a second larger capacitor (30.4 pF) is ready to be automatically inserted in parallel whenever the input signal exceeds a given threshold. A switch is used to reset periodically the preamplifier. To minimize the power consumption and guarantee a fast response, the comparator of the feedback control circuit is implemented with a Schmidt trigger with a value of the threshold of about 2 pC.
Since the charge integrated on the small capacitor is not destroyed when the system switches to low gain,  a precise calibration of the total range can always be performed. 
\begin{figure}[htbp]
  \centering
 \includegraphics[width=.8\textwidth]{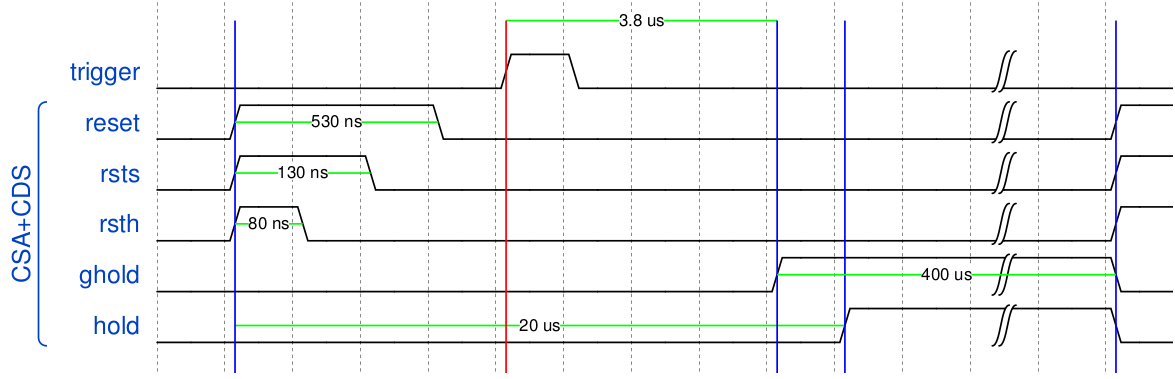}
  \caption{\label{fig:CASIS02} Timing sequences for the CSA and CDS blocks.}
\end{figure}

The voltage swing at the CSA output is about 1.5 V and the nominal sensitivities for the high and low gains are 0.67 mV/fC and 0.03 mV/fC, corresponding respectively to roughly 50 pC of maximum charge on the input before saturation sets in. This large range coupled to the previously mentioned  low power consumption, is one of the main benefits of the switched integrating capacitor design. This feature coupled to the dual PD readout (large area and small area) ensures the 10$^7$ dynamic range that we aimed for in the Calocube project.
\begin{figure}[htbp]
  \centering
\includegraphics[width=.8\textwidth]{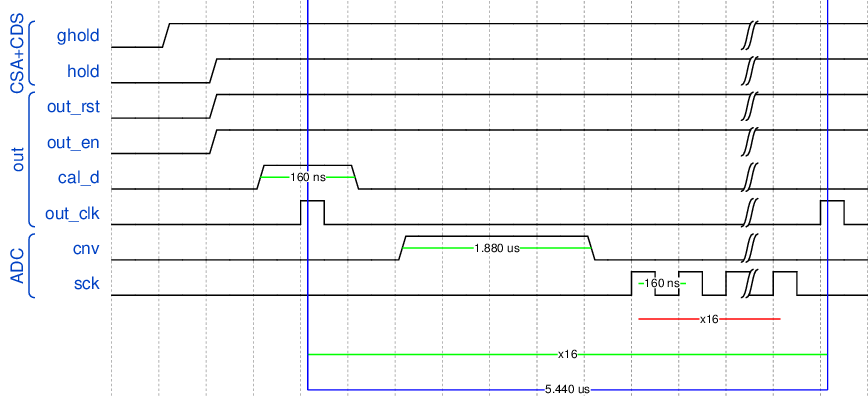}
  \caption{\label{fig:CASIS02b} Timing sequences fot the Hold and ADC part.}
\end{figure}
The CSA output is sampled by the CDS filter. This element executes a baseline substraction, using two samples for each event. The first one acquired before the trigger (periodically on every reset pulse falling edge)  and the second one with the trigger itself. When a trigger is present, the reset is inhibited, a HOLD signal is generated after a fixed (adjustable) delay, and the CSA outputs the difference between the baseline and the trigger events. The DC pedestal of the CDS output (normally set at 900 mV ) can be adjusted externally by means of the Vped bias. The timing sequences are shown in figure \ref{fig:CASIS02}. 

The signals for the CSA+CDS control are: RST H, RST S, RESET, GHOLD, and HOLD. The first tree signals perform a periodic reset of both the CSA (RST H, RST S) and of the CDS (RESET). In fact the chip must be reset periodically to remove the charge accumulated on the capacitors. 
% RST H drives a hard reset of the CSA discharging the feedback capacitors. Because of the enhanced dynamic range of the chip, the transistor for this operation should be quite large to perform a fast reset operation. As a result, this transistor has increased leakage current and injects a lot of charge when it’s switched off. The second transistor of smaller size is connected in parallel to the first one used to remove this charge. This second transistor performs a ”soft” reset of the CSA and is piloted by the RST S signal. These two signals act on the CSA Reset switch shown on Fig. 5.1. Signal RESET discharges the capacitors of the CDS part, and drives the CDS Reset switch (Fig. 5.1). 
All three signals are synchronized on the rising edge whereas their width are determined by the CASIS setup times. The HOLD and GHOLD lock the sampled voltage and automatic gain switch, respectively, before clocking the ouptut and ADC circuitry. The chip has one analogue output that provides a multiplexed output for all 16 channels. This multiplexing operation is driven by the the OUT CLK and the OUT RST lines. Thus it is possible to connect to the output all 16 channels, one by one, with 16 clock pulses. The OUT EN should be high to enable chip output. 

The complete diagram of the output control signals is shown in figure  \ref{fig:CASIS02b}. The SCK (ADC clock) width is 80 ns with a period of 5.44 µs and drives the conversion of each single output. These values were chosen on the basis of the ADC timing characteristics. More details are available both in  \cite{Casis01} and in  \cite{tesi_olek}.

\subsection{The calorimeter assemblies}

\begin{figure}[htbp]
  \centering
  \subfigure[\label{subfig:CASISboard}]{\includegraphics[width=.41\textwidth]{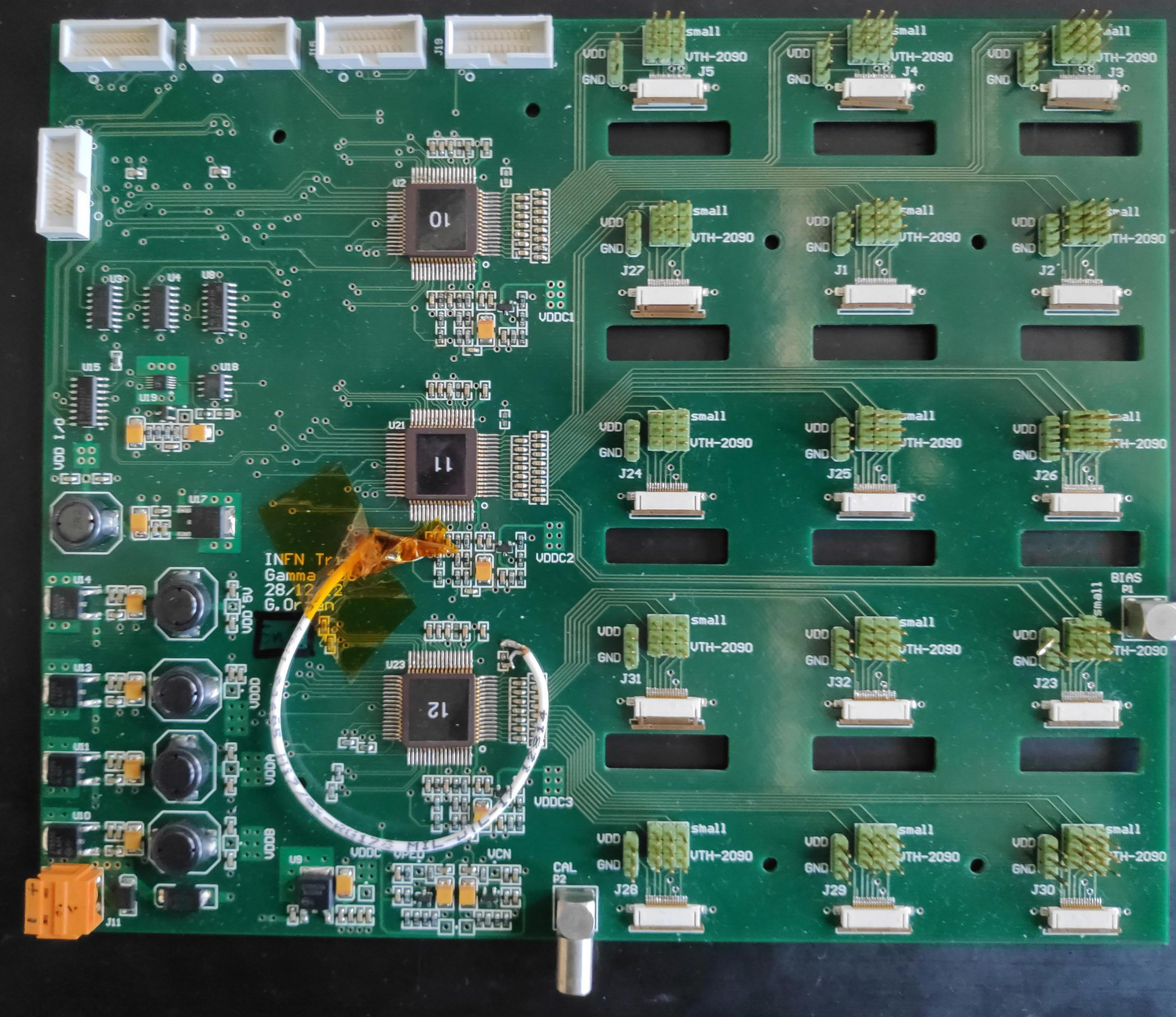}}
  \subfigure[\label{subfig:delrintray}]{\includegraphics[width=.476\textwidth]{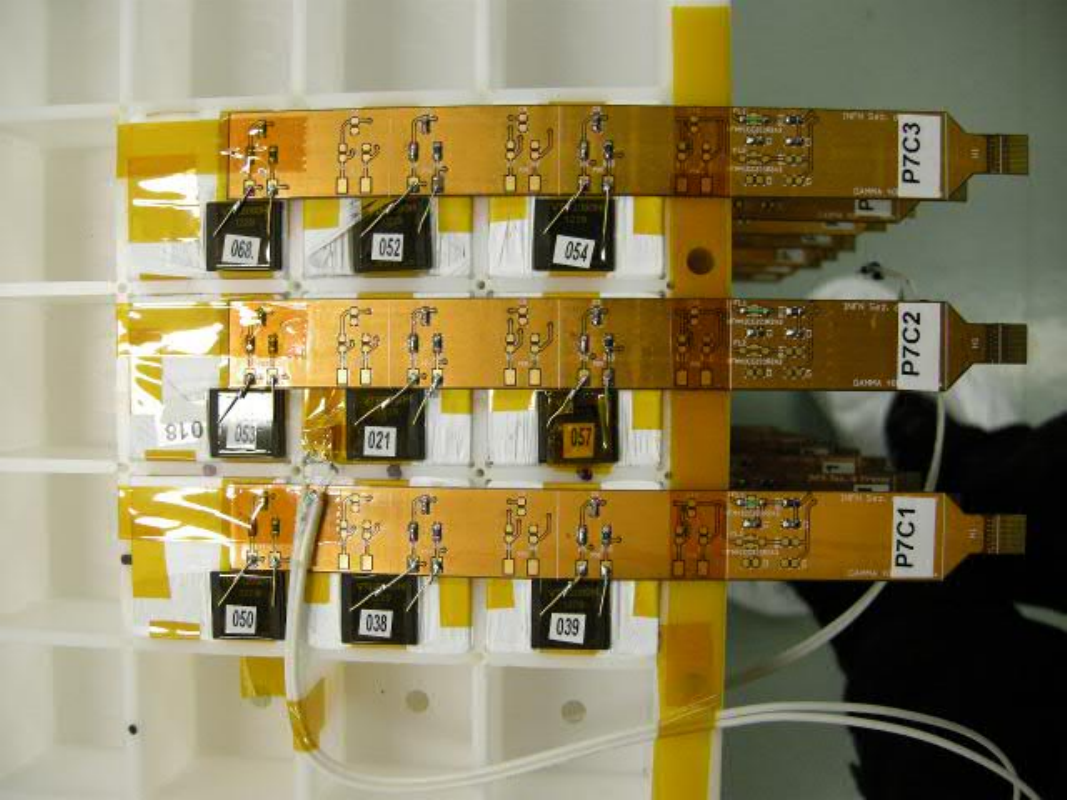}}
  \caption{\label{fig:CASIS03}Photo of a CASIS board with three  chips (a). The board was designed to allow the connection of the chips to the PDs of the scintillator crystals with flexible circuits (see text), which were threaded through holes in the PCB itself.  Front view of a prototype assembly with an early version (3+3 PDs) of the flexible circuits in full view with their tails ready for insertion (b).}
\end{figure}

For the first prototypes we designed and built a few boards using the CASIS chips (see figure  \ref{subfig:CASISboard})  which were coupled to the PDs through custom built flexible PCBs designed by us. A first version of these circuits are shown in figure \ref{subfig:delrintray}).
Once decided to opt for CsI(Tl) 36x36x36mm$^3$ crystals, we used the  polyoxymethylene trays described in section \ref{sec:mechanics} for the prototype mechanics. The crystals, wrapped in Vikuiti, are  loaded inside the trays with their PDs glued to them. 

\begin{figure}[htbp]
  \centering
  \subfigure[\label{subfig:scosch}]{\includegraphics[width=.8\textwidth]{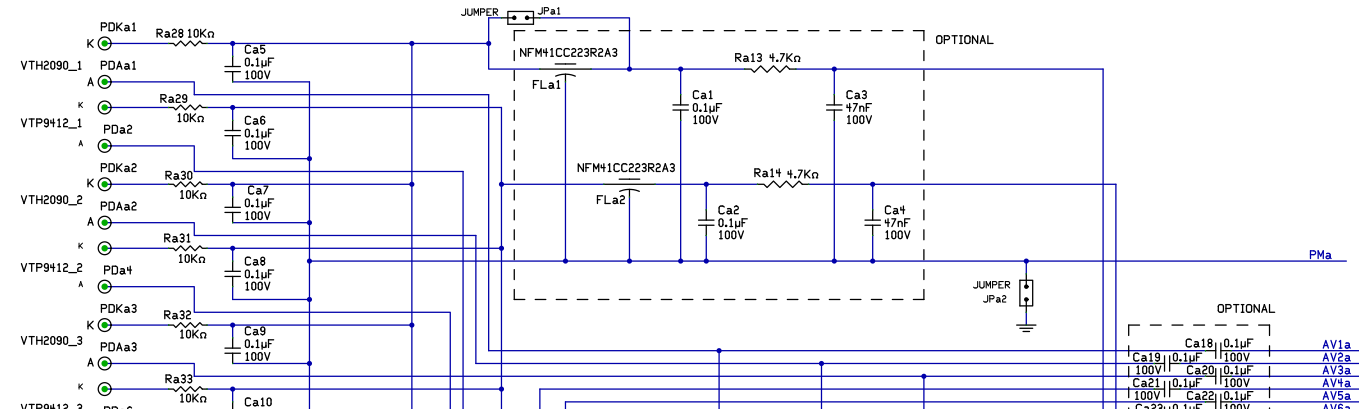}}
  \subfigure[\label{subfig:scolay}]{\includegraphics[width=.8\textwidth]{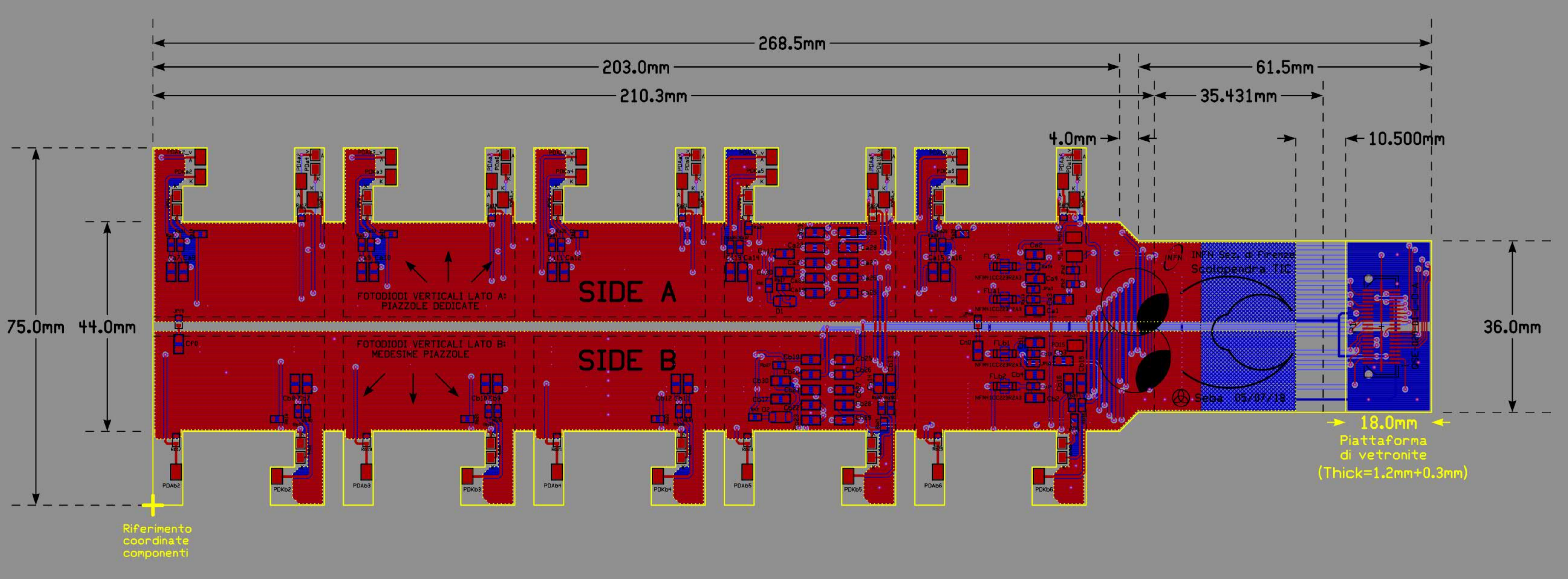}}
  \caption{\label{fig:CASIS03} Partial schematic of the flexible PD circuits (top). PCB layout of the components (below)}
\end{figure}

The connection to the F.E. electronics is realised with flexible PCBs, which provide not only a shielded connection from the PDs to the F.E. electronics, but have also on board components for the biasing and decoupling of the High Voltage used by the diodes. A partial circuit diagram that exemplifies this is shown in figure  \ref{subfig:scosch}.
The circuit referes to our latest implementation with each diode being biased through a 10K$\Omega$ resistor with a 100nF blocking capacitor placed in proximity of the PD itself. Two common biasing section are shared between 12+12 diodes, one for the large area and one for the small area ones. The anode signal from the diodes is sent to the F.E. electronics. The common reference is not ground but the 3.3 V power plane. This is due to internal details of the chip and is the configuration that minimises common mode noise (CMN) pick up. The component layout for our final version is  shown in figure  \ref{subfig:scolay} using a 10 + 10 design, where we kept two channels of the F.E. electronics connected to normal diodes soldered on the flex circuit, used exclusively for CMN subtraction. 
The total measured ADC value can be written as:
            \begin{equation}\label{ComNoise1}
              ADC_i^j=S_i^j+PED_i+CN^j
            \end{equation}
            where $i$ is the number of the CASIS cannel ($i=0,..., 15$), $j$ is the event number ($j=0,...,N$, where $N$ is a total number of acquired events), $S_i^j$ is a physical signal of the interacting particle, $PED_i$ is the pedestal of the channel $i$ and $CN^j$ is the {\itshape common noise} corresponding to the event $j$. Usually  $CN^j$ is calculated by using channels of the chip with no signal present (i.e. not invested by a shower signal).
\begin{figure}[htbp]
  \centering
  \subfigure[\label{subfig:ped}]{\includegraphics[width=.4\textwidth]{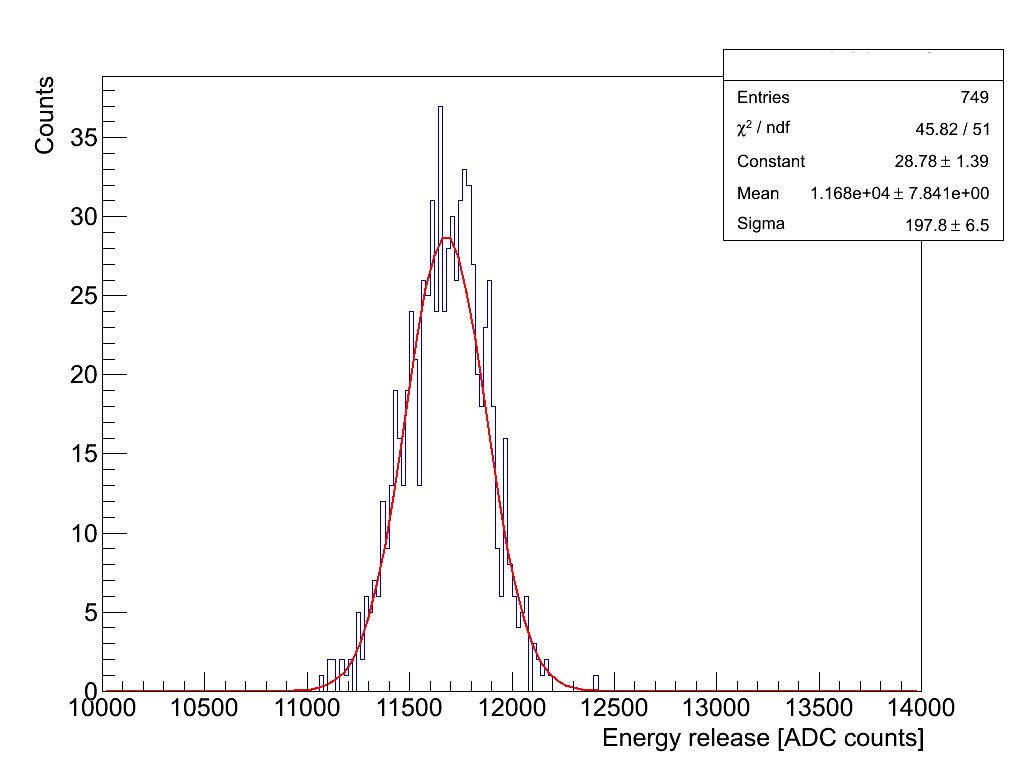}}
  \subfigure[\label{subfig:pedCMN}]{\includegraphics[width=.4\textwidth]{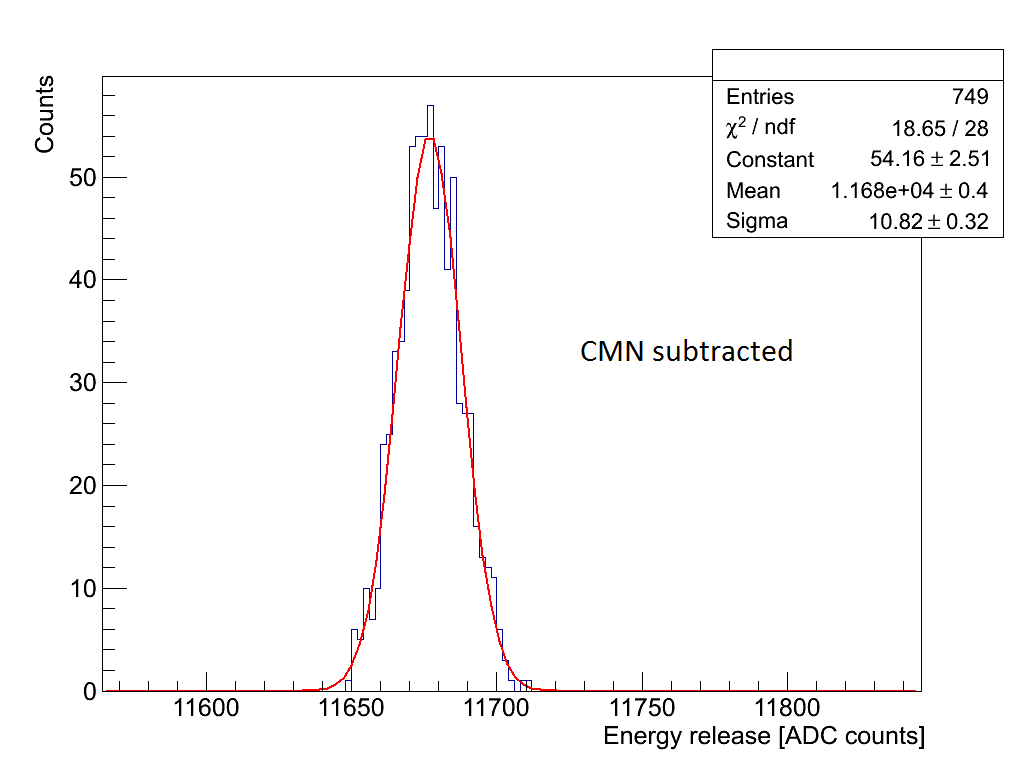}}
  \caption{\label{fig:CMNCASIS}Pedestal distribution for a typical CASIS channel connected to a PD. Before (a) and after CMN subtraction (b). The horizontal scales are different for the two plots. For comparison, with this setup, a MIP signal peaks at around 180 AC channels.}
\end{figure}
In a calorimeter this is not always the case and  it's the reason we have sacrificed at least one channel of each chip (even in the latest designs) to sample the  $CN^j$  value.
Figure  \ref{fig:CMNCASIS} shows the significant improvement obtained by subtracting the CMN on a typical  CASIS channel pedestal distribution. 

\begin{figure}[htbp]
\centering
\includegraphics[width=.7\textwidth]{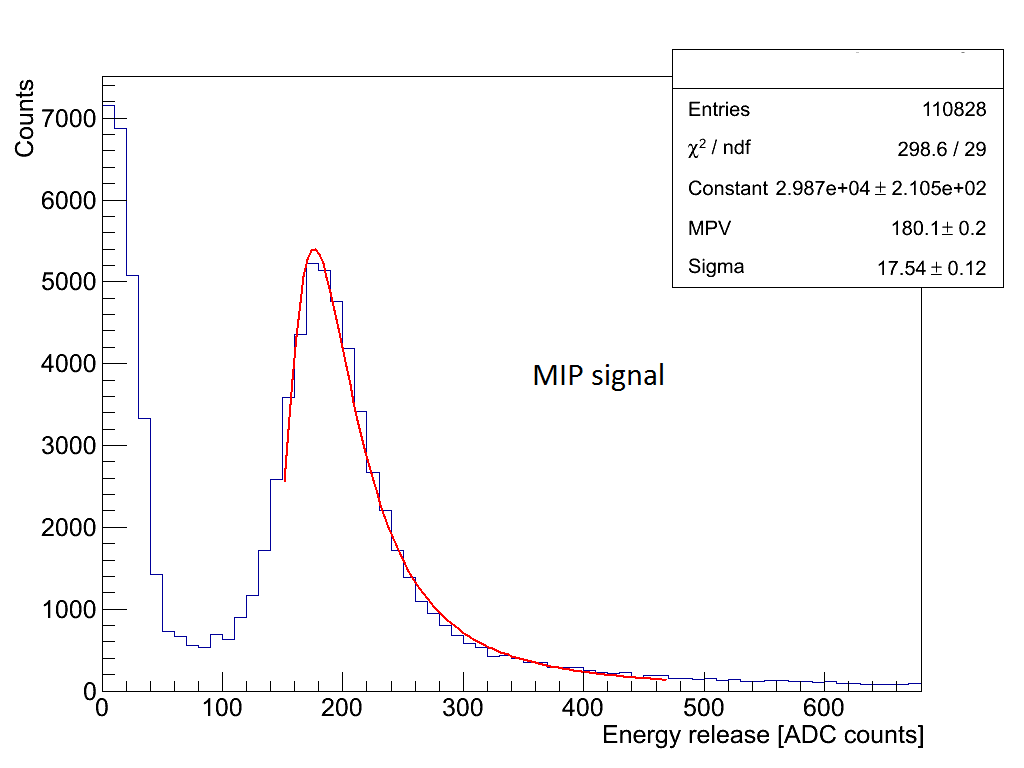}
\caption{\label{fig:MUONCASIS}   Cosmic ray spectra for one of the crystals inside a tray read out with the CASIS board after pedestal and CMN subtraction.}
\end{figure}

Since a MIP signal in our CsI(Tl) peaks at roughly 180 ADC channels (see figure \ref{fig:MUONCASIS}) this translates into an increase in the signal to noise ratio for MIPs, from 1 to roughly 18, which is more than enough to allow the use of MIPs for calibration purposes. This result also testifies to the excellent dynamic range achieved with this proptotype design and electronics. In fact, by changing CASIS settings and integration time sequences, the actual MIP signal can vary even by 50\% , while the signal to noise ratio remains more or less constant. A further improvement was obtained by using the Vikuiti wrapping (as stated previously) with increases in the signal up to 40\% . 

\subsection{Prototype Assembly and Readout}
\label{sec:ROC}
\begin{figure}[htbp]
  \centering
  \subfigure[\label{subfig:calotower}]{\includegraphics[width=.2\textwidth]{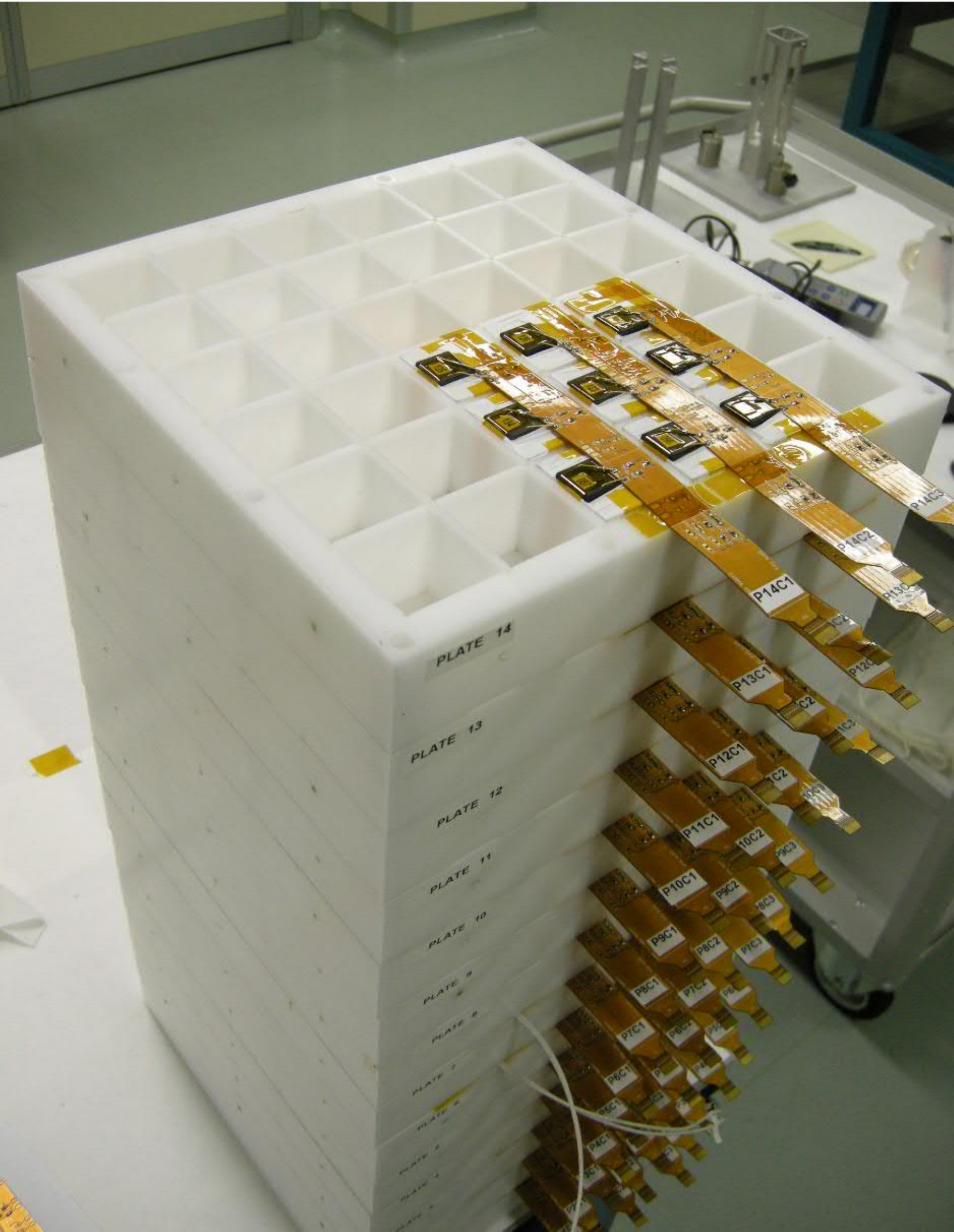}}
  \subfigure[\label{subfig:calotowerCASIS}]{\includegraphics[width=.6\textwidth]{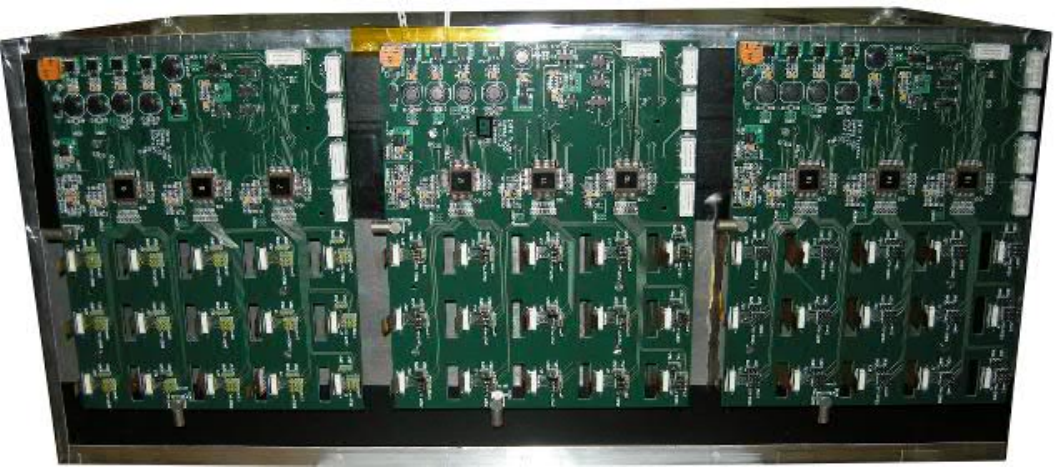}}
  \caption{\label{fig:proto01}The stack of trays each with 9 crystals (a). The assembled calorimter (126 crystals) read out by three CASIS boards placed on top of the stacked trays (b).}
\end{figure}

We began a first prototype construction by assembling 14 trays each one with a 3x3 crystal matrix (see figure \ref{subfig:calotower}), for a total of 126 crystals. 
We used a total of three CASIS boards for a total of nine Front End chips to readout the crystals. The whole was placed in an aluminium box that provided electrical and light shielding from the outside (see figure \ref{subfig:calotowerCASIS}.
 \begin{figure}[htbp]
  \centering
  \subfigure[\label{subfig:gainL}]{\includegraphics[width=.38\textwidth]{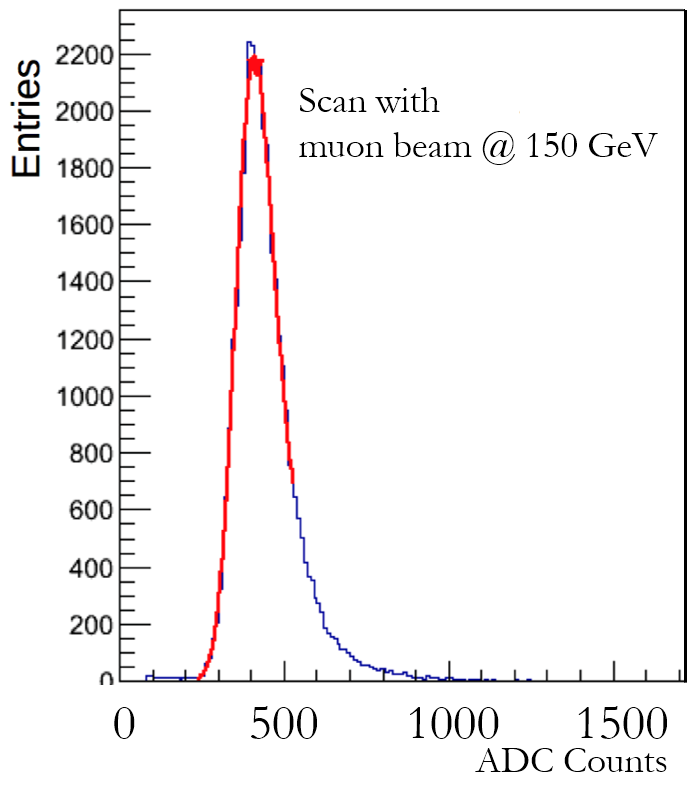}}
  \subfigure[\label{subfig:gaindispersion}]{\includegraphics[width=.55\textwidth]{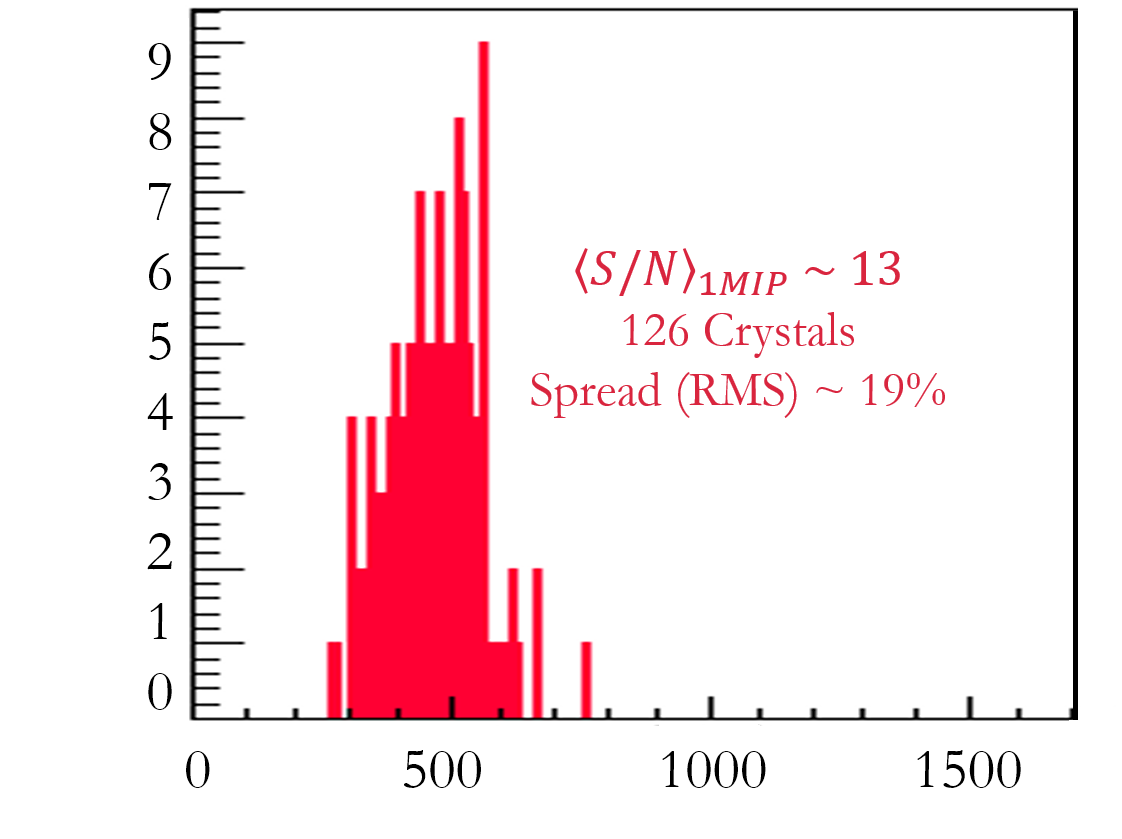}}
  \caption{\label{fig:gainmuon150} Typical crystal respons to a 150 GeV muon beam (a). The histogram (b) shows the ditribution of the peak values of the 126 Landau fits.}
\end{figure}

This first step was used by us to validate the whole assembly procedure and the PDs coupling to the electronics. We checked  the response to MIPs with the CASIS Front End (the analogue part has been kept in all subsequent developments, i.e. like the HIDRA chip mentioned later), and the uniformity of response of the wrapped crystals. Figure \ref{subfig:gainL} shows the excellent S/N ratio we have managed to obtain for MIPs, without sacrificing  the very high dynamic range of the calorimeter, proving that MIPs can be easily identified and used for calibration. Figure \ref{subfig:gaindispersion} shows the most probable values of the fitted Landau of each of the 126 crystals used in the first assembly. The dispersion is contsined within 19\% showing excellent reproducibility of the wrapping and optical coupling of the crystal to the PDs.

The F.E. electronics was driven by a back end board, the Read Out Controller (ROC) developed by us as a generic board  for DAQ applications. This board is based on a Xilinx Virtex 4 FPGA that can be programmed to provide the relevant signals needed by the F.E. and to transfer tha acquired data via fast USB (QuickUSB) or Ethernet links.
\begin{figure}[htbp]
  \centering
  \subfigure{\includegraphics[width=.45\textwidth]{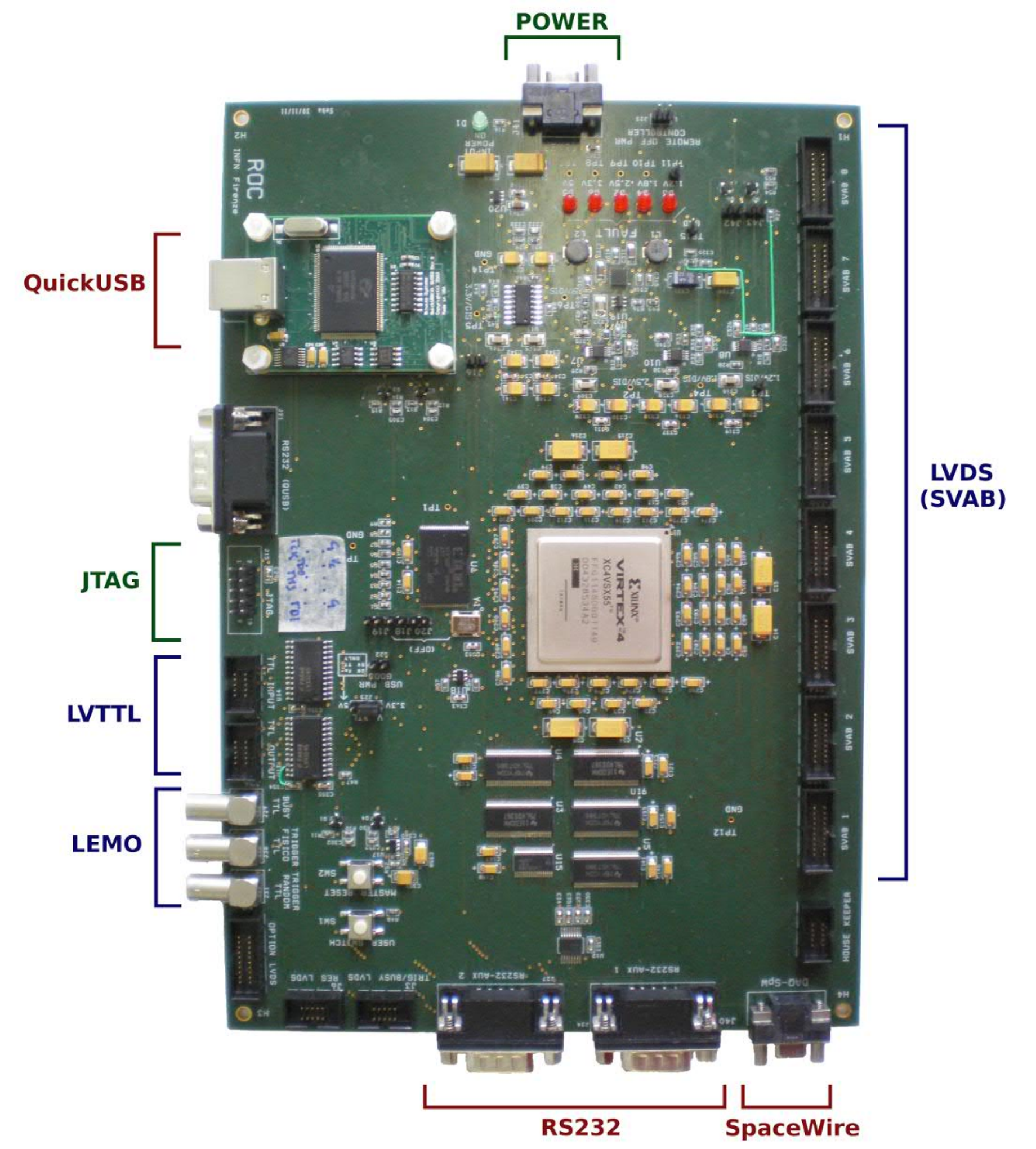}}~~
  \subfigure{\includegraphics[width=.45\textwidth]{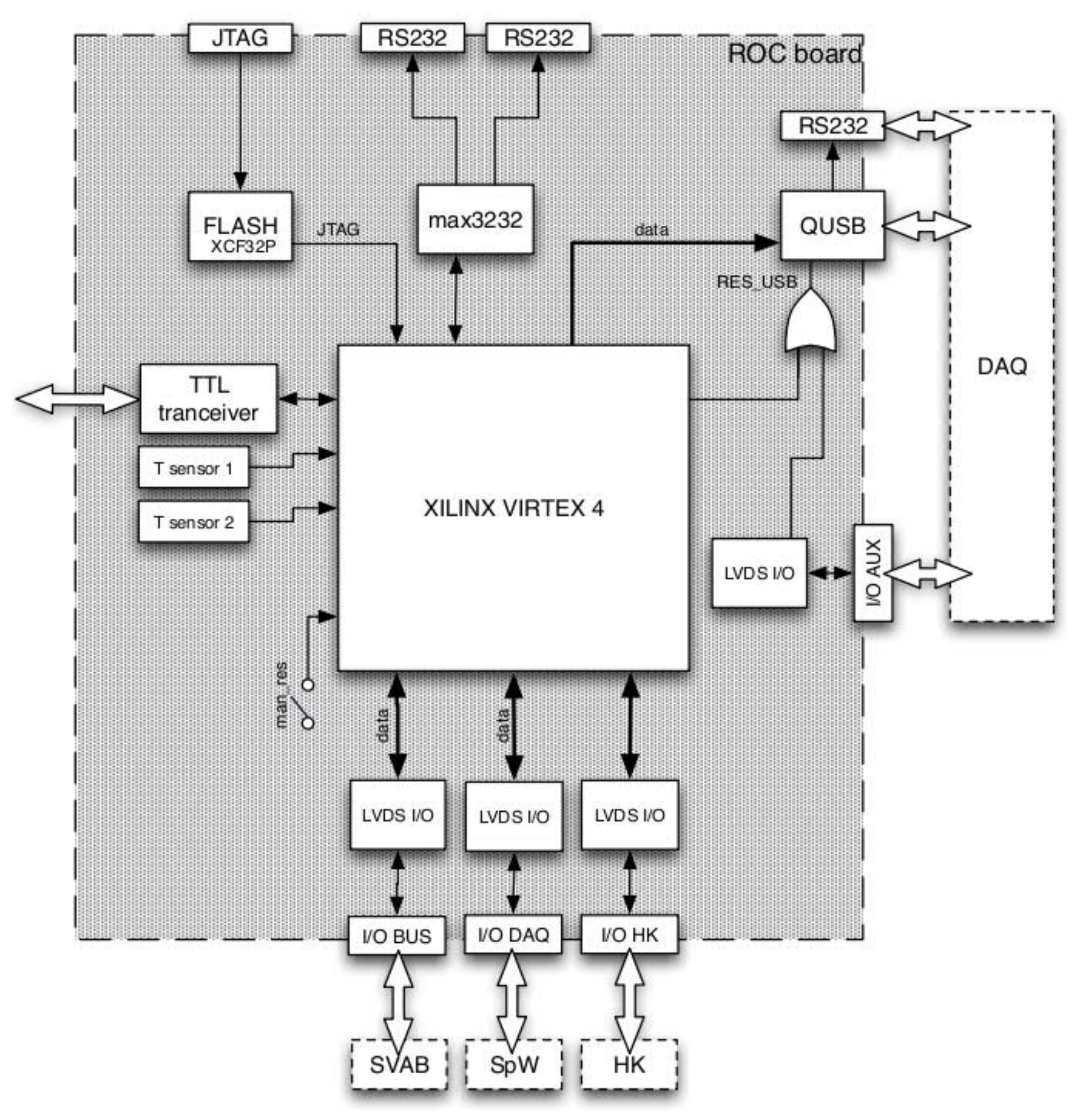}}
  \caption{\label{fig:proto01}Back end board (ROC) used for readout (left). Block diagram of the ROC (right).}
\end{figure}
The board of which a photo is shown in figure \ref{fig:proto01} has many LVDS I/O which we used to pilot the CASIS chips while we used a Quick USB piggy back to transfer the data from the CASIS ADCs to an external computer and storage. 
This board has been used by us for all the data taking and with all the Calocube prototypes including the final one with the HIDRA chips (an evolution of the CASIS design, next section), highlighting its extreme versatility. More information on the ROC board and the Finite State Machine used to drive the CASIS chips can be found in \cite{Berti01}.

\begin{figure}[htbp]
  \centering
  \subfigure[\label{subfig:muon}]{\includegraphics[width=.4\textwidth]{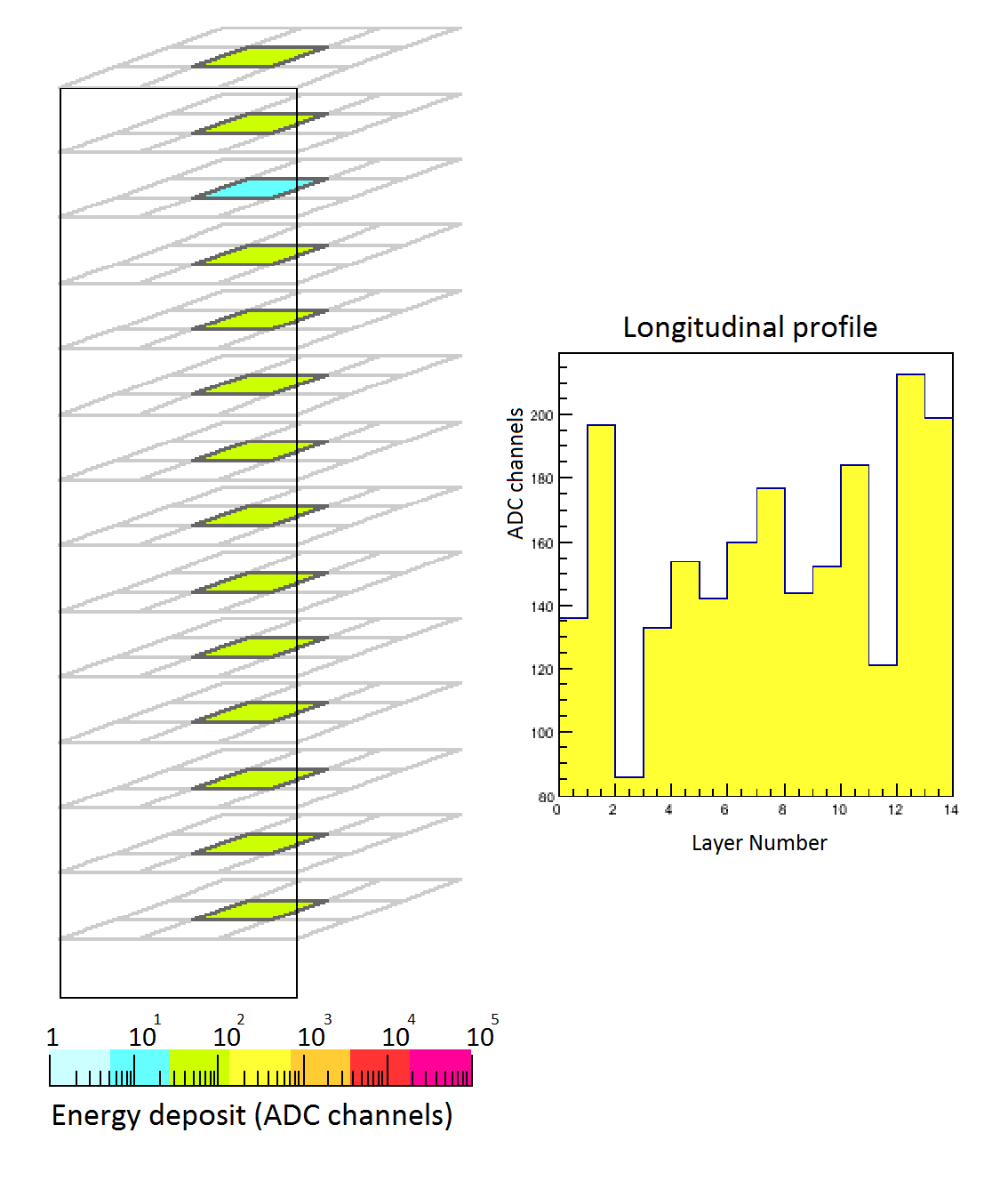}}
  \subfigure[\label{subfig:interactingMIP}]{\includegraphics[width=.4\textwidth]{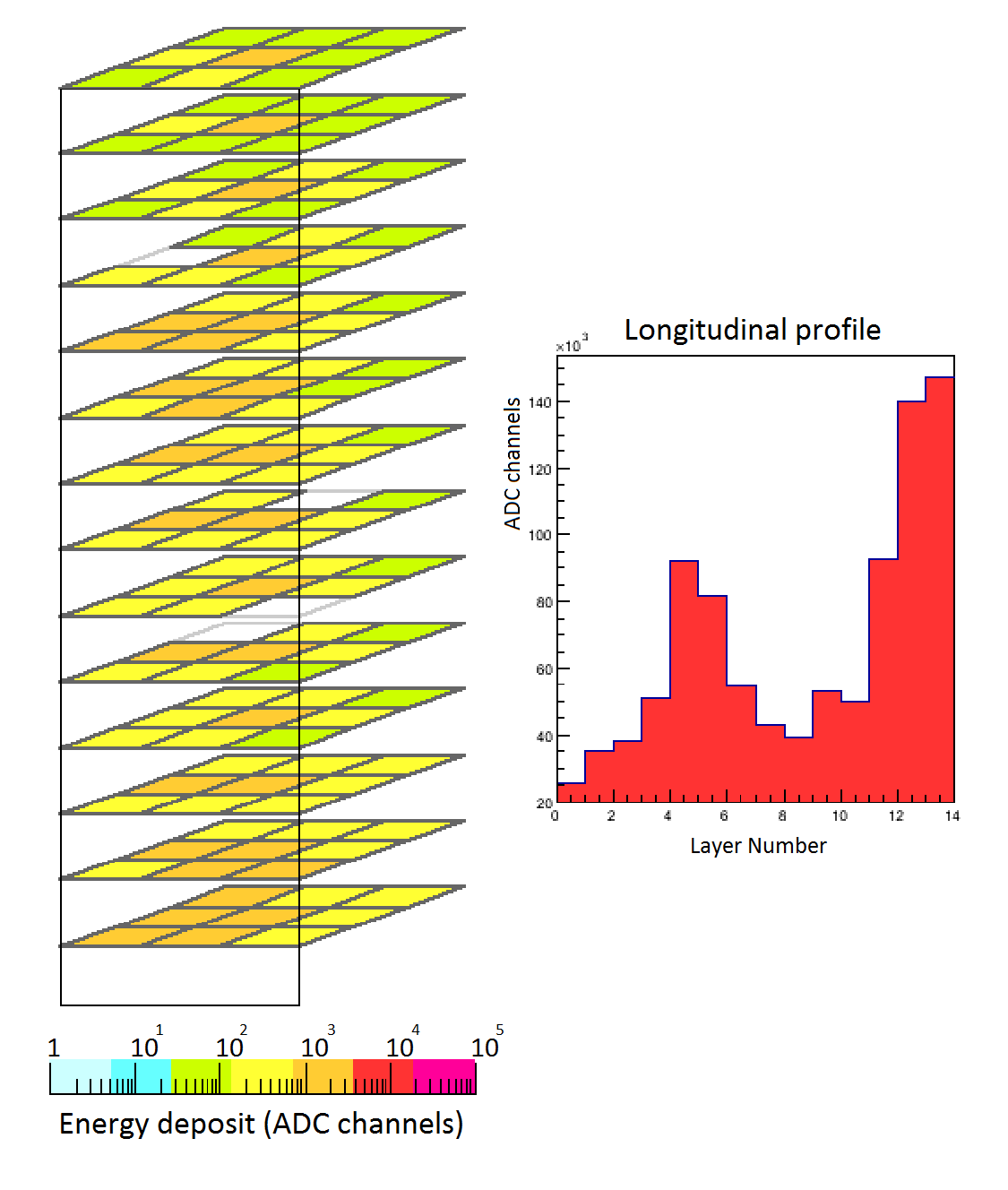}}
  \caption{Graphic visualisation of the reconstructed events in the Calocube prototype. A muon (a) and a proton (b). See text for details.}
  \label{fig:ev01}
\end{figure}

Finally we show (figure \ref{subfig:muon}, etc.) the imaging properties of the calorimeter highlighting some reconstructed events from the various test beams performed at SPS-CERN with Calocube. The pictures underline the ease with which particle discrimination can be performed. The detailed lateral and longitudinal profiles obtained with this detector allow us also to compensate for the energy leakage present in hadronic showers. All figures show the deposited energy for every crystal, with a longitudinal energy profile segmented layer by layer (14 in total,  corresponding to 27 X$_0$ and 1.3 $\lambda_I$).  The beam is aligned with the centre cube.
The first figure (\ref{subfig:muon}) shows the profile of a muon (a MIP) that produces a relatively low signal in all the centre cubes and only in them. The energy release is more or less the same from one layer to the other, with the small residual fluctuations due to the Landau process and differences in response for the different channels of the chips. Events of this type are in fact chosen for calibration purposes usually by fitting the Landau peak.  
The next figure  (\ref{subfig:interactingMIP}) shows an event where a MIP (a proton) interacts in the centre cube of  layer 4, inititating a shower in the next layers.  The shower develops rapidly and releases a significant amount of energy in the second part of the calorimeter. The interesting fact is that there is a very visible, high energy release aligned with the centre cubes in layers 4 to 7.  This behaviour is compatible with a formation, within the hadronic shower, of an electromagnetic shower caused by a $\pi^0$ decay.
\begin{figure}[htbp]
  \centering
  \subfigure[\label{subfig:inter1}]{\includegraphics[width=.4\textwidth]{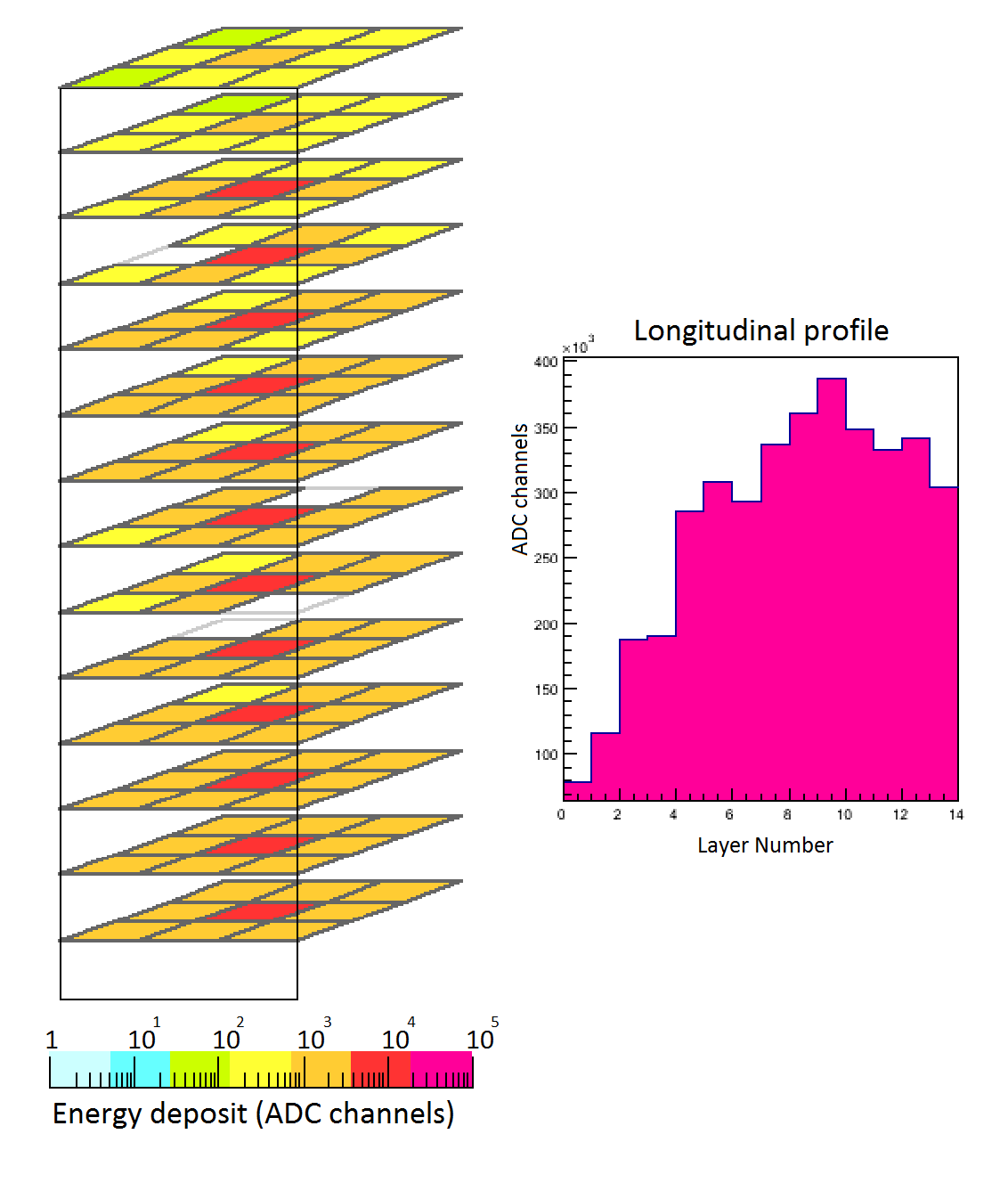}}
  \subfigure[\label{subfig:inter2}]{\includegraphics[width=.4\textwidth]{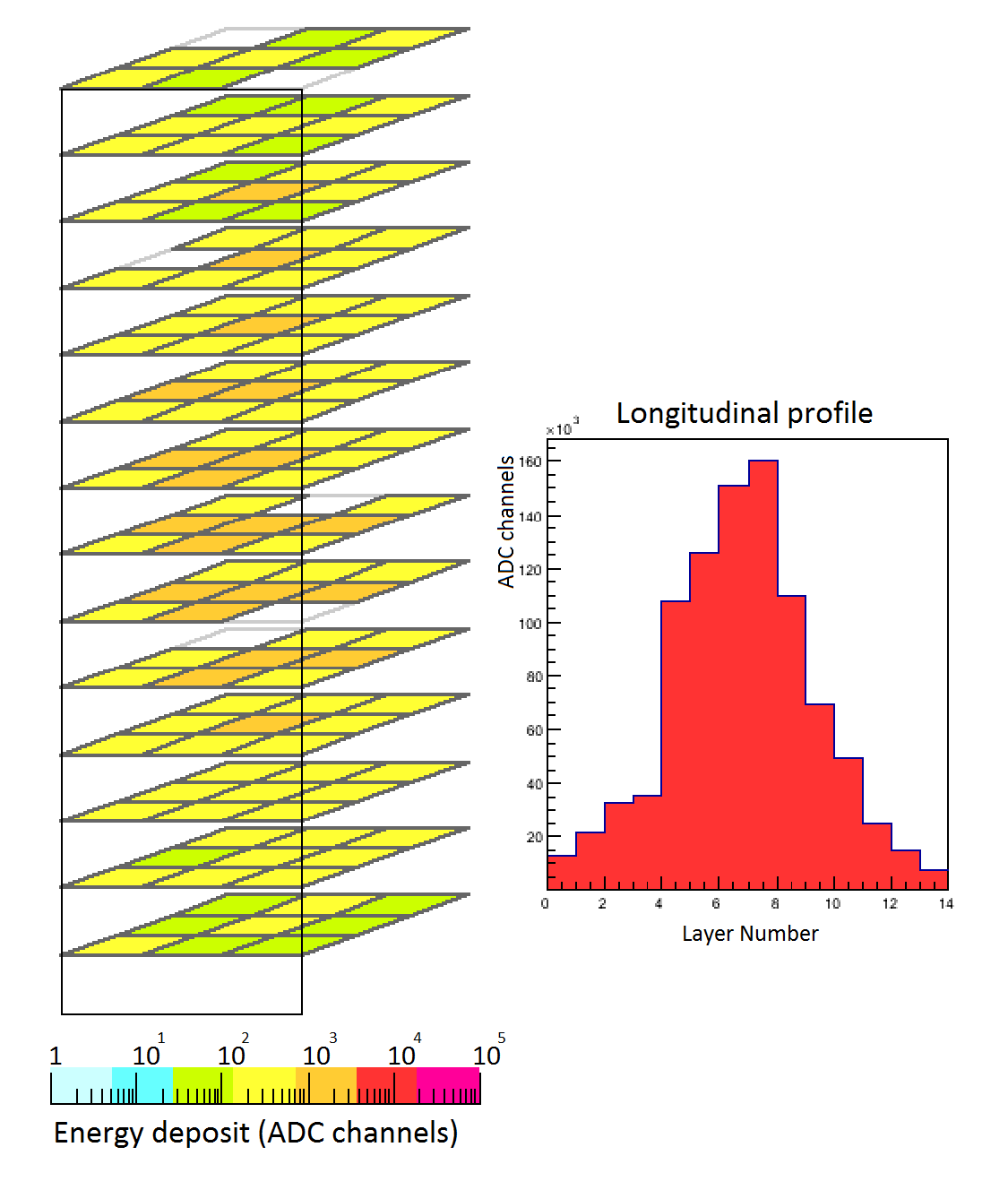}}
  \caption{Graphic visualisation of the reconstructed events in the Calocube prototype. A proton (a) and an electron (b). See text for details.}
  \label{fig:ev02}
\end{figure}  
The next two images (figures \ref{subfig:inter1} and \ref{subfig:inter2} ) are of two particles that interact straight away in the first layer. The first one has a lateral profile typical of a hadronic shower.  The shower development is relatively slow and reaches the maximum between layers 9 and 10, where the maximum deposited energy is observed. An electromagnetic component is distinguishable from the significant energy release in the central row of the crystals between layers 2 and 14. The last one is an electron shower profile that is fully contained.

%           Fig.~\ref{figBerti5.13}b shows hadronic shower with a very strong electromagnetic part generated by $\pi_0$ decay.            One can see a very strong and energetic core along the beam direction that runs from layer 2 up to layer 13.            This electromagnetic component provides intensive and almost constant energy release in the rear layers of the calorimeter.           As described in $\S$ \ref{chapCalCub}, the possibility of longitudinal profile reconstruction is very important for the scientific aims of the CaloCube project.           Together with the charge identification system it allows to reconstruct the energy of the primary interacting nucleus.           In the measurements described above, the charge identification provided by CALET tracker allows to perform this study with the acquired data.

\subsection{Current and final development}

\begin{figure}[htbp]
  \centering
  \subfigure{\includegraphics[width=.33\textwidth]{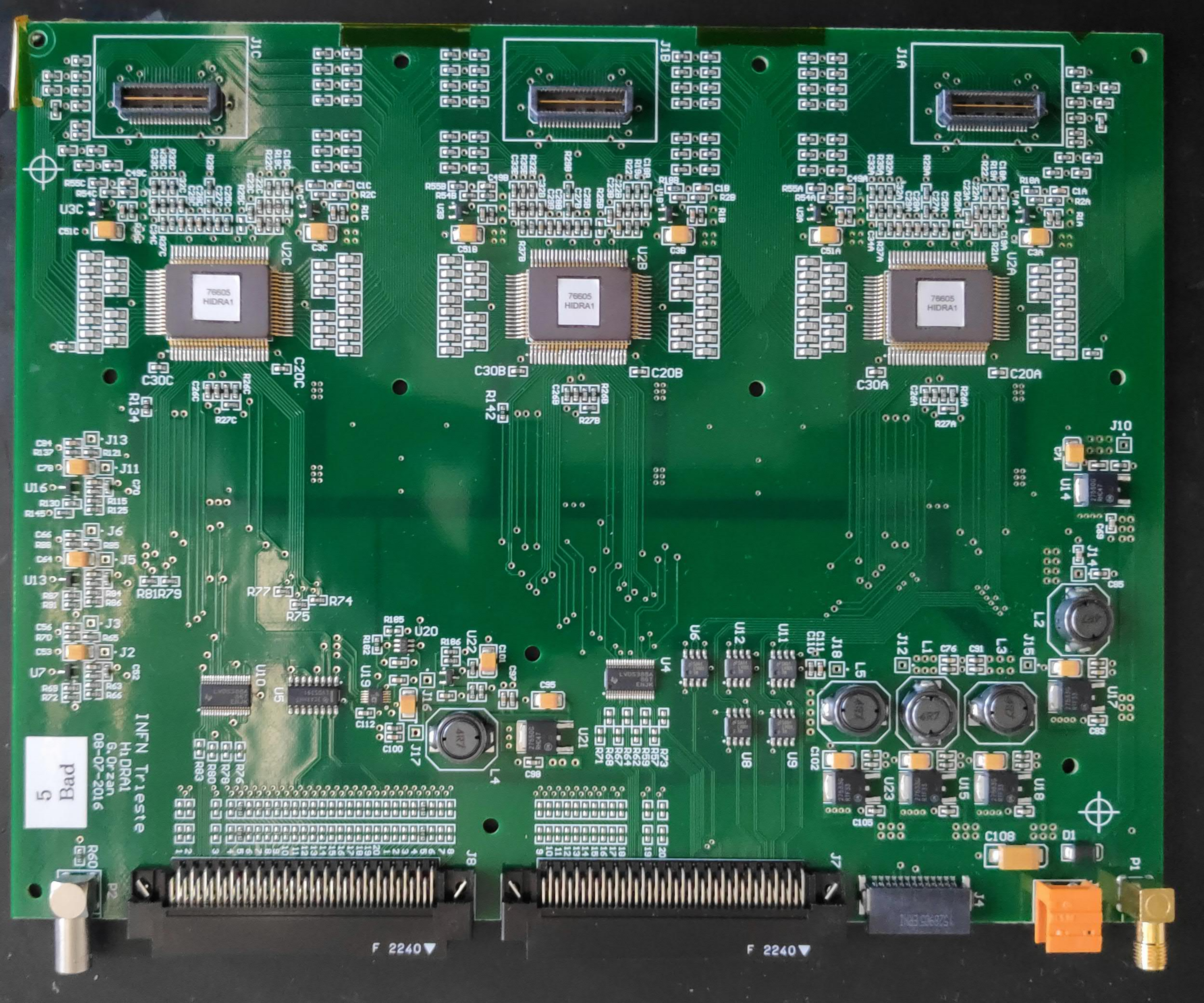}}~~
  \subfigure{\includegraphics[width=.49\textwidth]{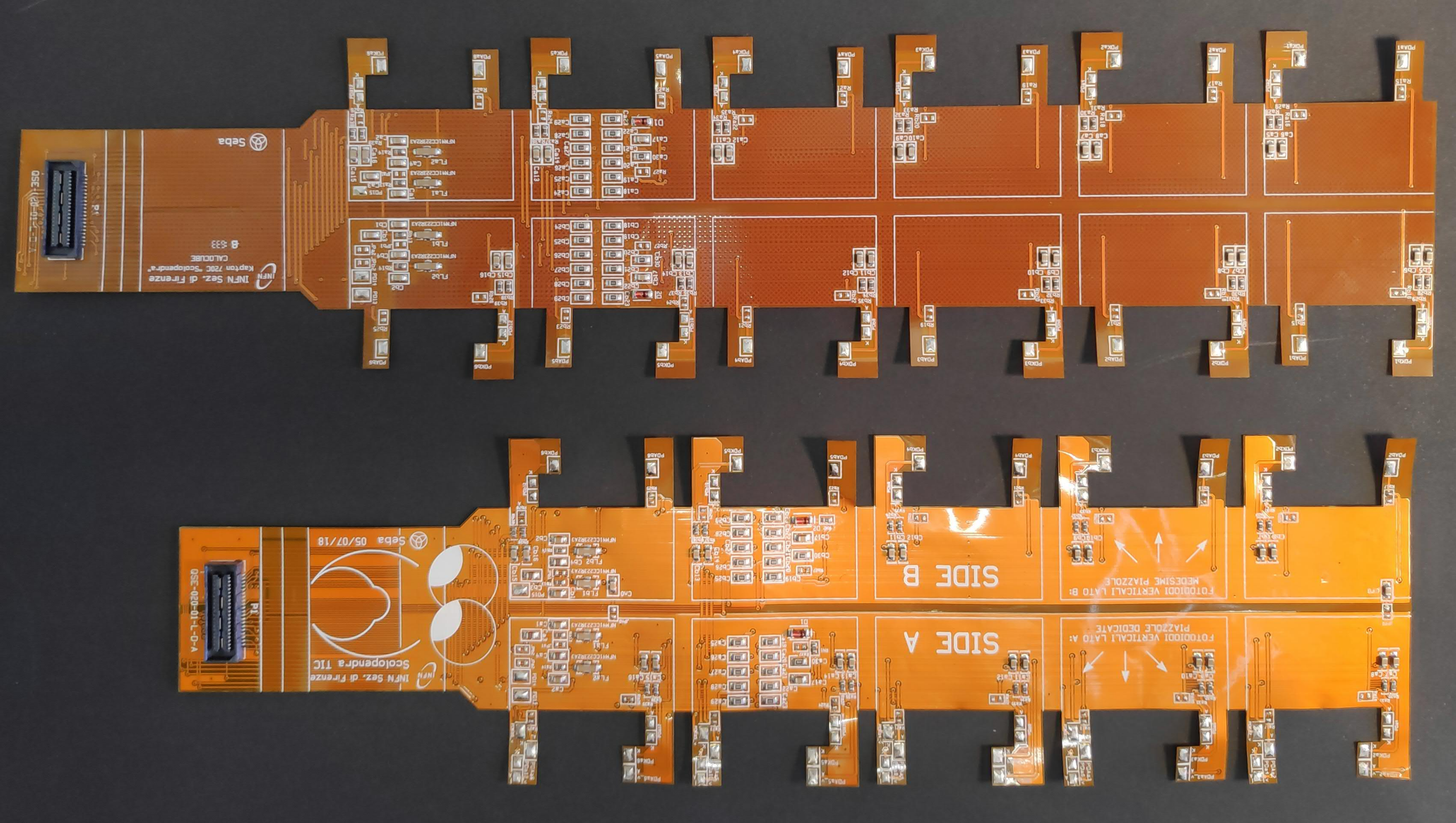}}
  \caption{\label{fig:CASIS04}The new HIDRA board (left) and the flexible circuits that connect the crystals to it (right). Each tray as explained in section \ref{sec:mechanics} has bolted on an aluminum section to which a HIDRA board is fixed.}
\end{figure}

\begin{figure}[htbp]
  \centering
  \subfigure[\label{subfig:calobello}]{\includegraphics[width=.55\textwidth]{Images/derlin_tray_assembly_Optimized.pdf}}~
  \subfigure[\label{subfig:calobellobello}]{\includegraphics[width=.3\textwidth]{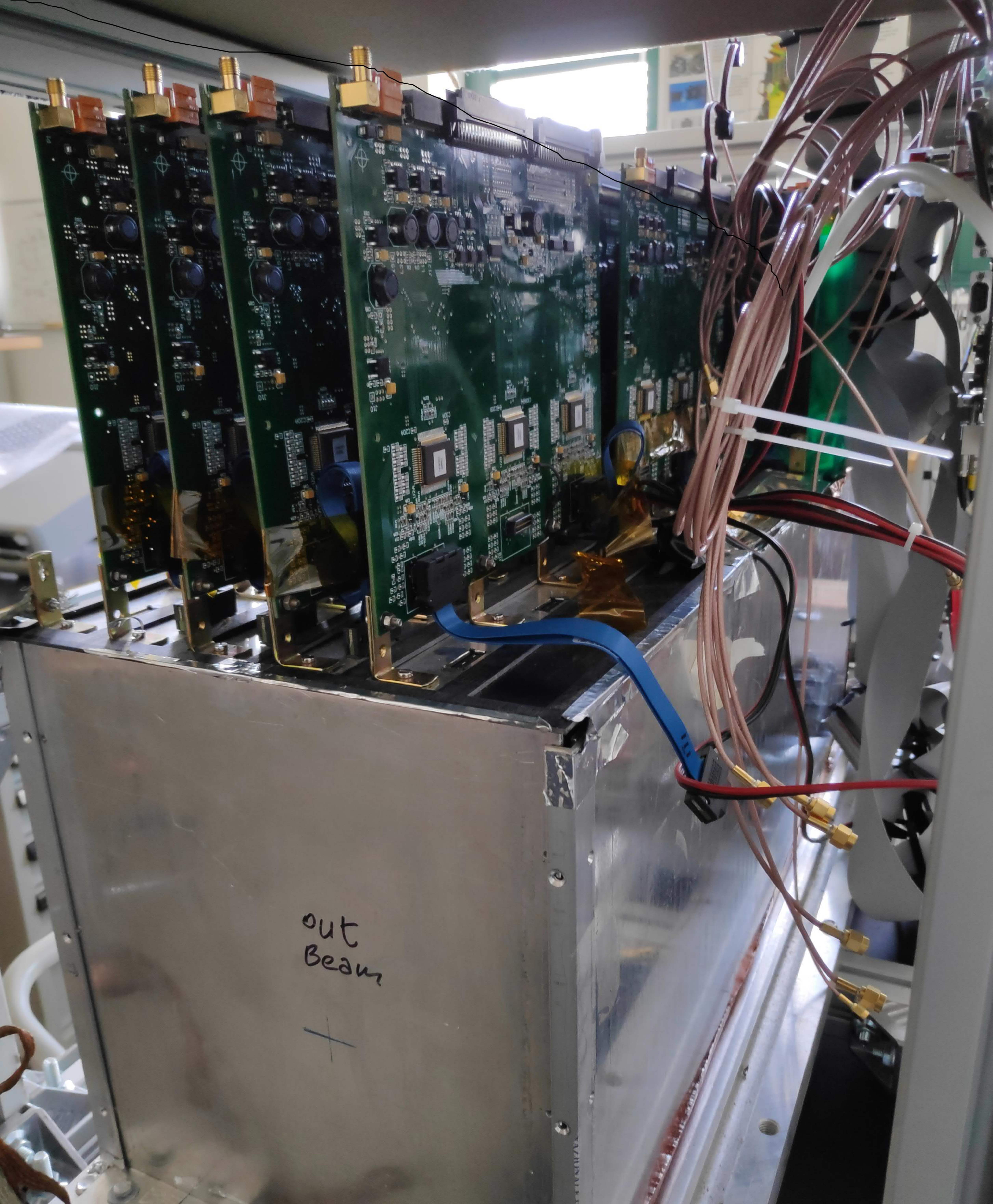}}
  \caption{\label{fig:KALOS}The final assembled Calocube prototype with the latest F.E.  boards (a). A detail showing an intervention taking place and the modular structure of the final prototype (b).}
\end{figure}

The CASIS chip itself has now evolved in a new version called HIDRA. The analogue part is the same but we have further improved its behaviour and have added self triggering capabilities and logical output registers for the trigger logic. With this new chip we have realised new boards (see figure  \ref{fig:CASIS04}) and flexible connection circuits for our final Calocube prototype.
The final prototype design uses the same mechanical trays but has a more modular approach with one F.E. board serving only one tray. This allows easier intervention on the prototype whenever repairs or substitutions need to be done on the crystals in the trays.
\begin{figure}[htbp]
  \centering
  \includegraphics[width=.5\textwidth]{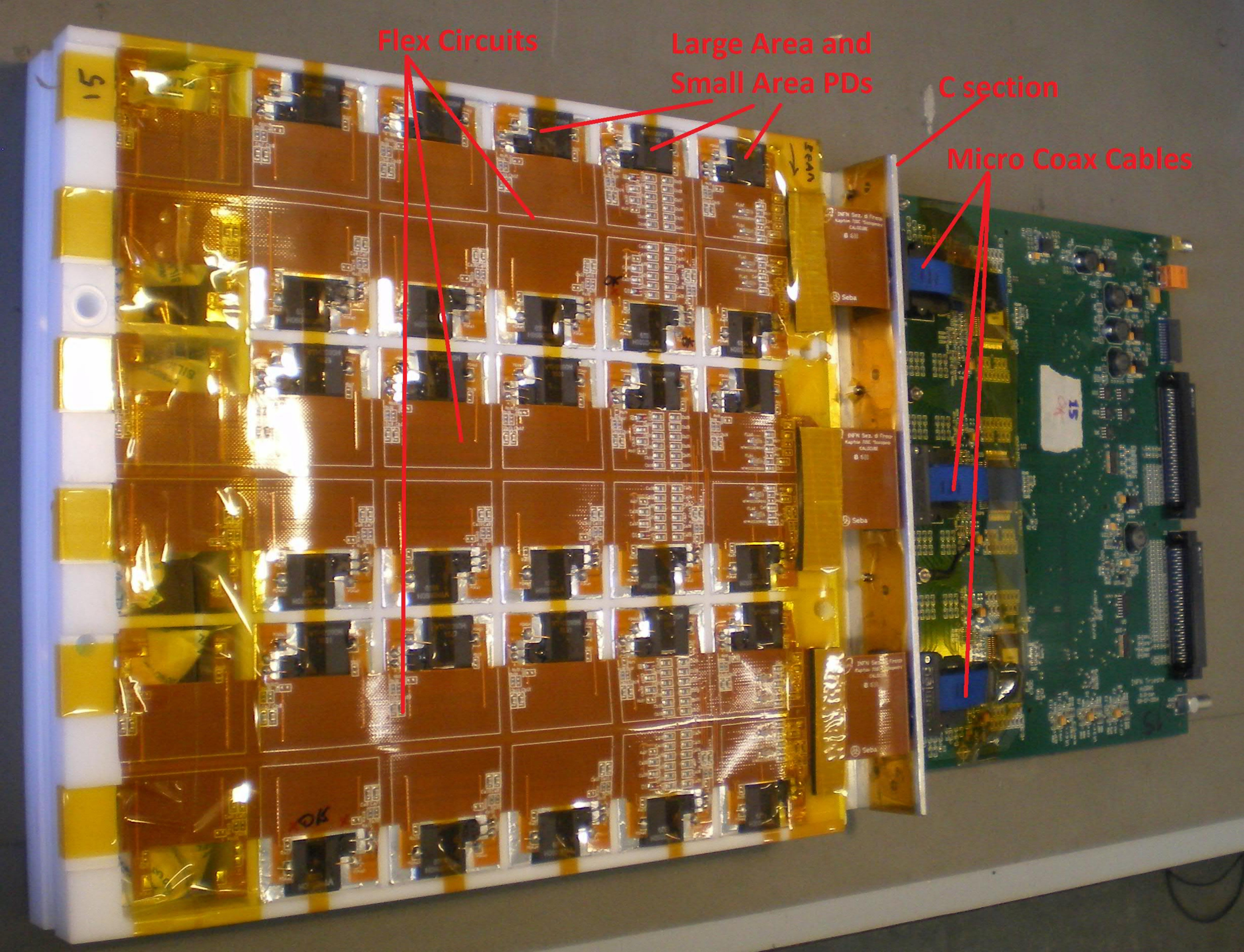}
  \caption{\label{fig:newtray}The current tray assembly. A C section aluminium piece holds the HIDRA board to the tray. Final connections between the Flex circuits and the F.E. board are realised with  SAMTEK microcoax cables.}
\end{figure}
 This is exemplified by the photo of a fully equipped tray shown in figure \ref{fig:newtray}. The 18 fully equipped trays (36 crystals each) for a total of 648 CsI(Tl) crystals constitute the larger Calocube prototype with which data has been taken with protons, pions, electrons, muons, and above all ions at various test beam facilities. The Calocube prototype (figure \ref{fig:KALOS}) even though finished is still an object of R\& D which we modify accordingly to test new ideas or concepts. Currently we are modifying the front part to make place for a few layers of silicon tracking modules to test a gamma ray application evolution (TIC, Tracking In Calorimeter~\cite{TIC}) that would still be capable of performing cosmic ray physics in orbit mantaining the highest performance for charged particles, while allowing accurate directional gamma ray physics down to 1 GeV.  

\subsection{Dual readout studies}
\label{sec:DualR}
Although as stated previously, detailed results from the various test beams we performed belong to a future paper, we end this section reporting on a very interesting calorimetry technique that we have tested with CsI(Tl) crystals during our prototype development. This is a self contained project within our R\&D that has shown promise but is currently too expensive to pursue further.

\begin{figure}[htbp]
  \centering
  \subfigure[\label{subfig:Dream1}]{\includegraphics[width=.46\textwidth]{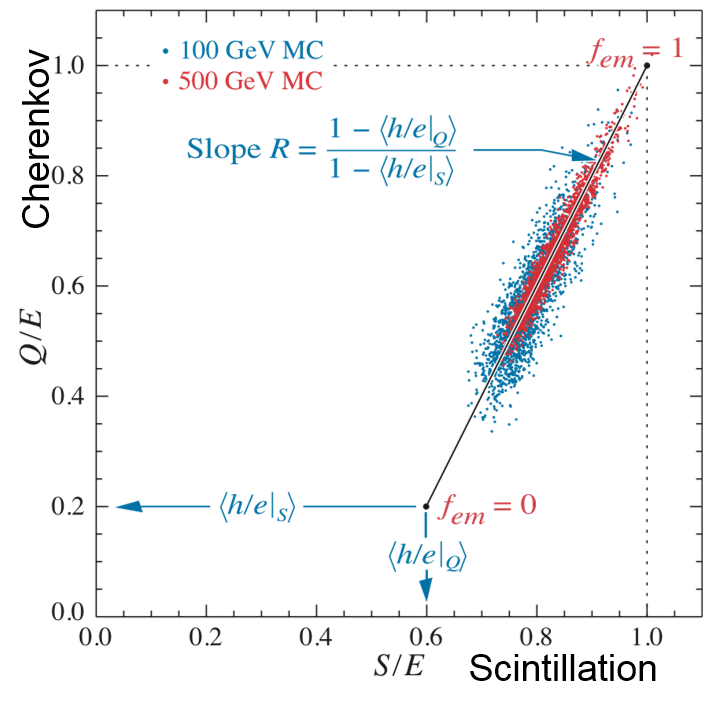}}
  \subfigure[\label{subfig:Dream2}]{\includegraphics[width=.43\textwidth]{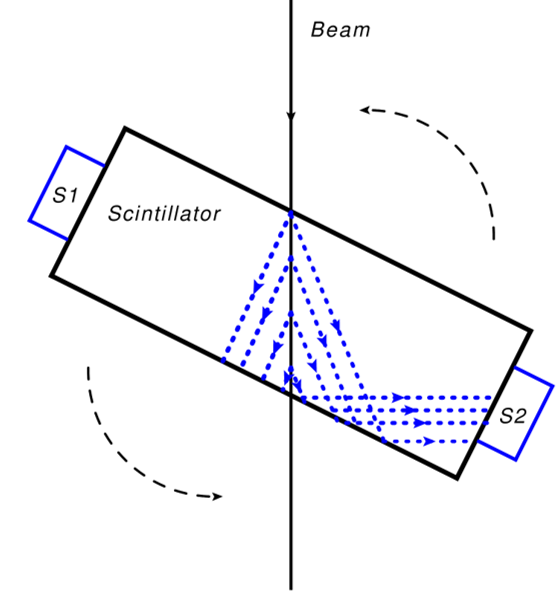}}
  \caption{\label{fig:Dream}Simulations of hadronic showers at two different energies. The figure (a) shows the scintillation signal vs Cherenkov signal scatter plot. The extra information provided by the Cherenkov light helps reducing overall fluctuations in the calorimeter signal. Schematic of  the scintillator test setup, with a PM at each end for readout, showing the principle of operation for Cherenkov light separation (b).}
\end{figure}

One of the main causes of the hadron energy measurement low resolution, are the fluctuations in the fraction of E.M. showers within the hadronic shower. This fraction not only fluctuates but also depends on the energy of the incident hadron. A long standing effort by the DREAM collaboration~\cite{Dream1,Dream2} has shown that a Dual Readout paradigm, where both scintillation and Cherenkov light are read out thorough two independent channels, can be used to correct the calorimeter response on an event basis. This can bring about a dramatic improvement in resolution at the price of a more complex readout scheme.  Basically while the scintillation light is sensitive to both hadronic and E.M. shower components, Cherenkov radiation is mainly produced in E.M. showers. Figure \ref{subfig:Dream1} shows the correlation between the two signals in simulations of 100 GeV and 500 GeV protons. %~\cite{dreamcorr} 
When both signals are available, the overall signal fluctuation (Cherenkov vs Scintillation, given by the projection on the line of slope R, see figure) is much reduced respect to fluctuations present in each signal.

We performed various tests to see if we could tease out the Cherenkov signal from CsI(Tl) scintillator. 
Due to the physics process underlying the two light production mechanisms, there are important differences in the spectra and time development of the two components that can be exploited to measure each one separately.
In fact Cherenkov photons are produced  concurrently with the particle passage through the crystal and as such last only a ns or less, also Cherenkov light peaks at short wave lengths in the ultraviolet region and can be easily separated from the green peaking scintillation one. We tested these assumptions with a special setup shown in figure~\ref{subfig:Dream2}, where a scintillator bar (CsI(Tl) is read out at its ends with two fast, UV sensitive phototubes.
Since Cherenkov is directional along the particle path, turning the crystal respect to the beam favours one PM respect to the other. Scintillation light on the other hand is isotropic and as such no dependence on the angle should be seen on the two PMs.
\begin{figure}[htbp]
  \centering
  \subfigure[\label{subfig:Dream3}]{\includegraphics[width=.42\textwidth]{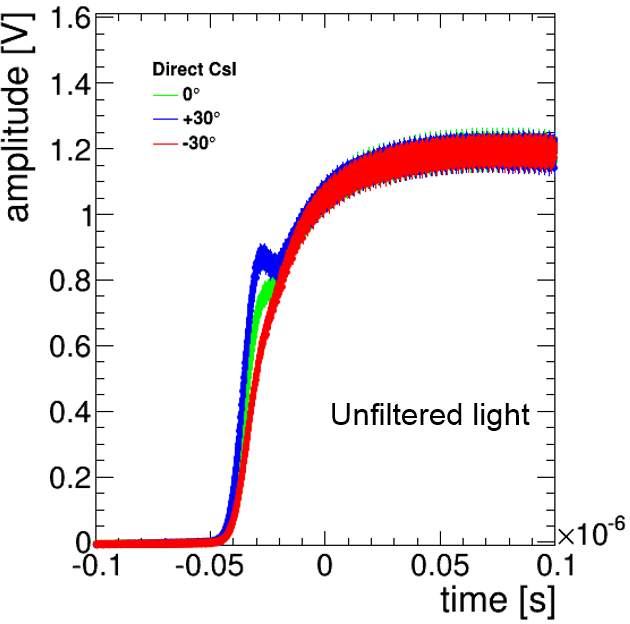}}
  \subfigure[\label{subfig:Dream4}]{\includegraphics[width=.39\textwidth]{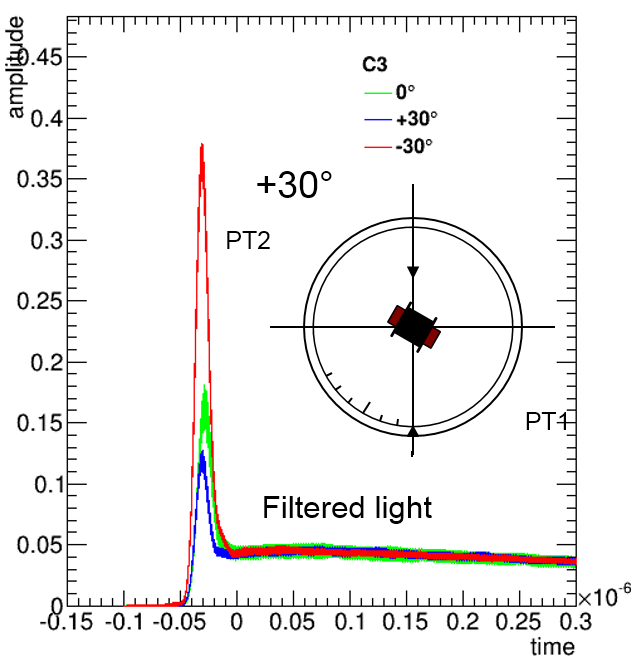}}
  \caption{\label{fig:Dream34}CsI(tl) signals acquired with a fast digitizer (a). The three angles refer to the crystal rotation respect to the beam axis that can increase or decrease the Cherenkov light on the PM. Cherenkov light is evident in the first few ns but is anyway swamped by the much more abundant scintillation component. If a filter blocking visible light is applied in front of the PM the Cherenkov component becomes prominent (b).}
\end{figure} 
As shown in figure~\ref{fig:Dream34} we have managed to find clear evidence for Cherenkov light production in CsI(Tl) crystals. The Cherenkov signal is much higher when the crystal is rotated at 30 degrees and is evident only in the first few ns. Unfortunately, even though CsI(Tl) scintillation light has a time constant of 600ns, its output is so abundant that it anyway swamps the Cherenkov photons. Using interference filters, we have managed to block the visible light (i.e. scintillation) leaving only the UV component. Figure~\ref{subfig:Dream4} shows how the signal behaves exactly as a directional Cherenkok light would behave when the crystal is rotated. 

As a final remark on Dual Readout, we would like to point out that while the technique is very promising for a full sized calorimeter, we have shown (with simulations) in \cite{calosimu} that the relative improvement in energy resolution that can be expected for a typical CaloCube depth of 2 $\lambda_I$ is only 10\%. This is to be compared to a 30\% or more improvement that can be obtained for a full containment calorimeter. The reason for such a difference, is to be found in the large longitudinal leakage and
its relative fluctuations which become very significant for a thin calorimeter and that dominate the energy resolution performance.

\section{Conclusions}

The Calocube project has lasted more than four years. During this period, we designed and constructed a prototype calorimeter to study its performance in view of a possible future space experiment. We have demonstrated that a highly segmented calorimeter design can be built using inorganic crystal scintillators and a photodiode readout. Our high segmentation design allows for excellent particle ID and energy resolution, while increasing significantly the geometrical acceptance of the detector. 

The project is now evolving both as a full scale space proposal (HERD), and as a test bench to experiment novel ideas for space based cosmic ray detectors (LAPUTA and TIC projects financed by INFN). The test beam data has been shown at various meetings and conferences and will be published in the near future by our collaboration.

% \appendix
% \section{Appendi(cite)}
% Please always give a title also for appendices.

\acknowledgments
We wish to acknowledge the financial support of INFN, in the framework of the CSN5 2013 Calls, that funded the Calocube project.

%\paragraph{Note added.} This is also a good position for notes added
%after the paper has been written.

% We suggest to always provide author, title and journal data:
% in short all the informations that clearly identify a document.

\end{document}